\documentclass[letterpaper,12pt,numbers,3p,sort&compress]{elsarticle}

\usepackage{siunitx}
\usepackage{amsmath}
\usepackage{mathtools}
\usepackage{graphicx}
\usepackage{wasysym}
\usepackage{amssymb}
\usepackage{array, multirow, bigdelim, makecell, booktabs} 
\usepackage{bm}% bold math
\usepackage[normalem]{ulem}

\usepackage[colorlinks=true, allcolors=blue]{hyperref}
\usepackage{tikz}
\usepackage{color}
\usepackage{float}
\usepackage{subfigure}

\definecolor{niceblue}{rgb}{0.388235, 0.627451, 0.847059}
\definecolor{Zpurple}{RGB}{119, 50, 168}
\definecolor{nicered}{rgb}{0.7,0.1,0.1}
\definecolor{nicegreen}{rgb}{0.1,0.5,0.1}
\definecolor{JMBcomment}{RGB}{28, 186, 70}

\makeatletter
\def\ps@pprintTitle{%
 \let\@oddhead\@empty
 \let\@evenhead\@empty
 \def\@oddfoot{}%
 \let\@evenfoot\@oddfoot}
\makeatother

\begin{document}

\allowdisplaybreaks

\renewcommand{\arraystretch}{1.5}

\begin{frontmatter}
\title{Neutron Stars with Baryon Number Violation, \\ Probing Dark Sectors}

\author[1,2]{Jeffrey M. Berryman}\ead{jeffberryman@berkeley.edu}
\author[3]{Susan Gardner}\ead{gardner@pa.uky.edu}
\author[3]{Mohammadreza Zakeri}\ead{m.zakeri@uky.edu}

\address[1]{Department of Physics, University of California, Berkeley, CA 94720, USA}
\address[2]{Institute for Nuclear Theory, University of Washington, Seattle, WA 98195, USA}
\address[3]{Department of Physics and Astronomy, University of Kentucky, Lexington, Kentucky 40506-0055 USA}

%%%%%%%%%%%%%%%%%%%%%%%%%%%%%%%%%%%%%%%%%%%%%%%%%%%%%%%%%%%%%%%%%%%%%%%%%%%%%%%
\begin{abstract}
The neutron lifetime anomaly has been used to motivate the introduction of new physics with hidden-sector particles coupled to baryon number, and on which neutron stars provide powerful constraints. Although the neutron lifetime anomaly may eventually prove to be of mundane origin, we use it as motivation for a broader review of the ways that baryon number violation, be it real or apparent, and dark sectors can intertwine and how neutron star observables, both present and future, can constrain them.
\vspace{5mm}

\noindent Preprint numbers: INT-PUB-22-001, N3AS-22-004

\end{abstract}

\end{frontmatter}
%%%%%%%%%%%%%%%%%%%%%%%%%%%%%%%%%%%%%%%%%%%%%%%%%%%%%%%%%%%%%%%%%%%%%%%%%%%%

%%%%%%%%%%%%%%%%%%%%%%%%%%%%%%%%%%%%%%%%%%%%
\section{Introduction}
\label{sec:introduction}
%%%%%%%%%%%%%%%%%%%%%%%%%%%%%%%%%%%%%%%%%%%%

 stars are remarkable for their very existence: they are the densest objects known in the observable Universe. With an upper mass limit of about $2\,M_{{\astrosun}}$ and a typical radius of about $12\,\rm km$, their central density can be in excess of a few times nuclear matter saturation density, where $n_{\rm sat} \approx 0.16\, \rm  fm^{-3}$~\cite{Baym:1971ax,Baym:1975mf,Lattimer:2004pg,Lattimer:2012nd,Landry:2020vaw,Legred:2021hdx}. A typical neutron star can contain in excess of $10^{57}$ baryons~\cite{Baym:1975mf}. Thus a neutron star is an exquisitely sensitive environment in which to study the possibility of new sources of baryon number violation (BNV). 

That a neutron star (NS) can serve as a ``graveyard'' of different theoretical extensions of the Standard Model (SM), particularly in regards to suggested solutions to the dark matter problem, is long known~\cite{Gould:1989gw,McDermott:2011jp}. In these scenarios, the capture of the suggested dark matter candidate by a NS alters the latter so severely that the existence of the model tested is precluded by that of the NS. The environment of a proto-NS is also long-known to be a sensitive discriminant of light new physics, such as axions~\cite{Iwamoto1984PhRvL..53.1198I, Turner:1987by, Brinkmann1988PhRvD..38.2338B, Burrows1989PhRvD..39.1020B, Janka1996PhRvL..76.2621J, Hanhart:2000ae, Sedrakian:2015krq} or dark photons~\cite{Dent2012arXiv1201.2683D, Dreiner2014PhRvD..89j5015D, Kazanas2015NuPhB.890...17K}, through its impact on the observed cooling of the star. Yet the advent of gravitational wave (GW) observations of compact object mergers~\cite{LIGOScientific:2016aoc}, and other observational facilities for the realization of multi-messenger probes of these objects, offer increasingly sensitive probes of new physics. In this article we focus on extensions of the SM with BNV, that may also entwine with dark sectors.\footnote{Dark or hidden sectors are comprised of particles that are uncharged under the SM gauge groups.} 

Our focus emerges from two connected ideas: that the long-standing neutron lifetime anomaly~\cite{Wietfeldt2011RvMP...83.1173W} could be resolved through ``dark'' decay channels of the neutron~\cite{Fornal2018PhRvL.120s1801F} and that the structure -- and even existence -- of neutron stars is extremely sensitive to the existence of such decay channels, at least at the strength required to explain the neutron lifetime anomaly~\cite{Mckeen2018PhRvL.121f1802M,Motta:2018rxp,Baym2018PhRvL.121f1801B}. This connection begs for a more systematic study. The long-standing theoretical problem of the origin of the baryon asymmetry of the universe (BAU) would also seem to require the existence of BNV~\cite{Sakharov:1967dj}. Our complete ignorance of the nature of BNV at low energies, for we have not established that it exists, and that the SM violates baryon number appreciably only at extremely high temperatures~\cite{Klinkhamer:1984di, Kuzmin:1985mm, Arnold:1987mh, Arnold:1987zg}, makes searching for traces of BNV a well-motivated endeavor. It has long been noted that the limits on the proton lifetime, and indeed on processes that violate baryon number $B$ by one unit, are severe~\cite{Zyla:2020zbs}. Yet the experiments that have established these limits are trivial in scale relative to that of the baryon reservoir in a neutron star,\footnote{The most severe limit on proton decay yet established comes from the Super-Kamiokande experiment~\cite{Super-Kamiokande:2002weg}, which holds about 50,000 metric tons of ultrapure water or some $5\times 10^{33}$ 
protons.} albeit the terrestrial detector can detect a proton's decay products directly.

\begin{figure}[t!]
    \centering
    \includegraphics[width=0.75\linewidth]{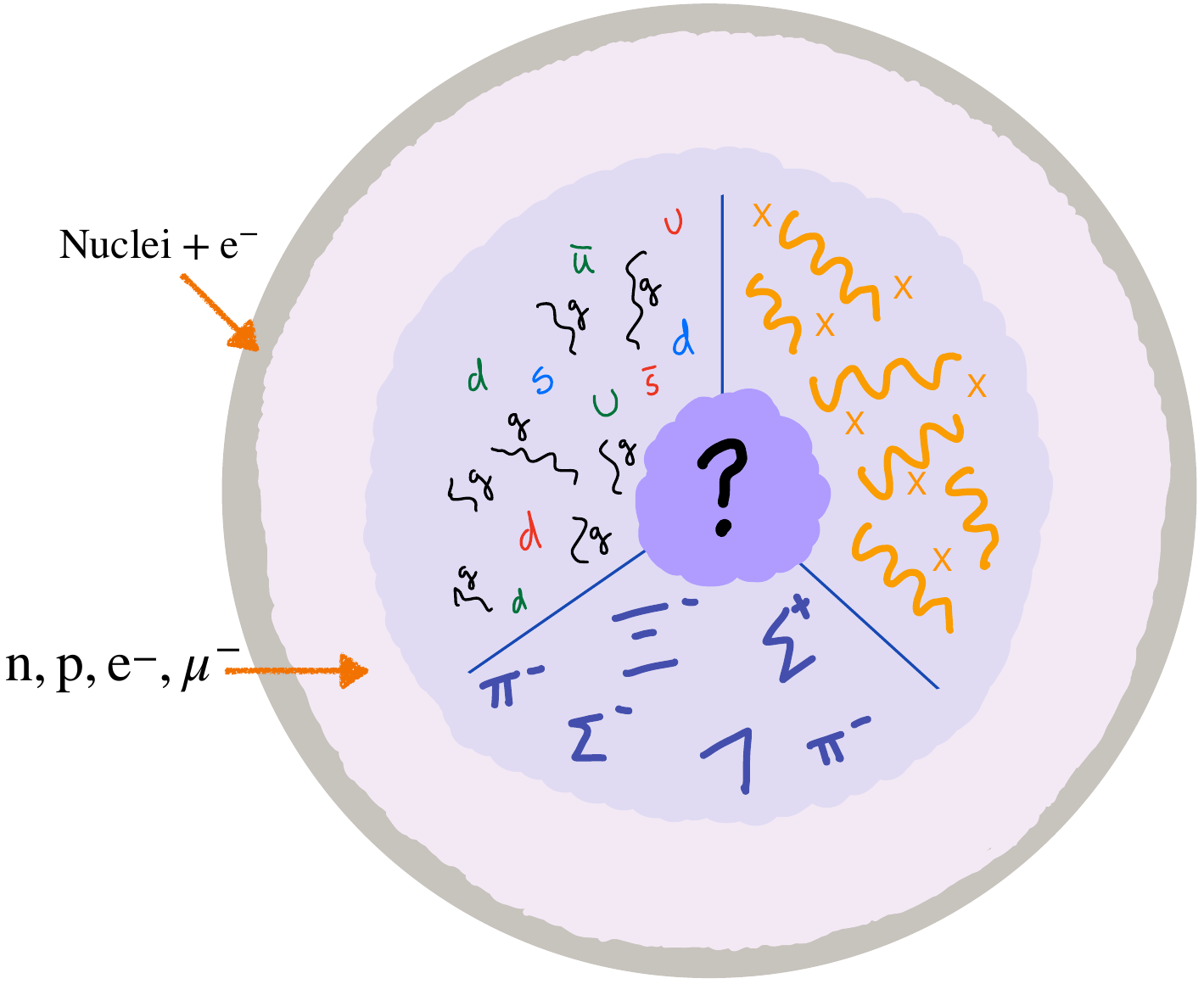}
    \caption{In this schematic, we update earlier notions regarding the interior structure of a neutron star~\cite{Baym:1975mf} to include not only the possibility of hyperonic matter in the inner regions of the star, but also the possibility of either a quark or a mixed quark/hadron phase, as well as the possibility of new matter or force mediators, ``$X$", that impact its structure, terrestrial constraints on neutron decay to dark final states (Sec.~\ref{sec:Nlifetime}). Baryon-number violation impacts the thermodynamics of the star (Sec.~\ref{sec:BV in NS}) and, be it apparent (Sec.~\ref{sec:apparent BV}), explicit (Sec.~\ref{sec:explicit BV}), or spontaneous (Sec.~\ref{sec:spont bv}), acts as a source of $X$.  For $X$ of still lighter mass, the cooling of the neutron star, as well as its merger dynamics, can be modified. The nature of the innermost region of the neutron star is unknown, but a dark-matter core or a dark-mediator condensate figure among the possibilities (Sec.~\ref{sec:DM}). Neutron stars typically have masses of ${\cal O}(1\,M_{\astrosun})$ and radii of ${\cal O}(10\,\rm km)$. }
    \label{fig:NS_interiors}
\end{figure}

The extreme conditions in neutron stars can also act to enhance baryon-number-violating (BNV) processes beyond those possible in terrestrial environments. The interior of a neutron star may exceed the density of nuclear matter by a factor of a few -- or perhaps by as much as a factor of ten -- and it may also contain significant strangeness, in either quark or hadron degrees of freedom, opening BNV channels with strangeness. Dinucleon decays can also be enhanced due to the greater overlap of the nucleons' wave functions, and multi-nucleon processes can help to mediate processes that are kinematically suppressed in nuclei~\cite{Ellis:2018bkr}. We should caution, however, that processes to final states with fermions already present in the neutron star would be highly suppressed by Pauli blocking. Other exotic processes can also appear. For example, neutron stars, by dint of their large mass, attract dark matter particles and thus dark-matter induced processes are also possible~\cite{Gould:1987ju,Gould:1989gw}. To give a sense of the sweep of the possibilities, and to provide some context, we illustrate and compare them with long-discussed, ongoing possibilities within the SM~\cite{Baym:1975mf} in Fig.~\ref{fig:NS_interiors}, though we also emphasize that much has been learned in recent years~\cite{Lattimer:2004pg,Lattimer:2012nd,Ozel:2016oaf,Landry:2020vaw,Legred:2021hdx}.

We conclude this section with an outline of our article. We first turn to the neutron lifetime anomaly, summarizing the outcome of interlocking SM tests in neutron beta-decay to revisit estimates of the maximum possible neutron decay branch to dark sector particles~\cite{Czarnecki:2018okw}. We then turn to an analysis of the thermodynamics of a neutron star in the presence of BNV, considering BNV processes that may potentially be either be an appreciable fraction of or much slower than the timescale of ordinary neutron beta-decay, as well as their implications. We then turn to a discussion of a broad sweep of new physics models with BNV, analyzing the consequences, in turn, of models with apparent BNV, explicit BNV, or spontaneous BNV for the structure of neutron stars and the broader environments in which they occur, as the latter are also subject to observational studies. Finally, we turn to a brief assessment of the broader ways in which dark sectors can impact neutron star observables before offering our final summary. 

%%%%%%%%%%%%%%%%%%%%%%%%%%%%%%%%%%%%%%%%%%%%
\section{Setting the Stage --- The Neutron Lifetime Anomaly} 
\label{sec:Nlifetime}
%%%%%%%%%%%%%%%%%%%%%%%%%%%%%%%%%%%%%%%%%%%%

Measuring the neutron lifetime with ever-increasing precision has been the ongoing work of decades~\cite{Wietfeldt2011RvMP...83.1173W}. This work has been motivated, in part, by anticipating the needs of precision cosmology: from the recognition that the neutron lifetime not only fixes the effective weak-interaction rate in standard Big-Bang nucleosynthesis~\cite{Wagoner:1972jh,Schramm:1977ne}, but it also represents the largest experimental uncertainty in the prediction of the $^4$He yield~\cite{Steigman:2007xt}. Over the last decade, or so, a significant disagreement has appeared in the determination of the neutron lifetime via two distinct methods~\cite{Wietfeldt2011RvMP...83.1173W}. This is the neutron lifetime anomaly --- namely, that the neutron lifetime determined from the detection of its decay products~\cite{Byrne:1996zz, Nico:2004ie, Yue:2013qrc}, as studied in neutron beam experiments, is longer than that inferred from counting the surviving neutrons~\cite{Mampe:1993an, Serebrov:2004zf, Serebrov:2007ve, Pichlmaier:2010zz, Steyerl:2012zz, Arzumanov:2015tea, Serebrov:2017bzo, Ezhou:2018, Pattie:2017vsj, UCNt:2021pcg}, as studied in neutron bottle or trap experiments. That the lifetime inferred from ``counting the living'' is smaller than that from ``counting the dead'' is evocative~\cite{Wietfeldt2011RvMP...83.1173W} --- perhaps the neutron decays to exotic, weakly-coupled final states and that could explain the difference. To our knowledge, the first work along these lines is due to Berezhiani~\cite{Berezhiani:2018eds}.\footnote{See references therein.} Fornal and Grinstein have developed new-physics models particular to the anomaly, yielding exotic final states in which no proton appears but containing particles that carry baryon number; and they have noted that these models can be probed through ancillary empirical tests~\cite{Fornal2018PhRvL.120s1801F}, which have been made~\cite{Tang:2018eln,UCNA:2018hup,Klopf:2019afh}, with null results thus far. It is quite possible that the anomaly could yet be explained through a subtle combination of experimental systematic effects in either or both types of experiments. In addition, different lines of evidence suggest that the entirety of the anomaly would not reasonably arise from new-physics effects. Powerful constraints come from the existence of neutron stars~\cite{Mckeen2018PhRvL.121f1802M, Motta:2018rxp, Baym2018PhRvL.121f1801B}, as well as from the connection to precision measurements of $\beta$-decay correlations in the Standard Model~\cite{Czarnecki:2018okw}. Yet the possibility of new-physics effects remain, and we use this as a springboard to consider the interconnections between neutron-star physics with new-physics models that contain baryon number, and its violation, and dark sectors in a broad way. 

The perspective on the interconnections between the neutron lifetime, $\beta$-decay observables, and SM radiative corrections (RCs) has shifted, due to new theoretical and experimental results, and we pause to consider these developments before proceeding to the main body of our article.

\subsection{Constraints from empirical studies of neutron $\beta$ decay within the SM}

\begin{figure}[t!]
    \centering
    \includegraphics[width=0.8\linewidth]{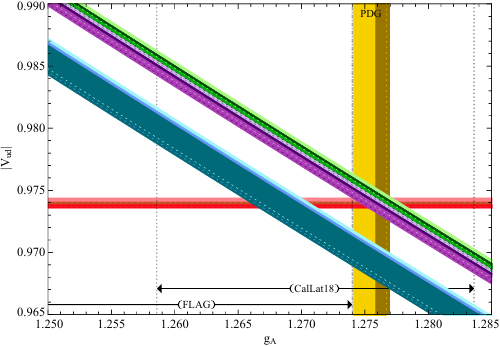}
    \caption{The SM relationship between the CKM element $|V_{ud}|$ and the axial coupling constant $g_{A}$, with different values of the neutron lifetime and estimated RCs, with values taken at 68\% C.L. throughout. We consider the beam lifetime~\cite{Yue:2013qrc} (cyan), as well as the 2020 PDG average of bottle/trap lifetimes~\cite{Zyla:2020zbs} (purple) and the latest magnetic trap result~\cite{Gonzalez:2017fku} (green), with the RCs of Refs.~\cite{Marciano:2005ec,Czarnecki:2019mwq,Seng:2018qru}, respectively, applied in each case, realized from the top to the bottom throughout. Assessments of $g_A$ from neutron $\beta$-decay from both decay-correlation measurements and lattice QCD calculations are also shown, as is the value of $|V_{ud}|$ from $0^+ \to 0^+$ nuclear $\beta$ decays, which is also sensitive to the precise value of the RCs. The lattice values of $g_A$ are the 2021 FLAG average for $N_f =2+1+1$ flavors~\cite{Aoki:2021kgd} and the 2018 CalLat result~\cite{Chang:2018uxx}. The decay correlation determinations of $|\lambda|$ are from the 2020 PDG compilation~\cite{Zyla:2020zbs} and from Ref.~\cite{Markisch:2018ndu} (ochre). We refer to the text for all details.}
    \label{fig:SM_test_beta_decay}
\end{figure}

Precision measurements of $\beta$-decay observables, along with accurate calculations of electroweak RCs, yield precision tests of the SM~\cite{Sirlin:1974ni, Sirlin:1977sv, Sirlin:1980nh, Sirlin:1981ie, Marciano:1985pd, Czarnecki:2004cw, Marciano:2005ec}. For example, the 
unitarity test stemming from the first row of the Cabibbo-Kobayashi-Maskawa (CKM) matrix, namely, 
\begin{equation}
    \Sigma_{\rm CKM} = |V_{ud}|^2 + |V_{us}|^2 + |V_{ub}|^2 - 1\,
    \label{firstrow}
\end{equation}
is the most precise known~\cite{Zyla:2020zbs}. A significantly nonzero value of $\Sigma_{\rm CKM}$ would establish the existence of physics beyond the SM, and there has been much discussion of the accuracy of the ingredients needed to determine $V_{ud}$ and $V_{us}$, since $|V_{ub}|^2 \sim {\cal O}(10^{-5})$ and is thus relatively negligible. An important ingredient is the electroweak RCs, which includes the evaluation of the $\gamma W$ box diagram and in which non-perturbative effects appear to be significantly larger~\cite{Seng:2018qru} than earlier estimated~\cite{Marciano:2005ec}. An updated analysis~\cite{Czarnecki:2019mwq} of the latter agrees with the sense of the shift but finds a result intermediate to that of Refs.~\cite{Marciano:2005ec, Seng:2018qru}. Moreover, both electromagnetic and isospin corrections are also key to determining consistent values of $|V_{us}|$ from $K_{\ell 2}$ and $K_{\ell 3}$ decays~\cite{FlaviaNetWorkingGrouponKaonDecays:2010lot,Cirigliano:2011ny}, with the current scale factor of $S=2.7$~\cite{Zyla:2020zbs} seemingly indicating the need for further theoretical and experimental work~\cite{Seng:2019lxf,Seng:2020jtz,Seng:2021boy,Seng:2021wcf}. Here we focus on the $V-A$ structure of neutron decay in the SM, and the concomitant ties between its observables, as this constrains the possibility that the neutron lifetime anomaly comes from new physics~\cite{Czarnecki:2018okw} --- and we refer to Refs.~\cite{Abele:2008zz, Dubbers:2011ns, Cirigliano:2013xha, Dubbers:2021wqv} for reviews. To that end, we consider~\cite{Czarnecki:2018okw} 
\begin{equation}
    \frac{1}{\tau_n} = \frac{G_F^2 |V_{ud}|^2}{2\pi^3} m_e^5 (1 + 3 g_A^2)(1 +\delta_{\rm RC}) f \,,
\end{equation}
where $\tau_n$ is the neutron lifetime, $G_F$ is the Fermi constant determined from $\mu$ decay after QED radiative corrections are subtracted, $G_F =1.1663787(6)\times 10^{-5}\,{\rm GeV}^{-2}$~\cite{MuLan:2012sih}, $m_e$ is the electron mass, $m_e =0.5109989461 (31)\,{\rm MeV}$~\cite{Zyla:2020zbs}, $g_A$ is the axial vector coupling constant of the nucleon,\footnote{Unlike Ref.~\cite{Czarnecki:2018okw}, we regard $g_A$ as a quantity amenable to direct theoretical calculation, as through lattice quantum chromodynamics (QCD) techniques, and thus it need not be a phenomenological parameter per se.} $\delta_{\rm RC}$ is the electroweak radiative correction, and $f$ is the statistical rate function~\cite{Wilkinson:1982hu}. The last follows from the allowed phase space, the recoil corrections assessed in the isospin symmetric limit, and the Coulomb correction in the $e-p$ final state as encoded in the Fermi function --- and it has been reevaluated to yield $f=1.6887(1)$~\cite{Czarnecki:2004cw}. In the SM, the combination 
\begin{equation}
    |V_{ud}|^2 \tau_n ( 1 + 3 g_A^2) = \frac{2\pi^3}{G_F^2 m_e^5 (1+ \delta_{\rm RC})f} \equiv \eta 
    \label{taun_tie}
\end{equation}
is tightly constrained, because $\eta$ is $4908.6 (1.9)\,{\rm s}$~\cite{Marciano:2005ec}, or $4903.6 (1.0)\,{\rm s}$~\cite{Seng:2018qru}, $4905.7 (1.5)\,{\rm s}$~\cite{Czarnecki:2019mwq}, depending on the calculation of $\delta_{\rm RC}$ used. The uncertainties are dominated by that in $\delta_{\rm RC}$; thus we use that to determine the reported errors in $\eta$. Using Eq.~(\ref{taun_tie}), different measurements of the neutron lifetime at 68\% confidence level (C.L.) give the diagonal bands in the $|V_{ud}|$ versus $g_A$ plot shown in Fig.~\ref{fig:SM_test_beta_decay}. We combine statistical and systematic errors in quadrature, assuming uncorrelated errors, to realize the bands shown in Fig.~\ref{fig:SM_test_beta_decay}.

We now turn to the inputs used to generate Fig.~\ref{fig:SM_test_beta_decay}. We consider the most precise beam lifetime result, $887.7\pm 1.2 \, {(\rm stat)} \pm 1.9\, {(\rm sys)}\,{\rm s}$~\cite{Yue:2013qrc}, the most recent PDG average of the neutron lifetime determined from bottle and trap experiments, $879.4\pm 0.6\,{\rm s}$~\cite{Zyla:2020zbs}, and the most precisely measured neutron lifetime, determined in a magnetic trap experiment, $877.75\pm 0.28\, {(\rm stat)} {{\stackrel{{}_{+0.22}}{{}_{-0.16}}}\, {(\rm sys)}}\,{\rm s}$~\cite{UCNt:2021pcg}. The numerical difference between the beam and bottle/trap measurements constitutes the neutron lifetime anomaly. The latest magnetic trap result~\cite{UCNt:2021pcg} is compatible with an earlier measurement with the same method~\cite{Pattie:2017vsj} but is not included in the PDG average. We pull it out for explicit study because of the possibility of significantly large and/or underestimated systematic errors in the older experiments; e.g., the scale factor in the average reported by the PDG is 1.6~\cite{Zyla:2020zbs}. 

We also include the value of $g_A$ from both measurements of neutron decay correlation coefficients and computations within lattice QCD, as well as the value of $|V_{\rm ud}|$ from superallowed $0^+ \to 0^+$ nuclear decays, in which the effect of the axial vector current enters in RCs, which can be modified by nuclear structure as well~\cite{Towner:2002rg}. The empirical determination of $g_A$ comes from that of $\lambda\equiv |g_A/g_V|$, where the interpretation of the measured $A$ and $a$ correlation coefficients in terms of $\lambda$ requires the application of radiative and recoil corrections, and in the latter additional hadronic matrix elements appear~\cite{Holstein:1974zf,Gardner:2000nk}. The weak magnetism contribution therein, in the isospin limit, is fixed by the determination of the isovector nucleon magnetic moment~\cite{Zyla:2020zbs}, and the matrix elements that are nonzero only if the $u$ and $d$ quarks differ in mass~\cite{Weinberg:1958ut} are set to zero. We note that the PDG average is $\lambda=1.2754(13)$ with a scale factor of 2.7, whereas the most precise determination, from that of the $A$ decay correlation in $n$ decay, is $\lambda=1.27641(56)$~\cite{Markisch:2018ndu}. This last result is consistent with the other two most precise determinations of $\lambda$, which are also determined from $A$~\cite{Mund:2012fq,UCNA:2017obv}. In what follows, following common practice, we employ $\lambda$ as $g_A$, which is strongly supported by an analysis of its RCs~\cite{Gorchtein:2021fce}. As for the lattice QCD results, we note the recent FLAG average from simulations with $N_f = 2 + 1 +1$ flavors, $g_A=1.246 (28)$, as well as the most precise lattice result, $g_A=1.271 (13)$~\cite{Chang:2018uxx}. Both calculations have significant errors, but the latter result is compatible with the most precise empirical determination of $\lambda$. We note that the $g_2$ form factor, which vanishes in the isospin limit, and/or the possibility of scalar and/or tensor currents can make the two assessments differ~\cite{Gardner:2013aya, Cirigliano:2013xha, Ivanov:2018uuk}. We refer to Refs.~\cite{Gardner:2013aya, Cirigliano:2013xha} for complete expressions for the hadronic vector and axial-vector currents. These contributions can also modify the relationship of Eq.~(\ref{taun_tie})~\cite{Gardner:2013aya}. Finally, we note the value of $V_{ud}$ from superallowed decays, for which the precise value depends on the assessment of the $\gamma W$ box and thus the RCs we have already noted. Assuming uncorrelated errors, and combining them in quadrature, we have $|V_{ud}|=0.97420(20)$~\cite{ParticleDataGroup:2018ovx, Hardy:2016vhg, Marciano:2005ec} (with $|V_{ud}| = 0.97373 (31)$ given in the update~\cite{Hardy:2020qwl}, arising from the theoretical developments in the radiative corrections we consider), $|V_{ud}|=0.97370(14)$~\cite{Seng:2018qru}, and $|V_{ud}|=0.97389(18)$~\cite{Czarnecki:2019mwq}. 

Turning to Fig.~\ref{fig:SM_test_beta_decay}, we see that the bottle/trap lifetime measurements are in better agreement with the determinations of $|V_{ud}|$ and the empirical assessments of $\lambda$ from $\beta$-decay correlations, thus limiting the phenomenological role of dark neutron decays. Interestingly, the most precise determinations of $\tau_n$ and $\lambda$ are also in good agreement with each other, supporting a $V-A$ theory of the weak interactions, albeit that first row CKM unitarity, Eq.~(\ref{firstrow}), may well be violated, making $\Sigma_{\rm CKM}$ non-zero. Although it would seem that the bulk of the neutron lifetime anomaly cannot come from BSM physics,  this conclusion is not definite, given the significant uncertainty in the lattice QCD assessments of $g_A$, as both theory and experiment should agree on its value if the SM gives a complete description of $\beta$ decay at current levels of precision. Indeed, the precision of the various measurements in Fig.~\ref{fig:SM_test_beta_decay} prompts further thought in regards to the size of neglected contributions. Interestingly, for example, the inclusion of the poorly known $g_2$ matrix element, which appears in the SM in the recoil corrections to $A$ and to the relationship in Eq.~(\ref{taun_tie}), modifies the intersection point of Eq.~(\ref{taun_tie}) and $g_A$ from $A$ in Fig.~\ref{fig:SM_test_beta_decay}, spoiling their mutual intersection with $|V_{ud}|$ from superallowed decays if the QCD sum rule estimate of $g_2/g_A = -0.0152\pm 0.0053$~\cite{Shiomi:1996np} is employed. Although both $g_2$ and $R$, the parameter that controls the size of the recoil corrections in $\beta$ decay~\cite{Bender:1968zz}, which evaluates to $\sim 1.4\times 10^{-3}$ for neutron decay, are both ${\cal O}(m_d - m_u)$ effects, they need not be of comparable numerical size. As illustrated concretely in Ref.~\cite{Shiomi:1996np}, $g_2$ is some ten times larger, making its inclusion part of the leading-order recoil correction~\cite{Gardner:2013aya}. This effect thus impacts the ability to detect or limit new physics in this case, as noted earlier in regards to the possibility of a non-zero tensor current~\cite{Gardner:2013aya}. It would be helpful if $g_2$ could be calculated or bounded with lattice QCD techniques, if not from $\beta$ decay studies. A recent global analysis of subleading corrections in $\beta$ decay can be found in Ref.~\cite{Falkowski:2021vdg}.

The putative agreement between the $|V_{ud}|$ and $g_A$ also limits the possibility of dark decays of the neutron, including that of neutron--mirror-neutron oscillations~\cite{Czarnecki:2018okw,Dubbers:2018kgh}. To give a sense of this effect, we neglect the possibility of BSM contributions to the breaking of the $V-A$ law, such as tensor currents, as well as the role of subleading SM effects such as the $g_2$ matrix element. In this context, we note the estimate ${\rm Br}({\rm exotic\, neutron\, decays}) < 0.16\%$ (95\% one-sided C.L.)~\cite{Czarnecki:2019mwq}, in which the average value of $g_A$ from post-2002 decay-correlation experiments and the average trap neutron lifetime were employed, as an estimate of the sensitivity of current $\beta$-decay studies. 

%%%%%%%%%%%%%%%%%%%%%%%%%%%%%%%%%%%%%%%%%%%%
\section{Neutron stars with baryon number violation}
\label{sec:BV in NS}
%%%%%%%%%%%%%%%%%%%%%%%%%%%%%%%%%%%%%%%%%%%%

It is typically assumed that baryon number is conserved in a neutron star during the span of its lifetime, whereas its  strangeness changes on the much shorter timescale of the weak interactions, because the kaons that are produced through the strong interactions decay to leptons and photons and the hyperons do not. The conservation of baryon number and electric charge are used as constraints in determining the state of matter inside neutron stars. This is achieved by finding the ground state of electrically neutral ($Q=0$), cold ($T=0$) matter at a given fixed baryon number density\footnote{We will suppress the baryon number density's index (``$B$'') in the rest of this section, and denote it by ``$n$'' instead.} ($n_B$), i.e., by defining a function
\begin{equation}
\label{eq:ns:state}
    \Phi \equiv \sum_i \varepsilon_i \left(\{n_j\}\right) +  \alpha \sum_i Q_i\, n_i + \beta \left( n_B- \sum_i  B_i\, n_i\right),
\end{equation}
and minimizing it with respect to the individual number densities ($n_i$), i.e., $\partial \Phi /\partial n_i = 0$. The sum in Eq.~\eqref{eq:ns:state} is over all of the particles present in the matter, $\alpha$ and $\beta$ are the Lagrange multipliers enforcing the electric charge and baryon number conservation respectively, $\varepsilon_i$ is the energy density that depends on the density of each species $j$, $Q_i$ is the electric charge, and $B_i$ is the baryon number of a particle of type $i$. The two constraints ($\alpha$ and $\beta$) relate the chemical potentials of particles present in a neutron star at chemical equilibrium. The chemical potential for a particle of type $i$ (if present and in equilibrium with the matter inside the star) is given by $\mu_i = B_i \mu_n - Q_i \mu_e$, in which $\mu_e$ and $\mu_n$ are the chemical potentials for the electron and neutron, respectively. In general a violation of baryon number conservation would change the chemical equilibrium and the composition of the star in a model-dependent manner. However, as we show later (\ref{subsec:BV in NS:gen}), a model-independent probe is feasible for a class of sufficiently slow BNV processes ($\tau_{\rm BNV} \gg \tau_{\rm weak},\tau_{\rm hyd}$) which act as small baryon perturbations over time. In response to each of these out-of-equilibrium perturbations, the star regains its chemical equilibrium using the standard (faster) baryon-number conserving (BNC) reactions (e.g., weak interactions) and ends up with a lower total baryon number. Here $\tau_{\rm weak}$ is the timescale for weak interactions in the neutron star medium (e.g., Urca reactions) which may be different from the free neutron lifetime ($\tau_n)$, and $\tau_{\rm hyd}$ is the time needed to adjust to and maintain hydrostatic equilibrium. We will explain these timescales further in Sec.~\ref{subsec:BV in NS:gen}. We will study the generic effects of BNV in this section, and defer a discussion of specific models to the following sections.

We assume that the neutron star matter (without BNV) has spherical symmetry.\footnote{Rapid rotations of pulsars cause oblateness and thus breaks this into a residual axial symmetry.} This spherical symmetry would remain intact after the inclusion of BNV processes, because BNV processes would be sourced by the matter already present in the star. Furthermore, we work in a quasi-static regime in which the changes to the metric ($g_{\mu\nu}$) are very slow in time. This warrants the use of the line element for a static spherically symmetric system~\cite{tolman1987relativity}
\begin{equation}
\label{eq:g:def}
    d\tau^2 = g_{\mu\nu} dx^{\mu} dx^{\nu} = e^{2\nu(r)} \, dt^2 - e^{2\lambda(r)}\, dr^2 - r^2\, d\theta^2 - r^2\, \sin^2\theta\, d\phi^2,
\end{equation}
in which $\nu(r), \lambda(r)$ are solutions to the Einstein field equations~\cite{Einstein:1916vd}, $G^{\mu\nu} = -8 \pi G T^{\mu\nu}$, in which $G^{\mu\nu}$ is Einstein's tensor, $G$ is the gravitational constant, and $T^{\mu\nu}$ is the stress-energy tensor. We use a geometric unit system in which the speed of light ($c$) and $G$ are both set to unity, i.e.,  $G = c = 1$. For a perfect fluid $T^{\mu\nu}$ has the form
\begin{equation}
    \label{eq:T:perfect}
    T^{\mu\nu} = -p\, g^{\mu\nu} + (\varepsilon + p) u^{\mu} u^{\nu},
\end{equation}
in which $p$ and $\varepsilon$ are the local pressure and energy density of the fluid respectively, and $u^{\mu}$ is the 4-velocity of the fluid, which has zero 3-velocity ($u^{i} = 0, u^0 \neq 0$) in a static star. The time component of $u^{\mu}$ is calculated (from the normalization condition: $g_{\mu\nu} u^{\nu} u^{\mu} = 1$) to be
\begin{equation}
u^{0} = 1 / \sqrt{g_{00}} =  e^{-\nu(r)}.   
\end{equation}
Therefore, for a static perfect fluid the only non-zero components of the stress-energy tensor are given by
\begin{equation}
    % \label{eq:T:perfect}
    T_{0}^{\,0} = \varepsilon, \quad T_{i}^{\,i} = -p \qquad (i = 1,2,3).
\end{equation}
Moreover, $\lambda(r)$ is found to be $g_{11}(r) = -\exp(2\lambda(r)) = -(1 - 2M(r)/r)^{-1}$, with $M(r)$ being the total mass included within radius $r$:
\begin{equation}
\label{eq:m:def}
M(r') = 4\pi \int_0^{r'} \varepsilon(r) r^2 dr.
\end{equation}
Further simplification of the Einstein field equations yields a differential equation for the pressure inside the star:  
\begin{equation}
    \label{eq:p:der}
    \frac{dp}{dr} = -\frac{\left[p(r) + \varepsilon(r)\right] \left[M(r) + 4\pi r^3 p(r)\right]}{r \left[ r - 2M(r) \right]}.
\end{equation}
Eq.~\eqref{eq:m:def} together with Eq.~\eqref{eq:p:der} are known as the Tolman-Oppenheimer-Volkoff (TOV)~\cite{Tolman:1939jz,Oppenheimer:1939ne} equations. The pressure ($p$) and energy density ($\varepsilon$) are in general functions of the number density of baryons in the rest-frame of the fluid ($n$) and temperature ($T$). Neutron stars cool down to $T \ll E_{\rm Fermi}\approx 30$ MeV within a few seconds after formation. Therefore, the thermal contribution to the pressure and energy density can be neglected, i.e., $p(n, T) = p(n, T = 0) = p(n)$ and $\varepsilon(n, T) = \varepsilon(n, T = 0) = \varepsilon(n)$. We can then deduce both $\varepsilon$ and $p$ from the knowledge of $n$.

The TOV equations~\eqref{eq:m:def} and~\eqref{eq:p:der} can be integrated with the initial conditions $M(0) = 0$ and $\varepsilon (0) = \varepsilon_c$ up to $p(r) = 0$ (surface of the star). Therefore, for any given equation of state (EoS), there is a unique family of stars parameterized by the central energy density ($\varepsilon_c$) also known as the \textit{single parameter sequence}~\cite{Glendenning:1997wn} of stars. We note in passing that in the case of a rotating neutron star~\cite{Hartle:1967he}, or a neutron star with a dark matter core~\cite{Sandin:2008db,Ciarcelluti:2010ji}, extra parameters in addition to $\varepsilon_c$ are needed to  describe the star uniquely. We discuss the possibility of generalizing our analysis to the rotating case in Sec.~\ref{subsec:BV in NS:effects}.

The baryon number current is given in terms of the fluid velocity ($u^{\mu}$) and the baryon number density ($n$) by~\cite{Misner:1973prb} $j^{\mu}(r) = n(r) u^{\mu}$. Bearing in mind that the invariant 4-volume is given by $\sqrt{-g}\, d^4x$ ($g \equiv {\rm det}|g_{\mu\nu}|$), the total baryon number in a static, spherically symmetric neutron star is given by~\cite{Glendenning:1997wn}
\begin{equation}
\label{eq:B:def}
B = \int j^{0}(r)\, \sqrt{-g}\, d^4x = 4\pi \int_0^R \left[1 - \frac{2M(r)}{r}\right]^{-\frac{1}{2}} r^2 n(r) dr,
\end{equation}
in which we used $\sqrt{-g} = \exp(\nu(r)+\lambda(r))\, r^2\, \sin\theta$.

%%%%%%%%%%%%%%%%%%%%%%%%%%%%%%%%%%%%%%%%%%%%

\subsection{General conditions}
\label{subsec:BV in NS:gen}

The exact consequences of BNV processes for neutron stars depend on the modeling of neutron stars' structure (in the absence of BNV) and the particle physics model producing those specific BNV reactions. Although the details of neutron star models may change the numerical results in this section, we expect that their order of magnitude and qualitative behaviour remain intact. On the other hand, the particle physics modeling of BNV could have drastic effects, and a generic study would require imposing some simplifying assumptions on the BNV models. We attempt to find a minimal set of conditions that makes such a broad investigation viable. There are two major effects that the inclusion of a BNV process in the star can generate. The first one is caused by the relaxation of the baryon number conservation constraint (i.e., $\beta = 0$ in Eq.~\eqref{eq:ns:state}). The system is now allowed to transition into more energetically favorable states subject to electric charge conservation (only). The EoS would be different from the standard BNC EoS, and this could cause drastic changes to the composition of the matter inside stars. As we will explain below, we are interested in slow BNV processes for which this effect is eliminated and the BNC EoS is revived. The second effect is because of the production of new particles in the BNV process that are not otherwise present in the star. The modification in the pressure and energy density of the matter (i.e., the EoS) would depend on the specific final states produced in the BNV process. For example, a fermionic final state would exert a Fermi pressure and its production would be Pauli suppressed, whereas a scalar final state (with negligible self-interaction) would significantly reduce the pressure of the system. This is an obstacle to our model-independent analysis objective. Accordingly, we set forth the following essential condition for the BNV processes that we consider in this section:

\textit{The final states are either already present (via BNC processes) or if they are not already present, then they maintain a negligible contribution to the EoS.}

In cases with new particles in the final states, the above condition can be realized if: 
\begin{enumerate}
    \item The new final state particles participate in annihilation or decay channels to yield particles already present plus neutrinos and photons.
    \item Their production rate ($\Gamma_{\rm BNV}$) is much less than their elimination rate via annihilation $\Gamma_{\rm ann}$ or decay $\Gamma_{\rm dec}$.
\end{enumerate}
We extend the above constraints by also demanding that: 

\textit{The BNV rate(s) are slower than the weak-interaction processes that they activate in the neutron star in response to their presence, i.e., $\Gamma_{\rm BNV} < \Gamma_{\rm weak}$.} 

\begin{figure}[t!]
    \centering
    \includegraphics[width=0.8\linewidth]{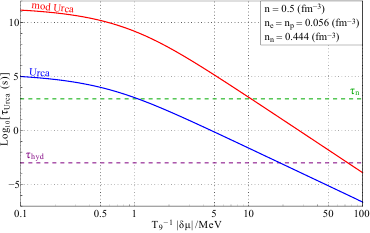}
    \caption{Timescales for the Urca (blue) and modified Urca (red) reactions in the $npe$ model as a function of temperature ($T_9$), and $\beta$-disequilibrium ($\delta \mu$). The free neutron life-time ($\tau_n$), and hydrodynamic response time ($\tau_{\rm hyd}$) are plotted in dashed green and purple for comparison.}
    \label{fig:Urca}
\end{figure}

We should elaborate on the nature of these responses and their timescales which vary greatly depending on the EoS, mass, and temperature of the neutron star. Reactions are generally suppressed by the small phase space available to fermions in chemical equilibrium in a cold, degenerate state. This is because the fraction of fermions on the edge of their Fermi surface that can undergo inelastic scattering is about $\sim k_B T/E_F \ll 1$. The slow BNV processes perturb the chemical equilibrium in the star. This imbalance in chemical equilibrium temporarily activates or enhances BNC reactions until a new chemical equilibrium is achieved. The exact timescale of the response to BNV reactions by the weak processes ($\tau_{\rm weak}$) would depend on the specific reaction, and the temperature of the star. For a general estimate of the timescales involved, let us consider the Urca processes~\cite{gamow1970my} which are extensively studied in the context of neutron star cooling theories~\cite{RevModPhys.64.1133, Yakovlev:1999sk, Anzuini:2021rjv}. Direct Urca processes involve baryon $\ell$-decays and lepton $(\ell = e, \mu)$ capture:
\begin{equation}
    B_1 \to B_2 + \ell + \overline{\nu}_{\ell}, \qquad B_2 + \ell \to B_1 + \nu_{\ell},
    \label{eq:urca}
\end{equation}
in which $B_{1,2}$ denotes nucleons or hyperons. The nucleonic direct Urca reactions ($B_{1,2} = n, p$) would be active if the proton fraction is above a minimum threshold ($n_p\gtrsim n_n/8$)~\cite{BOGUTA1981255, PhysRevLett.66.2701}, which is possible at the inner-core of a heavy neutron star with supranuclear densities ($\varepsilon \gtrsim 2\varepsilon_{\rm nucl}$). The hyperonic direct Urca processes ($B_{1,2} = \Lambda, \Sigma^-, \ldots$) can occur~\cite{1992ApJ...390L..77P} when the neutron chemical potential ($\mu_n$) surpasses the energy of the lowest state of a $\Lambda$, and if $\mu_n + \mu_e$ is greater than the energy of the lowest state of a $\Sigma^-$ in the neutron star. At lower densities (e.g., in the outer core), the direct Urca processes may be suppressed, but the following modified Urca reactions would still occur~\cite{PhysRevLett.12.413, PhysRev.140.B1452}:  
\begin{align}
    n + n &\to n + p + \ell + \overline{\nu}_{\ell}, \qquad n + p + \ell \to n + n + \nu_{\ell},\\
    n + p &\to p + p + \ell + \overline{\nu}_{\ell}, \qquad p + p + \ell \to n + p + \nu_{\ell}.
\end{align}
Since these modified Urca reactions have two extra degenerate fermions, their rate is suppressed by a factor of $(k_B T/E_F)^2$ compared to the direct processes. In the presence of so-called $\beta$-disequilibrium and processes involving the neutron, the sign of the disequilbrium parameter $\delta \mu \equiv \mu_n - \mu_p -\mu_e$ determines which of the reactions in Eq.~\eqref{eq:urca} dominates, so that if $\delta \mu>0 $, $\beta$ decay occurs and if $\delta\mu <0$, electron capture occurs. Thus for $\delta \mu > 0$, which yields net $\bar{\nu}_e$ production, the rates for direct and modified Urca processes in a simple $npe$ model (i.e., a degenerate Fermi gas consisting of neutrons, protons and electrons) are given by~\cite{1992A&A...262..131H}
\begin{align}
    \Gamma_{\rm Urca} =&\, 8.86 \times 10^{31} \left(\frac{n_e}{n_\textrm{sat}}\right)^{1/3} \, T_9^5\, G_d\left({\delta \mu} / {k_B T}\right) \, \left[ {\rm cm}^{-3}\, s^{-1}\right],\\
    \Gamma_\textrm{mod Urca} =&\, 5.91 \times 10^{23} \left(\frac{n_e}{n_{\textrm{sat}}}\right)^{1/3} \, T_9^7\, G_m\left({\delta \mu} / {k_B T}\right) \, \left[ {\rm cm}^{-3}\, s^{-1}\right],
\end{align}
in which $n_\textrm{sat} = 0.16\, {\rm fm}^{-3}$, $T_9 \equiv T/(10^9\, K)$, and the dimensionless functions $ G_d$ and  $G_m$ are defined as 
\begin{align}
  G_d(x)  \equiv& \int_0^{\infty} dy\, y^2 \left[\frac{\pi^2 + \left(y-x\right)^2}{1 + \exp(y-x)}\right],\\
  G_m(x)  \equiv& \int_0^{\infty} dy\, y^2 \left[\frac{9\pi^4 + 10\pi^2 \left(y-x\right)^2 + \left(y-x\right)^4}{1 + \exp(y-x)}\right].
\end{align}
For $\delta \mu <0$, yielding $\nu_e$ production, the rates evaluate to the same numerical value as in the $\delta \mu >0$ case, implying the replacement $\delta \mu \to -\delta \mu$. The various Urca timescales are plotted in Fig.~\ref{fig:Urca} as a function of $|\delta \mu|/T_9$. We take the baryon number density to be $n \approx 0.5\, {\rm fm}^{-3}$, and $(n_e/n) = (n_p/n) \approx 11.2 \%$, $n_n/n \approx 88.8\%$. We can see that the Urca rates are highly sensitive to $\delta \mu$. We expect BNV reactions to generate $\beta$-disequilibrium of order $|\delta \mu| \approx 10 - 100$ MeV from kinematics.

The other important timescale is the hydrodynamical relaxation time of the star ($\tau_{\rm hyd}$) for regaining hydrostatic equilibrium. We can approximate this timescale by finding the period of small oscillations of a uniform Newtonian fluid in hydrostatic equilibrium:
\begin{equation}
    \Ddot{r} = -\frac{G\, m(r)}{r^2} - \frac{1}{\varepsilon(r)} \frac{dP(r)}{dr} = 0,
\end{equation}
in which $\varepsilon(r) = \overline{\varepsilon}$, $m(r) = (4/3) \pi r^3 \, \overline{\varepsilon}$. The period of small adiabatic oscillations is given by $\omega^2 = 4 \pi (\overline{\Gamma}_1 - 4/3) G\, \overline{\varepsilon}$~\cite{Misner:1973prb}, in which $\overline{\Gamma}_1 \equiv \left(n/p\right)\left(\partial p/\partial n\right)_{s}$ is the mean adiabatic index. The hydrodynamical timescale ($\tau_{\rm hyd} \equiv 2\pi/\omega$) is then given by
\begin{equation}
        \tau_{\rm hyd} \approx \sqrt{\frac{3}{G\, \overline{\varepsilon}}} \approx 10^{-4} - 10^{-3}\, (s),
\end{equation}
in which $G$ is the Newton's constant, $\overline{\Gamma}_1 \in \left[ 2, 3\right]$~\cite{Haensel:2002qw} and we used $\overline{\varepsilon} \in \left[10^{14}, 10^{15}\right]\, (g/{\rm cm}^3)$.

The second condition on BNV (i.e., $\Gamma_{\rm BNV} < \Gamma_{\rm weak}$), permits the use of the standard BNC EoS. In essence, because baryon number is conserved on shorter timescales (compared to $\tau_{\rm BNV}$), the instantaneous states of neutron stars are governed by the same BNC EoS before the inclusion of the BNV reactions. Nevertheless, the star will be slowly leaking baryon number and its structure will change over time.

To better clarify these statements, let us consider a specific example: a process with $|\Delta B| = 2$ such as $n n \to e^- e^+$. In this case the positrons would start annihilating with the electrons that are already present at a rate $\Gamma_{\rm ann} \approx n_{e^-} \sigma_{e^+e^-} c$, in which $n_{e^-}$ and $\sigma_{e^+e^-}$ are the electron number density, and electron-positron annihilation cross section to two photons and $c = 1$ is the speed of light. In general, if the particles produced in the final states have a cross section $\sigma \approx 10^{-43}\, {\rm cm}^2$, and their counterpart in the annihilation has a density of at least $10^{-3}\, {\rm fm}^{-3}$, then we get  $\Gamma_{\rm ann} \approx 10^3 s^{-1}$. As long as the rate for the BNV processes under consideration is smaller than this annihilation (or decay) rate $\Gamma_{\rm BNV} \ll \Gamma_{\rm ann}$, our assumptions in this section are valid. In other words, we are considering the cases in which, through a chain of reactions, the final products end up being those particles that are already present in the star plus photons and neutrinos. Neutron stars become transparent to low-energy ($E_{\nu} \ll $ MeV) neutrinos as they cool down to temperatures $T\lesssim$ MeV. The mean-free path of electron-neutrinos ($\nu_e$) with $E_{\nu} \ll E_F(e)$ is given by~\cite{PhysRev.140.B1452}
\begin{equation}
    \label{eq:nu:mfp}
    \lambda_{\nu_e} \approx 5\times 10^{6}\, {\rm km} \left(\frac{\varepsilon_{\rm nucl}}{\varepsilon}\right)^{4/3} \left(\frac{100\, {\rm keV}}{E_{\nu}}\right)^3,
\end{equation}
in which $\varepsilon_{\rm nucl} = 3.7\times10^{14}\, g/{\rm cm}^3$. This mean-free path is much larger than the typical neutron star radius ($\sim 10$ km) for $E_{\nu} \lesssim 5$ MeV. The mean-free path of other neutrino flavors would be even larger in the SM. In comparison, photons have a much shorter mean-free path, and would deposit most of their energies before they can escape, which would result in heating of the neutron star.

%%%%%%%%%%%%%%%%%%%%%%%%%%%%%%%%%%%%%%%%%%%%

\subsection{Effects of Slow BNV perturbations}
\label{subsec:BV in NS:effects}

If the BNV processes are much slower than the weak interaction rates, then their sole effect is to change the baryon number density inside the star ($n$) and perturb the system out of equilibrium ($\Tilde{n})$.\footnote{The perturbations are Eulerian, i.e., they are changes measured by an observer at a fixed point $(t, r, \theta, \varphi)$.} The system will respond by adjusting the densities of each species via reactions that conserve baryon number and electric charge. Therefore, the final equilibrium state ($n'$) of the star would be the same as a star with a lower baryon number ($B'< B$) or equivalently a lower central energy density ($\varepsilon'_c < \varepsilon_c$) from the same single parameter sequence (Table~\ref{tab:b:pert}).

\begin{table}[H]
\centering
\begin{tabular}{cl cl cl cl}
$n(r)$ & \hspace{-1em}\rdelim\}{2}{*}[$\xrightarrow{ \text{ \,Perturb\, } }$] & \hspace{-1em}\ldelim\{{2}{*} $\Tilde{n}(r) = n(r) + \Tilde{\delta}n(r)$ & \hspace{-1em}\rdelim\}{2}{*}[$\xrightarrow{\text{ Equilibrate\, }}$] & \hspace{-1em}\ldelim\{{2}{*}  $n'(r) = n(r) + \delta n(r)$ &\\
$B$ & {\scriptsize \, \,\, BNV} & $B'= B + \delta B$ & {\scriptsize \quad\,\,\, BNC} & $B'$ \\
\end{tabular}
    \caption{The neutron star's response to local baryon number density perturbations.}
    \label{tab:b:pert}
\end{table}

The total baryon number after the perturbation generated by the BNV process ($B'$) is (from Eq.~\eqref{eq:B:def})
\begin{equation}
    \label{eq:barCons}
    \frac{B'}{4\pi} = \int_0^R \left[1 - \frac{2M(r)}{r}\right]^{-\frac{1}{2}} r^2 \Tilde{n}(r) dr = \int_0^{R'} \left[1 - \frac{2M'(r)}{r}\right]^{-\frac{1}{2}} r^2 n'(r) dr,
\end{equation}
in which $R'$ and $M'$ are the equilibrium radius and mass of the star after the perturbation. Since the structure of a spherically symmetric neutron star is fully determined (via Eq.~\eqref{eq:m:def} and Eq.~\eqref{eq:p:der}) by an EoS and the value of the energy density at the origin ($\varepsilon_c$), changes in the total baryon number $\delta B = B' - B$ can be uniquely mapped onto $\delta \varepsilon_c$. Therefore, we can quantify the changes in a neutron star observable $O$ along the single parameter sequence as
\begin{equation}
    \label{eq:obsPert:1}
    \delta O = \left(\frac{d O} {d \varepsilon_c}\right) \delta \varepsilon_c  = \delta B \left(\frac{d O / d \varepsilon_c}{d B / d \varepsilon_c}\right). 
\end{equation}
If we include the effects of rotations, then two parameters are needed: $(\varepsilon_c, \overline{\omega}_c)$, in which $\overline{\omega}_c\equiv \Omega - \omega(r)$, $\Omega$ is the angular frequency of the star, and $\omega(r)$ is the frequency of the local inertial frame~\cite{Hartle:1967he}. Since BNV does not change the angular momentum of the star, $L(\varepsilon_c, \overline{\omega}_c) = L'(\varepsilon'_c, \overline{\omega}'_c)$ and $B'(\varepsilon'_c, \overline{\omega}'_c) = B(\varepsilon_c, \overline{\omega}_c) + \delta B$ can still be solved for a unique set $(\delta\varepsilon_c, \delta\overline{\omega}_c)$. Therefore, the generalization to the rotating case is possible but we will consider the static non-rotating scenario for simplicity. Given the relative rate of change in the baryon number $\dot{B}/B \equiv  (dB / dt)/ B$, we find the relative rate of change in the observable $O$ to be
\begin{equation}
    \label{eq:obsPert:time}
    \frac{\dot{O}/O}{\dot{B}/B} = \left(\frac{d O / d \varepsilon_c}{O}\right)  \left(\frac{B}{d B / d \varepsilon_c}\right).
\end{equation}
The most stringent bound on BNV rates can be set with a careful choice of $O$. The relative change in $O$ is given by $\delta O/O \propto d \ln{O} / d \varepsilon_c$, such that quantities that are more sensitive to $\varepsilon_c$ will have a greater variation. At the same time, an observable quantity ($O$) that can be measured to a higher precision would yield a better constraint.

Thus far we have presented a global examination of a neutron star's  macroscopic observable quantities in terms of a given rate of change in the total baryon number ($\dot{B}$). Alternatively, we can inspect the local properties of Eq.~\eqref{eq:barCons} in terms of the baryon number density ($n(r)$). Since neutron stars are compact objects with $n(R) \ll n(r<R)$, the baryon conservation condition (Eq.~\eqref{eq:barCons}) can be expanded to leading order in the perturbation as
\begin{equation}
\label{eq:baryonPert}
\int_0^R \left[1 - \frac{2M(r)}{r}\right]^{-\frac{1}{2}} r^2 \left[ \Tilde{\delta} n(r)  - \delta n(r) - \frac{\delta M(r)}{r - 2M(r)} n(r) \right] dr = 0,
\end{equation}
in which $\delta M(r)$ is the variation of the mass function (Eq.~\eqref{eq:m:def}) and it is given by
\begin{equation}
\label{eq:m:var}
\delta M(r') = 4\pi \int_0^{r'} \delta \varepsilon(r) r^2 dr, 
\end{equation}
with $\delta \varepsilon(r) = (d \varepsilon / d n) \delta n(r)$. In principle the integral equation in~\eqref{eq:baryonPert} should be solved for $\delta n(r)$ given $\Tilde{\delta} n(r)$ and in a consistent manner with the TOV equations (Eq.~\eqref{eq:p:der}). The function $\Tilde{\delta} n(r)$ can be calculated from the rates of the BNV processes, which implicitly depend on time ($t$) and radial position ($r$). Specifically we have
\begin{equation}
\label{eq:deltan}
    \Tilde{\delta} n(r, t) = f(n) \times n(r, t) \times \Gamma_{\rm BNV}(n) \times \delta t \times \Delta B,
\end{equation}
in which $f(n)$ is the relative fraction of the decaying species ($i$) in the BNV process, i.e., $n_i (r, t) = f (n) \times n(r, t)$, and $\Delta B$ is the change in the baryon number per each decay. We expect a BNV process to slow down and eventually halt once the density falls below the threshold needed for that specific BNV reaction. For example, if only hyperons are involved in the BNV process, then as the baryon number density drops below a certain threshold the star will essentially be depleted of hyperons and the BNV reaction stops.

\begin{figure}[t!]
    \centering
    \includegraphics[width=0.8\linewidth]{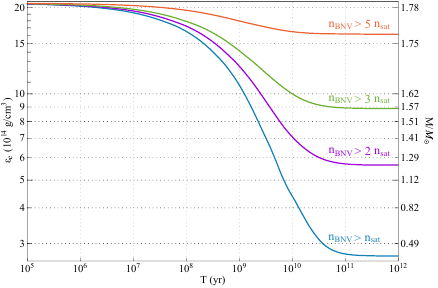}
    \caption{Evolution of neutron stars along their one-parameter sequence in the presence of BNV processes. The central energy density ($\varepsilon_c$) is plotted as a function of time for $f = \Delta B=1$, $\Gamma_{\rm BNV} = 10^{-10}\, {\rm yr}^{-1}$, and four different BNV number-density thresholds ($n_{\rm BNV}$). We have used the hyperonic EoS H3 from~\cite{Lackey:2005tk} together with the BPS EoS~\cite{Baym:1971pw} for lower densities.}
    \label{fig:eps:T}
\end{figure}

If we assume that the conditions from Sec.~\ref{subsec:BV in NS:gen} are satisfied by the BNV processes, then $n(r,t)$ is simply given by the solutions to the one-parameter sequence. Specifically, let us assume that BNV is only active in regions of the star that have a baryon number density greater than a certain threshold, i.e., $n_{\rm BNV} > N\, n_{\rm sat} = 0.16 N \, \rm  fm^{-3}$, in which $N$ is a ratio of order unity. In other words, we assume the form
\[ \Gamma_{\rm BNV}(n) = \begin{cases} 
      0 & n< N\, n_{\rm sat} \\
      \Gamma_{\rm BNV} & n\geq N\, n_{\rm sat}
   \end{cases}
\]
Starting from any point ($\varepsilon_c^i$) along the sequence in $\varepsilon_c$ %(Fig.~\ref{fig:odot}) 
with a known density ($n(r)$) and mass profile ($M(r)$), we can find the change in the total baryon number of the star $\delta B$ using Eq.~\eqref{eq:deltan} and
\begin{equation}
\label{eq:deltaB}
    \delta B  = 4\pi \int_0^R \left[1 - \frac{2M(r)}{r}\right]^{-\frac{1}{2}}\, \Tilde{\delta}n(r) \,r^2\, dr.
\end{equation}
We can then find the $\delta \varepsilon_c$ corresponding to this $\delta B$, and repeat this process for the new point ($\varepsilon_c^f = \varepsilon_c^i + \delta \varepsilon_c$) until the central density $n(r=0)$ falls below the threshold ($n_{\rm BNV}$), and the BNV process is deactivated. We have adopted the hyperonic EoS H3 ($K = 300$ MeV, $m^{\star}/m = 0.70$, $x_{\sigma} = 0.60$) from Ref.~\cite{Lackey:2005tk} and plotted the results for $f=\Delta B =1$ and $\Gamma_{\rm BNV} = 10^{-10}\, {\rm yr}^{-1}$ in Fig.~\ref{fig:eps:T}. We considered a set of four BNV threshold densities $n_{\rm BNV} >\{1,2,3,5\}n_{\rm sat}$ for illustration. As expected, BNV processes with a lower density threshold have a much more significant effect because the BNV is active in a larger region inside the star. We expect the BNV effects to be most notable for a timescale $T \sim 1/\Gamma_{\rm BNV}$, and we see that the epoch between $10^{8} ({\rm yr}) \lesssim T \lesssim 10^{10}\, ({\rm yr})$ shows the fastest evolution for the neutron star. Similar results can be found for other choices of parameters $f$, $\Delta B$, and $\Gamma_{\rm BNV}$ via
proper scaling of the timescale $T$.

%%%%%%%%%%%%%%%%%%%%%%%%%%%%%%%%%%%%%%%%%%%%

\subsection{Constraining BNV from neutron star observations:}
\label{subsec:BV in NS:results}

\begin{figure}[t!]
    \centering
    \subfigure[]{\includegraphics[width=0.482\linewidth]{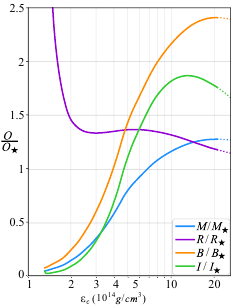}}
    \subfigure[]{\includegraphics[width=0.50\linewidth]{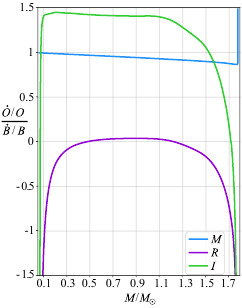}}
    \caption{(a) The set of observable quantities ($O$): mass $(M)$, radius $(R)$, baryon number $(B)$, and moment of inertia $(I)$ for a sequence of neutron stars as a function of the central energy density ($\varepsilon_c$) relative to their canonical values ($O_{\star}$): $M_{\star} = 1.4\, M_{\astrosun}$, $R_{\star} = 10\, {\rm km}$, $B_{\star} = 10^{57}$, $I_{\star} = 70\, ({\rm M_{\astrosun}\, {km}^2})$. (b) The relative rate of change in three observable parameters ($O = M, R, I$) divided by the relative rate of change in the total baryon number ($B$) as a function of the neutron stars' mass. We have chosen the hyperonic EoS H3 from~\cite{Lackey:2005tk} together with the BPS EoS~\cite{Baym:1971pw} for lower densities.}
    \label{fig:odot}
\end{figure}

We calculated the expression in Eq.~\eqref{eq:obsPert:time} for a sequence of neutron stars with the hyperonic EoS H3 from Ref.~\cite{Lackey:2005tk}. The values of mass ($M$), radius ($R$), baryon number ($B$) and moment of inertia ($I$) divided by a set of canonical values ($M_{\star} = 1.4\, M_{\astrosun}$, $R_{\star} = 10\, {\rm km}$, $B_{\star} = 10^{57}$, $I_{\star} = 70$ $M_{\astrosun}$ km$^2$) are plotted on the left side of Fig.~\ref{fig:odot}. Note that the sequence becomes unstable (dotted curves) 
%at both ends, i.e., 
where $dM/d\varepsilon = 0$~\cite{Misner:1964zz}. The rates of change for mass, radius, and moment of inertia from the expression in Eq.~\eqref{eq:obsPert:time} are plotted on the right side of Fig.~\ref{fig:odot}. We can see that for almost all the points along the sequence the ratio in Eq.~\eqref{eq:obsPert:time} is of ${\cal O}(1)$. This indicates that relative changes in a neutron star's observable would be at the same order as the relative changes in the baryon number. The exceptions to this statement are in the beginning and at the end of the sequence where $B$ has extrema, i.e.,  $dB/d\varepsilon_c = 0$. Although the exact locations of these extrema depend on the EoS, their existence is independent of it. Close to these points an infinitesimal change in the baryon number would result in a substantial variation in other observable quantities. As a consequence, the existence of BNV processes as we have outlined would shift the heavier neutron stars away from the maximum mass, and make light neutron stars close to the minimum mass unstable. 

\begin{figure}[t!]
    \centering
    \includegraphics[width=0.8\linewidth]{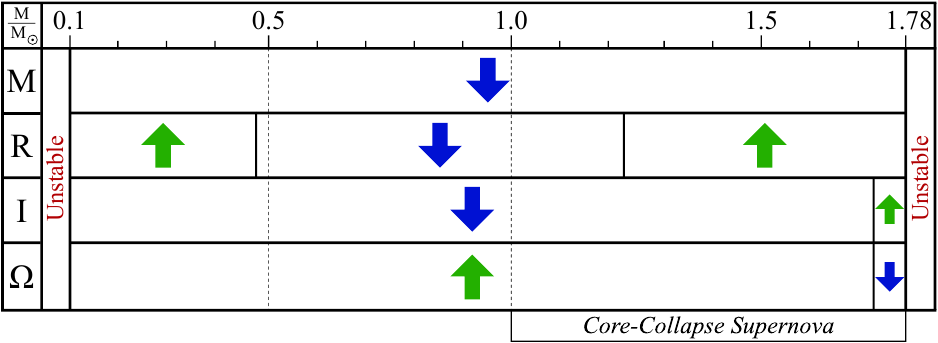}
    \caption{The generic effects of BNV on neutron star parameters: mass ($M$), radius ($R$), moment of Inertia ($I$), and angular frequency ($\Omega$) from Fig.~\ref{fig:odot} for $\dot{B}<0$. Neutron stars lose mass during their evolution, i.e., they evolve from the right side to the left. The arrows indicate if an observable is increasing (green) or decreasing (blue). Typical core-collapse supernova scenarios generate neutron stars with masses above $\approx M_{\astrosun}$. Those with masses below $M_{\rm min} \approx 0.1\, M_{\astrosun}$ and above $M_{\rm max} \approx 1.78\, M_{\astrosun}$ are unstable. We note that the value of the latter depends on the adopted EoS.}
    \label{fig:ns:evol}
\end{figure}

The main question regarding the destabilizing effects of BNV on low-mass neutron stars is how light neutron stars in nature can be. In order to answer this question, it is helpful to consider different aspects of neutron stars including their theoretical mass limit, observation, birth, and evolution. The absolute (theoretical) minimum mass for a stable cold neutron star is about $M_{\rm min} = 0.1\, M_{\astrosun}$~\cite{Baym:1971pw}. If the mass of a neutron star in hydrostatic equilibrium decreases below $M_{\rm min}$, then it will become unstable and explode~\cite{1990SvA....34..595B}. This will result in a burst of hard X-rays and soft gamma rays with a total energy of $10^{43} - 10^{47}$ erg~\cite{1990SvA....34..595B}. On the observation side, the current minimum neutron star mass is observed to be in the range $1 $ -- $1.1\, M_{\astrosun}$ ~\cite{Lattimer:2012nd,Martinez:2015mya}, albeit with significant errors, with one of the lightest neutron stars observed, PSR J0453+1559, having a more precisely determined mass of $M = 1.174 (4)\, M_{\astrosun}$\cite{Martinez:2015mya}. As for their birth, neutron stars can either be directly born in explosive death of a massive star ($>8 M_{\astrosun}$)~\cite{Potekhin2010, Burrows:2020qrp}, or as a result of white dwarf accretion-induced collapses~\cite{1991ApJ...367L..19N}. Typical core-collapse supernova creation scenarios for neutron stars predict that their masses must be at least $1\, M_{\astrosun}$~\cite{Lattimer:2004pg, Haensel:2002cia}. Furthermore, the minimum mass of hot proto-neutron stars predicted by the models considered in Ref.~\cite{1999A&A...350..497S} and formed from supernovae is in the range $0.89$ -- $1.13$ $M_{\astrosun}$. Therefore, the core-collapse supernova paradigm would appear to impose a lower limit on neutron star masses at birth~\cite{Lattimer:2012nd,Suwa:2018uni}. As an example of a non-standard scenario, the effects of dark matter accretion by a massive white dwarf and the subsequent core collapse into a light proto-neutron star is studied in Ref.~\cite{Leung:2019ctw}. They show that a dark matter admixed core with $0.01$ $M_{\astrosun}$ of dark matter can lower the minimum mass to $1$ $M_{\astrosun}$. On a distinct note, the discovery of ZTF J190132.9+145808.7~\cite{Caiazzo:2021xkk}, which is a nearby ($41$ pc) white dwarf with a mass $\sim1.3 M_{\astrosun}$, opens up the possibility of observing a new formation channel~\cite{Schwab:2017epw} for neutron stars, which may yield relatively lighter masses. Therefore, in the absence of a mass-loss mechanism, it is unlikely to have isolated neutron stars with masses below $1$ $M_{\astrosun}$~\cite{Lattimer:2012nd}. On the other hand, neutron stars may lose some of their mass during the course of their evolution. For example, in close binaries, accretion of mass from one component to another is possible. Since a decrease in a neutron star mass increases its radius (making it even more susceptible to mass loss in the binary), this accretion can be self-accelerating and lead to the explosion of the low-mass component~\cite{Blinnikov:2018boq, Sumiyoshi:1997pz}. The existence of BNV processes would only increase this acceleration, as in  Fig.~\ref{fig:ns:evol}. If BNV is active in light isolated neutron stars, then it can lower their masses below $1$ $M_{\astrosun}$, and even possibly result in a mass as small as $0.1$ $M_{\astrosun}$, in which case it would lead to an explosion which we may be able to detect. However, we should note that if a baryon number density of at least $n_{\rm sat}$ is required for the BNV reactions to occur, it seems implausible for the mass to decrease below $\sim 0.4$ -- $0.5\, M_{\astrosun}$ (see Fig.~\ref{fig:eps:T}). In any case, the observation of light isolated neutron stars with $M < 1\, M_{\astrosun}$ would point to a mass-loss mechanism which could be caused by BNV, or else a modification of neutron star genesis theories would be required. Moreover, the continued lack of such observations can be used to constrain BNV processes in neutron stars.

As pulsars emit radiation they lose rotational energy and their spin periods ($P_s = 2\pi/\Omega$) slow down over time~\cite{Goldreich:1969sb}. An accurate knowledge of this mechanism along with pulsar timing data can then be used to constrain additional non-standard contributions to this slow-down. A spherically symmetric BNV sink within the neutron star would not affect its angular momentum ($L = I \Omega$), but as we showed in Fig.~\ref{fig:odot}, the moment of inertia ($I$) changes and as a result the spin period  would also change. For heavier pulsars ($M \gtrsim 1.7\,M_{\astrosun}$) BNV will have a spin-down effect, whereas for lighter pulsars it will cause a spin-up effect, as indicated in Fig.~\ref{fig:ns:evol}. Assuming that the BNV processes (only) are responsible for a change in a pulsar's spin period, we have
\begin{equation}
    \frac{\dot{I}}{I} =  \frac{\dot{P_s}}{P_s} \approx \pm \frac{\dot{B}}{B}, 
\end{equation}
in which for $\dot{B}<0$, the plus (minus) sign corresponds to spin-up (down), and the last equality comes from $|(\dot{I}/I)/(\dot{B}/B)| \sim \mathcal{O}(1)$. The spin period and its first derivative have been measured to a remarkable precision for many pulsars. Taking advantage of these precise measurements would require a robust understanding of the pulsar spin-down due to magnetic dipole radiation. An independent measurement of the magnetic field would make it possible to separate the exotic (e.g., BNV) from the standard electromagnetic contributions to the spin-down rate. This is not currently possible and so the extreme precision in $\dot{P_s}/P_s$ can not be utilized. However, the total value can still be used for comparison, i.e., $|(\dot{P_s}/P_s)_{\rm BNV}| <  |(\dot{P_s}/P_s)_{\rm Obs}|$.  

Alternatively, stronger limits (compared to the pulsar spin-down bounds) can be inferred from the decay rate of a binary pulsar's orbital period ($\dot{P}_b$), which can be used to constrain changes in its components' parameters (e.g., in their masses~\cite{Goldman:2009th}). 
The observed relative rate of orbital period decay comprises of various intrinsic and extrinsic terms with the following dominant contributions~\cite{1991ApJ...366..501D}:
\begin{equation}
    \left(\frac{\dot{P}_b}{P_b}\right)^{\rm obs}  =  \underbrace{\, \left(\frac{\dot{P}_b}{P_b}\right)^{\rm GR} + \left(\frac{\dot{P}_b}{P_b}\right)^{\dot{E}}\, }_\textrm{intrinsic} +
    \left(\frac{\dot{P}_b}{P_b}\right)^{\rm ext}.
\end{equation}
The first term is due to gravitational radiation~\cite{PhysRev.136.B1224}, the second term is due to mass-energy loss, and the third term includes the extrinsic effects such as Doppler effects caused by the relative acceleration (due to the Galactic gravitational potential) of a binary pulsar with respect to the solar system.

The rate of change in $P_b$ due to gravitational radiation ($\dot{P}_b^{\rm GR}$) can be expanded as a series in powers of $(v/c)^2$ in the post-Newtonian (PN) approximation. To leading order (2.5PN),\footnote{Here 2.5PN, and 3.5PN refer to terms of order $(v/c)^5$, and $(v/c)^7$ respectively.} $\dot{P}_b^{\rm GR}$ is given by the quadrupole formula in Ref.~\cite{PhysRev.136.B1224}, and the next higher correction (3.5PN) is calculated in Ref.~\cite{10.1093/mnras/239.3.845} which would be needed for more accurate values of $\dot{P}_b^{\rm obs}$ (e.g., in the case of J0737$-$3039A/B~\cite{PhysRevX.11.041050}). We set limits on $\dot{P}_b^{\dot{E}}$ by subtracting this GR contribution ($\dot{P}_b^{\rm GR}$) from the intrinsic orbital-period decay rate, $\dot{P}_b^{\rm int} \equiv \dot{P}_b^{\rm obs} - \dot{P}_b^{\rm ext}$, for three binary pulsar examples (Table~\ref{tab:psrbinary}):

\begin{enumerate}
    \item \textbf{PSR B1913+16:} This binary system (Hulse-Taylor binary) is the first binary pulsar ever discovered~\cite{Hulse:1974eb}, and consists of a neutron star ($M_c = 1.39\, M_{\astrosun}$) and a pulsar ($M_p = 1.44\, M_{\astrosun}$) with a pulse period of $59$ ms. We use the results from the analysis in Ref.~\cite{Weisberg:2016jye} which is based on timing measurements performed over the last 35 years. 
    
    \item \textbf{PSR J0737$-$3039A/B:} The only known double pulsar was discovered in 2003~\cite{Burgay:2003jj}, and is comprised of two radio pulsars ($A$ and $B$) with masses $M_A = 1.34\, M_{\astrosun}$, $M_B = 1.25\, M_{\astrosun}$, and with pulse periods of $22.7$ ms and $2.8$ ms, respectively. We use Ref.~\cite{PhysRevX.11.041050} which is based on data acquired over 16 years of observation. As a result of the increased accuracy in measurements, the higher-order GR corrections (3.5PN) to $\dot{P}_b^{\rm GR}$, and contribution to $\dot{P}_b^{\dot{E}}$ from the spin-down of pulsar A are added~\cite{PhysRevX.11.041050}. 
    
    \item \textbf{PSR J1713+0747:} This binary system was discovered in 1993~\cite{1993ApJ...410L..91F}, and it contains a $4.6$ ms radio pulsar with $M= 1.3\, M_{\astrosun}$ and a companion white dwarf with $M = 0.29\, M_{\astrosun}$~\cite{Zhu:2018etc}. A sub-microsecond precision is achieved at measuring its pulse time of arrivals~\cite{Zhu:2018etc} owing to the short spin period and its narrow profile. It has a much longer orbital period ($P_b = 67.8$ day) compared to the other two binaries considered here. This is why the inferred limit on $(\dot{P}_b/P_b)^{\dot{E}}$ is one order of magnitude better than the limit from the Hulse-Taylor binary despite the higher precision in the latter. 

\end{enumerate}
The $2\sigma$ (98\% C.L.) bound on $(\dot{P}_b/P_b)^{\dot{E}}$ for each of these binaries is listed in Table~\ref{tab:psrbinary}. We now elaborate on the BNV contributions, via changes in $M$ and $I$, to $(\dot{P}_b/P_b)^{\dot{E}}$. 

\begin{table}[t!]
    \centering
    %\resizebox{\textwidth}{!}{%
    \begin{tabular}{|c|c|c|c|}
    \hline 
    Name & J0737$-$3039A/B  & B1913+16 & J1713+0747 \\
        \hline\hline
        $P_b \, ({\rm days})$ & $0.1022515592973(10)$ & $0.322997448918(3)$ & $ 67.8251299228(5)$\\\hline
        $\dot{P}_b^{\rm int} \, (\times 10^{-12})$ & $-1.247752(79)$ & $-2.398(4)$ & $0.03(15)$\\\hline
        $\dot{P}_b^{\rm GR} \, (\times 10^{-12})$ & $-1.247827(+6,-7)$ & $-2.40263(5)$ & $-6.3(6)\times 10^{-6}$\\\hline
         $(\frac{\dot{P}_b}{P_b})^{\dot{E}}_{2\sigma} \, ({\rm yr^{-1}})$ & $8.3\times 10^{-13}$  & $1.4\times 10^{-11}$  & $1.8\times 10^{-12}$\\\hline
         $(\frac{\dot{P}_b}{P_b})^{\dot{\Omega}}\, ({\rm yr^{-1}})$ & $1.04(7)\times 10^{-13}$ & $\lesssim 2.5 \times 10^{-13} $ & $\approx 8\times10^{-14}$\\\hline
        $(\frac{\dot{P}_b}{P_b})^{\rm BNV}_{2\sigma} \, ({\rm yr^{-1}})$ & $7.3\times 10^{-13}$  & $1.4\times 10^{-11}$  & $1.8\times 10^{-12}$\\\hline
        $|\frac{\dot{B}}{B}|_{2\sigma} \, ({\rm yr^{-1}})$ & $3.7\times 10^{-13}$  & $7\times 10^{-12}$  & $1.1\times 10^{-12}$\\\hline
    \end{tabular}%}
    \caption{The values of the orbital period ($P_b$), its intrinsic decay rate ($\dot{P}_b^{\rm int}$), and the gravitational wave radiation contributions ($\dot{P}_b^{\rm GR}$) to it for J0737-3039A/B~\cite{PhysRevX.11.041050}, B1913+16 (Hulse-Taylor)~\cite{Weisberg:2016jye}, and J1713+0747~\cite{Zhu:2018etc} binary pulsars. The bound on $(\dot{P}_b/P_b)^{\dot{E}}$ is found from the difference between $\dot{P}_b^{\rm int}$ and $\dot{P}_b^{\rm GR}$. The BNV ($(\dot{P}_b/P_b)^{\rm BNV}$), and spin-down contributions ($(\dot{P}_b/P_b)^{\dot{\Omega}}$) are given in Eq.~\eqref{eq:pbdot:bnv} and Eq.~\eqref{eq:pbdot:spindown} respectively. The last row is the $2\sigma$ bound (98\% C.L.) on the relative rate of change in baryon number ($\dot{B}/B$). }
    \label{tab:psrbinary}
\end{table}

We begin by noting that since the mass loss due to BNV is spherically symmetric, and it appears in the form of photons and neutrinos, implying that we have the very high velocity ejecta needed, we can thus apply the \textit{Jean's mode of mass ejection}~\cite{1963ApJ...138..471H}. In this mode the relative rate of change in the binary period is given by~\cite{10.1093/mnras/85.1.2, 10.1093/mnras/85.9.912}
 \begin{equation}
 \label{eq:jeans:mloss}
    \left(\frac{\dot{P}_b}{P_b}\right)^{\dot{E}} = - 2 \left(\frac{\dot{M}_1^{\rm eff} + \dot{M}_2^{\rm eff}}{M_1 + M_2}\right),
 \end{equation}
in which $M_{1,2}$ are the masses for each of the components in the binary system, and $\dot{M}_{1,2}^{\rm eff}$ is their respective mass loss which, by virtue of Einstein's mass–energy equivalence,  can be written as
\begin{equation}
\label{eq:mdot:eff}
    \dot{M}^{\rm eff} = \frac{d}{dt} \left( M + \frac{1}{2} I \Omega^2 \right) = \dot{M} + \frac{1}{2} \dot{I} \Omega^2 + I \Omega \dot{\Omega},
\end{equation}
in which we suppressed indices $1,2$. The first term is due to a direct (rest) mass loss, which could be caused by BNV. The second term is due to a change in the moment of inertia ($I$) which has a direct contribution from BNV ($\dot{I}_{\rm BNV}$) and an indirect contribution as a result of changes in the angular velocity ($\Omega$). We assume that the latter effect, i.e., $\dot{I}_{\Omega} = (dI/d\Omega) \dot{\Omega}$, is negligible. The third term is the energy loss due to the pulsar spin-down which arises from both BNV and electromagnetic radiation. Therefore, after defining $\eta^{(O)} \equiv (\dot{O}/O)/(\dot{B}/B) \approx \mathcal{O}(1)$ for an observable ($O$), we can rewrite Eq.~\eqref{eq:mdot:eff} in terms of the observed pulsar spin periods ($P_s$), and its observed rate of change ($\dot{P_s}$) as
\begin{equation}
\label{eq:mdot:eff2}
    \dot{M}^{\rm eff} = \underbrace{\, \eta^{(M)}\left(\frac{\dot{B}}{B}\right) M + \eta^{(I)}\left(\frac{\dot{B}}{B}\right) \left(\frac{2 \pi^2 I}{P_s^2}\right) \, }_\textrm{BNV} - \frac{4\pi^2 I \dot{P_s}}{P_s^3},
\end{equation}
in which the values for $\eta^{(M,I)}$ can be read from Fig.~\ref{fig:odot}. Equation \eqref{eq:jeans:mloss} can then be written in terms of the two separate contributions from BNV and spin-down effects ($\dot{\Omega}$) as
 \begin{equation}
 \label{eq:jeans:mloss:expanded}
    \left(\frac{\dot{P}_b}{P_b}\right)^{\dot{E}} = \left(\frac{\dot{P}_b}{P_b}\right)^{\rm BNV} + \left(\frac{\dot{P}_b}{P_b}\right)^{\dot{\Omega}},
 \end{equation}
with each of the contributions given by
\begin{align}
  \left(\frac{\dot{P}_b}{P_b}\right)^{\rm BNV}  =& \frac{- 2}{M_1 + M_2}\sum_{i=1,2}  \left(\frac{\dot{B}_i}{B_i}\right) \left[ \eta^{(M)}_i M_i + \eta^{(I)}_i \left(\frac{2 \pi^2 I_i}{P_{s,i}^2}\right) \right],\label{eq:pbdot:bnv}\\
  \left(\frac{\dot{P}_b}{P_b}\right)^{\dot{\Omega}}  =& \frac{8\pi^2}{M_1 + M_2} \left(\frac{ I_1 \dot{P}_{s,1}}{P_{s,1}^3} + \frac{ I_2 \dot{P}_{s,2}}{P_{s,2}^3}\right).\label{eq:pbdot:spindown}
\end{align}

We now turn to the assessment of the period decay rate due to pulsar spin-down $\dot{P}_b^{\dot{\Omega}}$ for the different binary systems of interest. 
As we will see, this is only relevant for PSR J0737$-$3039A/B. In that case, the rates of orbital decay for pulsars A and B are given by $2.3\times 10^{-17}\, I_A^{45}$ and $6.3\times 10^{-21}\, I_B^{45}$ respectively~\cite{10.1093/mnras/staa2107}, in which $I^{45}\equiv I / (10^{45}\, {\rm g\, {cm}^2})$, $I_A^{45} \approx 1.15-1.48$~\cite{PhysRevX.11.041050}, and the negligible contribution from pulsar B is ignored, giving the result reported in Table~\ref{tab:psrbinary}. In the case of the Hulse-Taylor binary we use the estimated value for the pulsar ($1$), and the 68\% C.L. bound on the companion neutron star ($2$) from Ref.~\cite{1991ApJ...366..501D}:
\begin{align}
    \left(\dot{P}_b/P_b\right)^{\dot{\Omega}_1} &\approx \left(2.1 \pm 0.6 \right) \times 10^{-14}\, ({\rm yr^{-1}}),\\
    \left(\dot{P}_b/P_b\right)^{\dot{\Omega}_2} &\lesssim 2.3  \times 10^{-13}\, ({\rm yr^{-1}}) \qquad \textrm{(\,at 68\%\,C.L.\,)}.
\end{align}
We see that the combined effect is two orders of magnitude smaller than the limit on $(\dot{P}_b/P_b)^{\dot{E}}$ shown in  Table~\ref{tab:psrbinary} and can be safely ignored. In the case of PSR J1713+0747, we estimate the spin-down effects using Eq.~\eqref{eq:pbdot:spindown} with $I^{45} \approx 1$~\cite{Lattimer_2005}, $\dot{P}_s = 8.96(3)\times 10^{-21}$~\cite{Zhu:2018etc} and ignore the companion white dwarf's contribution as it is much lighter in mass. We see that the spin-down effect is about two orders of magnitude smaller than the limit on $(\dot{P}_b/P_b)^{\dot{E}}$ for this binary and thus it can be neglected.

With these estimates, we subtract the spin-down contributions ($\dot{\Omega}$) from the energy-loss term in Eq.~\eqref{eq:jeans:mloss:expanded} to find 98\% C.L. limits on the BNV contributions and record them in  Table~\ref{tab:psrbinary}. In principle, given the values for all of the parameters in Eq.~\eqref{eq:pbdot:bnv}, we would be able to infer limits on a linear combination of $(\dot{B}/B)_{1,2}$, but not on each of them individually. One may attempt to resolve this degeneracy by measuring parameters other than binary period decay rate, such as the individual pulsar spin-down rate. Unfortunately, as we have already mentioned, our understanding of the electromagnetic contributions to the pulsar spin-down rate are not as precise as the GR contributions to the binary period decay rate, making the overall separation of the contributions challenging.

For a general estimate, however, we can make theoretical assumptions about the nature of BNV processes to resolve this issue. In the case of PSR J1713+0747, we assume that BNV would be only active in the pulsar and not the white dwarf companion. For the other two binaries, given that the masses of binary components are close to each other, we assume that $\dot{B}_1/B_1 \approx \dot{B}_2/B_2 \equiv \dot{B}/B$. The exact value of $\eta$ coefficients depends on the adopted EoS, but since they are of order unity (Fig.~\ref{fig:odot}), we can approximate them as $\eta^{(M)} \approx \eta^{(I)} \approx 1$. Equation \eqref{eq:pbdot:bnv} is then simplified as
\begin{align}
  \left(\dot{P}_b/P_b\right)^{\rm BNV}_\textrm{NS-NS}  &\approx -2 \left(\frac{\dot{B}}{B}\right) \left[  1 +  \frac{2 \pi^2}{M_1 + M_2}\left(\frac{I_1}{P_{s,1}^2} + \frac{I_2}{P_{s,2}^2}\right) \right],\label{eq:pbdot:bnv:simp:ns-ns}\\
 \left(\dot{P}_b/P_b\right)^{\rm BNV}_\textrm{NS-WD}  &\approx \frac{-2 M_1}{M_1 + M_2} \left(\frac{\dot{B}}{B}\right) \left[  1 +  \frac{2 \pi^2}{P_{s,1}^2} \left(\frac{I_1}{M_1}\right) \right].\label{eq:pbdot:bnv:simp:ns-wd}
\end{align}
The second term in both cases is more than two orders of magnitude smaller than $1$ and can be safely neglected. We translate the bounds on $(\dot{P}_b/P_b)^{\rm BNV}$ into limits on $\dot{B}/B$ and report our results in the last row of Table~\ref{tab:psrbinary}. 

We can write an expression for the derivative of baryon number, $\dot{B} = f \times B \times \Gamma_{\rm BNV}$, with $f$ being the proportion of the baryons involved in the BNV process. If we assume that most of the matter inside the neutron star has densities above the required BNV threshold we would have $f\approx 1$ for neutrons, and $f\approx 10^{-3}$ for hyperons. The limit on $\Gamma_{\rm BNV}$ is then given by
 \begin{equation}
 \label{eq:BNV rate:binary limit}
     \Gamma_{\rm BNV} \lesssim \alpha \left(\frac{0.01}{f}\right) \left(\frac{1} {10^{10}\, {\rm yr}}\right),
 \end{equation}
in which $\alpha = 0.4, 7, 1$ for PSR J0737$-$3039A/B, the Hulse-Taylor binary, and PSR J1713+0747 respectively. This indicates that if BNV is active throughout any of these binary pulsars, its rate must be less than one per $10^{10}\, {\rm yr}$, i.e., the characteristic lifetime for a typical pulsar.\footnote{The \emph{characteristic} age of a pulsar is defined as $\tau = P_s/(2\dot{P}_s)$. This is used as a proxy for the \emph{true} age, which is only known for two pulsars: NS1987A and Cassiopeia A, whose associated supernovae happened to have occurred during recorded human history.}

In this section, we have studied the generic consequences of BNV processes in neutron stars. Our only assumptions have been that the rates for such processes are slower than the weak interactions, and their final products will ultimately, either directly or via a cascade of interactions, turn into the particles already present in the neutron star and with structure dictated by the standard BNC EoS. We have demonstrated that these processes can relocate the neutron stars along their one-parameter sequence away from the maximum mass configuration and have analyzed the rate at which this can occur. We have also shown that observations of neutron star properties such as the orbital periods of pulsar binaries can lead to stringent constraints on this generic class of BNV processes. 

In the following sections we consider specific BNV mechanisms, as well as model realizations thereof, and their effects on neutron star physics. Three types of processes are  of interest. The first concerns models in which baryon number is not violated but transferred to a hidden sector. Since these baryon-number-carrying particles are not observed, these processes appear to be violating baryon number. We consider this apparent BNV and its implications in the next section. In subsequent sections we consider explicit and spontaneous BNV in turn. 

%%%%%%%%%%%%%%%%%%%%%%%%%%%%%%%%%%%%%%%%%%%%
\section{Implications of apparent BNV}
\label{sec:apparent BV}
%%%%%%%%%%%%%%%%%%%%%%%%%%%%%%%%%%%%%%%%%%%%

The possibility of apparent baryon-number violation, with the concomitant notion that baryon number can also be carried by particles of a hidden sector, emerges naturally from the idea that ordinary baryonic matter and dark matter could share a common origin~\cite{Hut:1979xw,Nussinov:1985xr,Barr:1990ca}, since the dark matter relic density is within a factor of a few of that of ordinary matter. Thus the dark matter relic density would be set by its cosmic relic asymmetry much as in the case of the relic density of ordinary baryonic matter. The earliest examples of such \emph{asymmetric dark matter} models were in the context of technicolor models~\cite{Nussinov:1985xr,Barr:1990ca}, and a number of mechanisms to explain the cosmic genesis of dark and visible matter have since been proposed --- and we direct the reader to the review of Ref.~\cite{Davoudiasl:2012uw} for a succinct summary. 

We have noted that the neutron lifetime anomaly could, in principle, be resolved through the existence of a dark decay channel of the neutron, in which one or more or all particles in the final state are uncharged under the SM gauge groups~\cite{Fornal2018PhRvL.120s1801F,Fornal:2020gto}. There are many such possibilities. The neutron, e.g., could oscillate to a mirror, or dark, neutron~\cite{Berezhiani:2005hv} or a neutron could be destroyed through its interaction with asymmetric DM~\cite{Davoudiasl:2011fj} --- or it could decay to an exotic final state~\cite{Fornal2018PhRvL.120s1801F}. Alternatively, if a neutron star were to capture a Q-ball~\cite{Coleman:1985ki,Frieman:1988ut}, noting that $B$- or $L$-carrying Q-balls can appear in supersymmetric extensions of the SM~\cite{Kusenko:1997si}, then neutrons can be consumed by the Q-ball, increasing its baryonic charge, shortening the lifetime of the star~\cite{Kusenko:2005du}. Although these scenarios are all realizations of neutron disappearance, we say that the baryon number violation is {\it apparent}, rather than explicit, because baryon number remains unbroken. That is, in these scenarios the concept of baryon number has been generalized, as appropriate, to include particles in a hidden sector~\cite{Davoudiasl:2012uw}. This construct is convenient in that (1) it permits the appearance of exotic decays of the neutron without incurring proton decay~\cite{Fornal2018PhRvL.120s1801F}, (2) it enables models of cosmic baryogenesis through visible-hidden sector interactions~\cite{Davoudiasl:2010am,Davoudiasl:2012uw}, and (3) it can also provide a natural way of stabilizing a DM candidate. Thus the possibility of apparent baryon violation can be used more broadly, and we note Ref.~\cite{Heeck:2020nbq} for a study of the possible decay channels in the presence of light particles with $B$ or $L$, such as $p \to \pi^+ \chi$ decay, which searches for $p\to \pi^+ \bar \nu$ limit severely, as we show in the next section. Earlier work concerning the possibility of dark neutron decay~\cite{Davoudiasl:2014gfa} via a mass-dimension-six ``neutron portal'' such as $udd \chi_L^c/\Lambda_n^2$~\cite{Davoudiasl:2010am} gives rise to decays such as $n\to \chi \gamma$ and $n\to \chi Z_d$ with $Z_d \to e^+ e^-$, though the predicted rates are too slow to explain the neutron lifetime anomaly~\cite{Fornal2018PhRvL.120s1801F}. It strikes us that apparent baryon violating processes stand out among the BNV processes we consider in that they are not immediately required to occur very much more slowly than ordinary weak processes within the SM. Thus in this section, we focus on models of this ilk and consider not only how the existence of neutron stars constrains them but also how such constraints can potentially be evaded. 

We open our discussion with a recap of made-to-measure models of the neutron lifetime anomaly~\cite{Fornal2018PhRvL.120s1801F}, and we refer to Ref.~\cite{Fornal:2020gto} for a review. A model with operator ${\cal O}$ that mediates an exotic decay of the neutron can potentially admit proton decay vis-\`{a}-vis the same operator, through $p\to n^\star + e^- + {\bar \nu}_e$, with the virtual neutron $n^*$ decaying via the exotic decay channel. This is at odds with proton lifetime constraints, as we detail in Sec.~\ref{sec:explicit BV}, but it can be eliminated altogether if the mass of the exotic {\it final state} $M_f$ exceeds that of $m_p - m_e$, with nuclear stability imposing a still stricter constraint~\cite{Fornal2018PhRvL.120s1801F}, namely that $937.993 \,{\rm MeV} < M_f < 939.565\, {\rm MeV}$ to  prevent $^9{\rm Be} \to f \alpha\alpha$ decay~\cite{Pfutzner:2018ieu}. The exotic final states contain at least one hidden sector particle $\chi$, and $\chi$ can be a dark matter candidate if its mass is less than $m_p + m_e$. Nuclear decays open the opportunity of studying neutron decays to invisible final states, and we return to this point later in this section. 

Herewith we note model realizations of ``fast'' dark decays $n\to \chi\gamma$~\cite{Fornal2018PhRvL.120s1801F}, $n\to \chi\phi$~\cite{Fornal2018PhRvL.120s1801F}, and $n \to \chi\chi\chi$~\cite{Strumia:2021ybk}. The particle $\chi$ is a Dirac fermion and a SM gauge singlet, whereas $\phi$ is a complex scalar. To realize $n\to\chi\gamma$, an additional particle is needed, and different choices are possible. For example, after introducing a massive, complex scalar $\Phi$ with $(3,1)_{-1/3}$, which is a color triplet and weak singlet diquark with hypercharge $Y=-1/3$, the Lagrangian
\begin{equation}
    {\cal L} \supset \lambda_q \epsilon^{ijk} \bar{u}_{Li}^c d_{Rj} \Phi_k + \lambda_{\chi} \Phi^{*\,i} \bar{\chi} d_{Ri} + h.c. 
\end{equation}
with baryon number assignments $B_\chi=1$ and $B_\Phi = -2/3$, permits $n\to \chi\gamma$ decay~\cite{Fornal2018PhRvL.120s1801F}. Alternatively, e.g., a $\Phi$ with $(3,1)_{2/3}$ would also work~\cite{Fornal2018PhRvL.120s1801F}, though it cannot couple to two first-generation $d$-like quarks; $n\to \chi\gamma$ can, however, appear via the strange quark content of the neutron~\cite{Fornal2018PhRvL.120s1801F} or through a one-loop weak process~\cite{Fajfer:2020tqf}. In contrast, the realization of the invisible decay $n\to\chi\phi$, with the particle content thus far noted, requires the introduction of another Dirac fermion $\tilde{\chi}$ as well~\cite{Fornal2018PhRvL.120s1801F}. Finally the decay $n\to\chi\chi\chi$ with $B_\chi = 1/3$ can be realized at the nucleon level via~\cite{Strumia:2021ybk}
\begin{equation}
    {\cal L} \supset \frac{1}{3!\,\Lambda_{\chi n}^2}
    ({\bar{\chi}}^c \Gamma \chi)({\bar n}\Gamma \chi)  + \rm h.c.\,,
\end{equation}
where $B_{\chi} = 1/3$ and $\Gamma$ is a combination of $V\pm A$ interactions. For Dirac fermions $\chi$, a choice which makes ``short-cut'' $|\Delta B|=2$ transitions via dark neutron decay impossible, empirical constraints from direct searches for $\Phi$ at colliders, allow a loci of parameters that would permit the resolution of the neutron lifetime anomaly with new physics, that is, a dark branching ratio of $\sim 1\%$ ~\cite{Fornal2018PhRvL.120s1801F}. However, direct searches for $n\to \chi\gamma$~\cite{Tang:2018eln} and $n\to \chi e^+ e^-$~\cite{UCNA:2018hup,Klopf:2019afh} decay processes significantly constrain the allowed possibilities, particularly if $\chi$ is light enough to be stable. In particular, the study of Ref.~\cite{Tang:2018eln} does not constrain $E_\gamma < 0.782\,\rm MeV$. However, studies at Borexino~\cite{Borexino:2015qij} can also be employed to the same end, limiting ${\rm Br}(n\to\chi\gamma) < 10^{-4}$, as noted by Ref.~\cite{McKeen:2020zni}, thus precluding this particular solution to the anomaly, though dark H decay is still possible~\cite{Berezhiani:2018udo,McKeen:2020vpf}. In addition, dark neutron decays, at the strength to explain the anomaly, particularly that of $n \to \chi \gamma$, can render a massive neutron star unstable~\cite{Mckeen2018PhRvL.121f1802M, Baym2018PhRvL.121f1801B, Motta:2018rxp}, limiting its maximum mass to $0.8\, M_{\astrosun}$, which is inconsistent with observations~\cite{Hulse:1974eb,Raaijmakers:2021uju}. If $\chi$ has repulsive self-interactions, then such effects can make dark neutron decays to final states with $\chi$ energetically less favorable and ultimately permitting  neutron star masses that are consistent with observations. We note Refs.~\cite{Mckeen2018PhRvL.121f1802M,Cline:2018ami} for models in which such self-interaction effects has been studied. Interestingly, Ref.~\cite{Strumia:2021ybk} has shown that it is possible to evade these constraints if only the dark decay mode $n\to \chi \chi \chi$ is permitted --- indicating that a new physics solution to the anomaly would be possible, though the $\beta$-decay constraints studied in Sec.~\ref{sec:Nlifetime} would seem to limit its role. Thinking broadly, we emphasize that dark decay models~\cite{Davoudiasl:2014gfa,Fornal2018PhRvL.120s1801F,Heeck:2020nbq,Elahi:2020urr,Fajfer:2020tqf}, even if they ultimately make a negligible contribution to the neutron lifetime anomaly, nevertheless allow for much larger apparent BNV effects than that permitted from direct searches for explicit BNV. For example, the minimal dark sector model with $\Phi$ in the $(3,1)_{2/3}$ representation can mediate $n\to \chi\gamma$ up to the $\sim 10^{-6}$ level, with a $\Phi$ at the TeV scale, opening the possibility for its discovery at the LHC~\cite{Fajfer:2020tqf}. Moreover, this model permits $\Lambda \to \chi \gamma$, for which there are no direct constraints, and it is possible to trade the size of $n\to\chi \gamma$ for $\Lambda\to \chi\gamma$, or vice versa, given the existing constraint from $D^0-\bar D^0$ oscillations~\cite{Fajfer:2020tqf}. In the current context, the notion of differing rates for $n\to\chi \gamma$ and $\Lambda\to \chi\gamma$, predict very different evolutionary effects within the neutron star, as we have studied in Sec.~\ref{sec:BV in NS}. In particular, the possibility of $\Lambda\to \chi \gamma$ decay is only appreciable at central densities for which a $\Lambda$ population appears, whereas $n\to\chi\gamma$ decay has no such requirement. 

Neutron decays into dark final states can also potentially be discovered or constrained through the study of nuclear decays~\cite{Fornal2018PhRvL.120s1801F,Pfutzner:2018ieu,Ejiri:2018dun}. One possibility concerns the study of $\beta$-delayed proton emission in $^{11}{\rm Be}$ decay, ${}^{11}{\rm Be}(\beta p)$~\cite{Pfutzner:2018ieu}, though the estimated $n^\ast\to\chi$ rate in the nucleus (the $\gamma$ is not needed in the nuclear process) appears to be significantly larger than the empirical width of ${}^{11}{\rm Be}$, significantly constraining this solution to the neutron lifetime anomaly~\cite{Ejiri:2018dun}. It is nevertheless the case that a surprisingly large branching ratio for quasi-neutron-like decay has been inferred from the detection of $^{10}\rm Be$ in $^{11}{\rm Be}\to {\rm ^{10}Be} p e^-\bar \nu_e$ decay~\cite{Riisager:2014gia}, perhaps the neutron also decays invisibly in this process~\cite{Pfutzner:2018ieu}. This process has been investigated further, with direct observation of the final-state protons~\cite{Ayyad:2019kna}  confirming the size of the branching ratio from the earlier indirect result~\cite{Riisager:2014gia}, even if a subsequent experiment~\cite{Riisager:2020glj} in the manner of Ref.~\cite{Riisager:2014gia} fails to do so. It has been noted that a new resonance state could explain the large branching ratio, and this appears to be possible theoretically, both from direct study of the resonance properties~\cite{Okolowicz:2019ifb} and from a study within halo-nucleus effective field theory~\cite{Elkamhawy:2019nxq}. Without a resonance, the decay rate would be very challenging to explain~\cite{Volya:2020ivu,Okolowicz:2021qgl}. These studies constrain the size of a possible dark decay of the neutron in reference to the neutron lifetime anomaly, but have little impact on the broader possibilities we consider here.

Alternative explanations for the neutron lifetime anomaly come from the possibility of dark-matter--neutron interactions in the bottle experiments~\cite{Rajendran:2020tmw}, the existence of a dark $Z_d$~\cite{Ivanov:2018uuk}, or from the possibility of neutron--mirror-neutron mixing~\cite{Berezhiani:2005hv,Berezhiani:2018eds,Tan:2019mrj}, where we note Ref.~\cite{Babu:2021mjg} for a theoretical review of the latter set of models. The last possibility is limited by neutron star heating constraints~\cite{McKeen:2020oyr, McKeen:2021jbh}, direct experimental searches~\cite{Ban:2007tp, Serebrov:2008her, Sarrazin:2012sc, Sarrazin:2016bsw, Stasser:2020jct, Almazan:2021fvo, Broussard:2021eyr}, as well as pulsar timing studies~\cite{Goldman:2019dbq}, though the possibility of ``hidden'' magnetic fields~\cite{Berezhiani:2009ldq, Berezhiani:2018eds} makes for a large phenomenological parameter space to explore \cite{Berezhiani:2017jkn, nEDM:2020ekj, Mohanmurthy:2022dbt}. We now turn to the study of explicit BNV. 

%%%%%%%%%%%%%%%%%%%%%%%%%%%%%%%%%%%%%%%%%%%%
\section{Implications of explicit BNV}
\label{sec:explicit BV}
%%%%%%%%%%%%%%%%%%%%%%%%%%%%%%%%%%%%%%%%%%%%

Baryon number is only an \emph{accidental} symmetry of the Standard Model: given its particle content, the complete set of possible renormalizable interactions conserves this quantity without requiring it a priori. This concept has been invoked to explain the apparent stability of matter, analogous to how the electron is stabilized by electric charge conservation \cite{Weyl:1929fm, Stueckelberg:1938hvi, 1952PNAS...38..449W, Wigner1995}. Neutron decays into new states, considered in the previous section, apparently violate $B$, but this accidental symmetry can be readily generalized to incorporate the new particles so that they do not. However, a generic new physics scenario need not conserve baryon number. Proton decay into SM states, in particular, is a generic consequence of grand unified theories (GUTs) \cite{Georgi:1974sy, Georgi:1974yf} and of models of supersymmetry \cite{Weinberg:1981wj, Sakai:1981pk}; see, e.g., Refs.~\cite{Langacker:1980js, Perkins:1984rg, Senjanovic:2009kr, Nath:2006ut, Babu:2013jba}. Given that baryon number must ostensibly be violated on some level in order to generate the observed baryon asymmetry of the universe (BAU) \cite{Sakharov:1967dj}, it is pertinent to consider how this might appear at low energies.\footnote{We note the existence of models in which the BAU is a product of \emph{apparent} BNV processes; see, e.g., Refs.~\cite{Agashe:2004bm, Davoudiasl:2010am, Aitken:2017wie, Barducci:2018rlx, Elor:2018twp, Alonso-Alvarez:2021oaj}. A generic feature of these models is that the excess of baryons in the SM sector is compensated by a dearth of baryon number in the dark sector; each also predicts some set of apparent BNV processes that one can observe or constrain in the laboratory.}

We adopt a model-independent perspective on low-energy baryon number violation, considering the operators, comprised solely of SM fields, that generate some subset of BNV processes with no assumed relationships between them. This approach dates to Refs.~\cite{Appelquist:1974tg, Weinberg:1979sa, Wilczek:1979hc, Abbott:1980zj, Rao:1982gt} and can be recast in the language of Standard Model Effective Field Theory (SMEFT) \cite{Buchmuller:1985jz, Grzadkowski:2010es, Henning:2014wua, Henning:2015alf, Brivio:2017vri}. If heavy new particles exist, then these can only manifest at low energies through the tower of operators that they engender, including nonrenormalizable ones. These operators must be invariant under the Lorentz group and the SM gauge group, but need not preserve accidental symmetries. The nonrenormalizable part of the SMEFT Lagrangian, $\mathcal{L}_{\rm NR}$, can be generically expressed as follows:
\begin{equation}
    \mathcal{L}_{\rm NR} = \sum_{d=5} \sum_{i} \frac{c_i}{\Lambda^{d-4}} \mathcal{O}^{(d)}_i,
\end{equation}
where $\mathcal{O}^{(d)}$ is an operator with mass-dimension $d$. The $c_i$ are defined to be dimensionless; the factor $1/\Lambda^{d-4}$ is required by dimensional consistency and reflects the intuition that higher-scale physics contributes with lesser strength to low-energy processes, i.e., that ultraviolet physics \emph{decouples} from physics in the infrared \cite{Appelquist:1974tg}. On one hand, operators at high mass-dimension will be suppressed relative to those with lower mass-dimension if they share a common scale $\Lambda$. On the other, processes with small rates may be connected to relatively low-scale physics if they are dominantly generated by higher mass-dimension operators.

We categorize these operators by the extent to which they break $B$, i.e., the $\Delta B$ of the operator, starting with operators with $|\Delta B| = 1$ before moving on to those with $|\Delta B| = 2$. Processes with larger violations have received significantly less theoretical and experimental interest, though some constraints do exist \cite{Hazama:1994zz, Bernabei:2006tw, EXO-200:2017hwz, Majorana:2018pdo}.\footnote{We note electroweak sphalerons, nonperturbative gauge configurations that can convert baryon number into lepton number (and vice versa) in units of three. These processes are only operative at high temperatures and naturally allow for BNV in the early universe; these are inoperative at the energy scales we consider, and we refer to  Ref.~\cite{DOnofrio:2014rug} for a detailed study in SM lattice gauge theory, though we also note the exploration of possible exceptions~\cite{Tye:2015tva,Ellis:2016ast,Tye:2017hfv}.} We note, however, that these operators would be suppressed by relatively high powers of the scale $\Lambda$, allowing for small rates of baryon-number violation to be connected to relatively low-scale new physics.

\subsection{Processes with $|\Delta B|=1$}

These first arise at mass-dimension six\footnote{One often reads about contributions that arise at dimension five \cite{Weinberg:1981wj, Sakai:1981pk} within supersymmetric models, not six. This is in a framework in which the superpartners are dynamical --- if one were to integrate these out, as would be the case in SMEFT, then the resulting effective interactions would be no lower than dimension six.} within SMEFT and are schematically of the form
\begin{equation}
    \label{eq:d6lag}
    \mathcal{L}^{(d=6)}_{|\Delta B|=1} \supset \sum_i \frac{c_i}{\Lambda_{|\Delta B| = 1}^2} (qqq\ell)_i \text{ + h.c.},
\end{equation}
where $q$ represents a quark operator and $\ell$ represents a lepton operator; the sum over $i$ represents the sum over possible chirality and gauge structures. It is this class of operators that can give rise to proton decay and nonstandard neutron decay. Restricting to the first quark generation, there are four possible operators for each generation of lepton $\ell$ (= $e, \, \mu$):
\begin{eqnarray*}
    \mathcal{O}_{LL}^{(\ell)} = \left(Q \varepsilon Q \right) \left(L_{\ell} \varepsilon Q\right) & \to & \left( \overline{u}^c P_L d \right) \left( \overline{\ell}_{e,\mu}^c P_L u - \overline{\nu}_{e,\mu}^c P_L d \right), \\
    \mathcal{O}_{LR}^{(\ell)} = \left(Q \varepsilon Q \right) \left(e_{\ell} u \right) & \to & \left( \overline{u}^c P_L d \right) \left( \overline{\ell}_{e,\mu}^c P_R u\right),\\
    \mathcal{O}_{RL}^{(\ell)} = \left( u d \right) \left(L_{\ell} \varepsilon Q\right) & \to & \left( \overline{u}^c P_R d \right) \left( \overline{\ell}_{e,\mu}^c P_L u - \overline{\nu}_{e,\mu}^c P_L d \right), \\
    \mathcal{O}_{RR}^{(\ell)} = \left(u d \right) \left(e_{\ell} u \right) & \to & \left( \overline{u}^c P_R d \right) \left( \overline{\ell}_{e,\mu}^c P_R u\right).
\end{eqnarray*}
On the left, we write the operators before electroweak symmetry is broken in terms of two-component (Weyl) spinors: the left-handed quark doublet, $Q$; the left-handed lepton doublet, $L_{\ell}$; the right-handed singlet up quark, $u$; the right-handed singlet down quark, $d$; and the right-handed singlet charged lepton $e_{\ell}$). Fermion fields enclosed in parentheses have been grouped into Lorentz scalar bilinears and $\varepsilon$ is the antisymmetric tensor, which is used to form weak singlets. On the right, we write them in the broken phase in terms of the more familiar four-component spinors; $\overline{\psi}^c = \psi^T \mathcal{C}$ is the charge conjugate of the field $\psi$ and $P_{R,L}$ are the standard projection operators. Color contractions have been left implicit; only the totally antisymmetric combination may be used. The task becomes constraining the coefficients $c_i/\Lambda_{|\Delta B|=1}^2$ for these operators.

\begin{figure}[!t]
    \centering
    \includegraphics[width=0.8\linewidth]{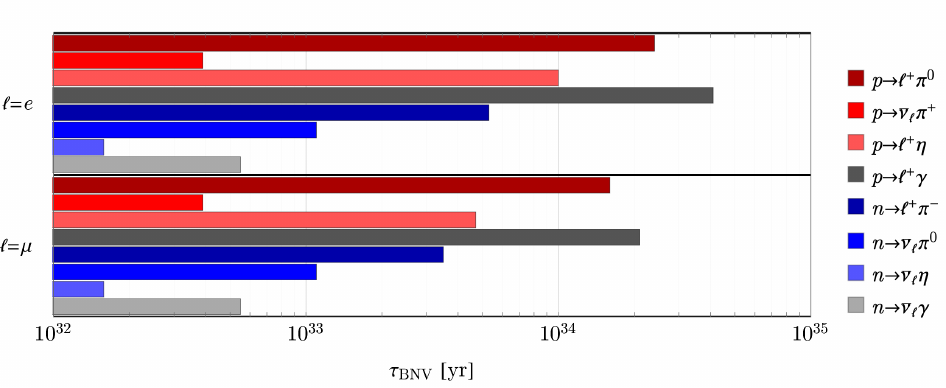}
    \caption{Constraints on the partial lifetimes of representative two-body BNV nucleon decays: $p\to\ell^+ \pi^0$ \cite{Super-Kamiokande:2020wjk}, $p\to\overline{\nu}_\ell \pi^+$ \cite{Super-Kamiokande:2013rwg}, $p\to\ell^+\eta$ \cite{Super-Kamiokande:2017gev}, $p\to\ell^+\gamma$ \cite{Super-Kamiokande:2018apg}, $n\to\ell^+\pi^-$ \cite{Super-Kamiokande:2017gev}, $n\to\overline{\nu}_\ell\pi^0$ \cite{Super-Kamiokande:2013rwg}, $n\to\overline{\nu}_\ell \eta$ \cite{McGrew:1999nd} and $n\to\overline{\nu}_\ell\gamma$ \cite{Super-Kamiokande:2015pys}, with $\ell = e, \, \mu$. All values represent 90\% C.L. upper limits.}
    \label{fig:B1_lifetimes}
\end{figure}

Constraints on the proton lifetime date back to the 1950s \cite{Perkins:1984rg} and the current limit on the total lifetime is $3.6 \times 10^{29}$ yr \cite{SNO:2018ydj}, though some partial lifetimes have been measured at the level $\mathcal{O}(10^{32}-10^{34})$ yr; we refer the reader to the comprehensive tabulation of partial widths in Ref.~\cite{Heeck:2019kgr} and references therein. A subset of two-body decay lifetimes are shown in Fig.~\ref{fig:B1_lifetimes}. One can interpret these limits in terms of the operators in Eq.~\eqref{eq:d6lag}; we briefly outline this analysis, following the formalism established in, e.g., Refs.~\cite{JLQCD:1999dld, Aoki:2017puj, Yoo:2021gql}. The decay width for $N \to \overline{\mathcal{L}}_\ell \Pi$, with $N$ a nucleon, $\overline{\mathcal{L}}_\ell$ an antilepton of flavor $\overline{\ell} = e, \, \mu$ and $\Pi$ a pseudoscalar meson, is given by\footnote{We write $\Lambda$ instead of the more cumbersome $\Lambda_{|\Delta B| = 1}$ here.}
\begin{eqnarray}
    &\Gamma\left(N \to \overline{\mathcal{L}}_\ell \Pi \right) = \dfrac{E_\mathcal{L} |\vec p_\mathcal{L}|}{8\pi m_N} \left( |W^L|^2 + |W^R|^2 - \dfrac{2m_\mathcal{L}}{E_\mathcal{L}} \Re\left[W^L (W^R)^*\right] \right),& \\
    &W^L = \sum_{\chi=R,L} \left[ \dfrac{c_{\chi L}^{(\ell)}}{\Lambda^2} W_0^{\chi L}(Q^2) - \dfrac{m_\mathcal{L}}{E_\mathcal{L}} \dfrac{c_{\chi R}^{(\ell)}}{\Lambda^2} W_1^{\chi R}(Q^2) \right]_{Q^2 = -m_\mathcal{L}^2},& \\
    &W^R = \sum_{\chi=R,L} \left[ \dfrac{c_{\chi R}^{(\ell)}}{\Lambda^2} W_0^{\chi R}(Q^2) - \dfrac{m_\mathcal{L}}{E_\mathcal{L}} \dfrac{c_{\chi L}^{(\ell)}}{\Lambda^2} W_1^{\chi L}(Q^2) \right]_{Q^2 = -m_\mathcal{L}^2},&
\end{eqnarray}
where $m_\mathcal{L}$, $\vec p_\mathcal{L}$, $E_\mathcal{L}$ are the antilepton mass, three-momentum and energy, respectively, $c_{\chi\chi^\prime}^{(\ell)}$ is the coefficient of the operator $\mathcal{O}_{\chi\chi^\prime}^{(\ell)}$ and $W_{0,1}^{\chi\chi^\prime}$ are form factors for the matrix element $\langle \overline{\mathcal{L}} \Pi| \mathcal{O}_{\chi\chi^\prime}^{(\ell)} | N\rangle$, which are calculated on the lattice. If $\overline{\mathcal{L}} = \overline{\nu}$, then contributions proportional to $c_{\chi R}$ must be taken to zero. We use the nucleon-pion form factors from lattice calculations with a physical pion mass of Ref.~\cite{Yoo:2021gql}; nucleon-eta form factors are taken from calculations with unphysical pion masses in Ref.~\cite{Aoki:2017puj}. Moreover, we calculate the decay widths for $N\to\overline{\mathcal{L}}\gamma$ to find 
\begin{eqnarray}
    &\Gamma(N\to\overline{\mathcal{L}}\gamma) = \dfrac{\alpha_{\rm EM} (\mu_N-Q_N)^2}{16m_N} \left( 1- \dfrac{m_\mathcal{L}^2}{m_N^2} \right) \times \qquad \qquad \qquad \qquad & \nonumber \\ 
    & \qquad \qquad \left[ \left( 1 + \dfrac{m_\mathcal{L}^2}{m_N^2} \right)\left(|\hat{W}^L|^2 + |\hat{W}^R|^2 \right) - 4\dfrac{m_\mathcal{L}}{m_N} \Re\left[\hat{W}^L(\hat{W}^R)^*\right] \right]& \\
    & \hat{W}^L = \dfrac{\alpha c_{RL}^{(\ell)} + \beta c_{LL}^{(\ell)}}{\Lambda^2}, \qquad \hat{W}^R = \dfrac{\alpha c_{LR}^{(\ell)} + \beta c_{RR}^{(\ell)}}{\Lambda^2}, &
\end{eqnarray}
where $Q_N$, $\mu_N$ are the nucleon charge and magnetic moment (the latter in units of the Bohr magneton) and $\alpha$, $\beta$ are low-energy constants that characterize the proton-to-vacuum transitions, which we obtain from Ref.~\cite{Yoo:2021gql}.

\begin{figure}[!t]
    \centering
    \includegraphics[width=0.8\linewidth]{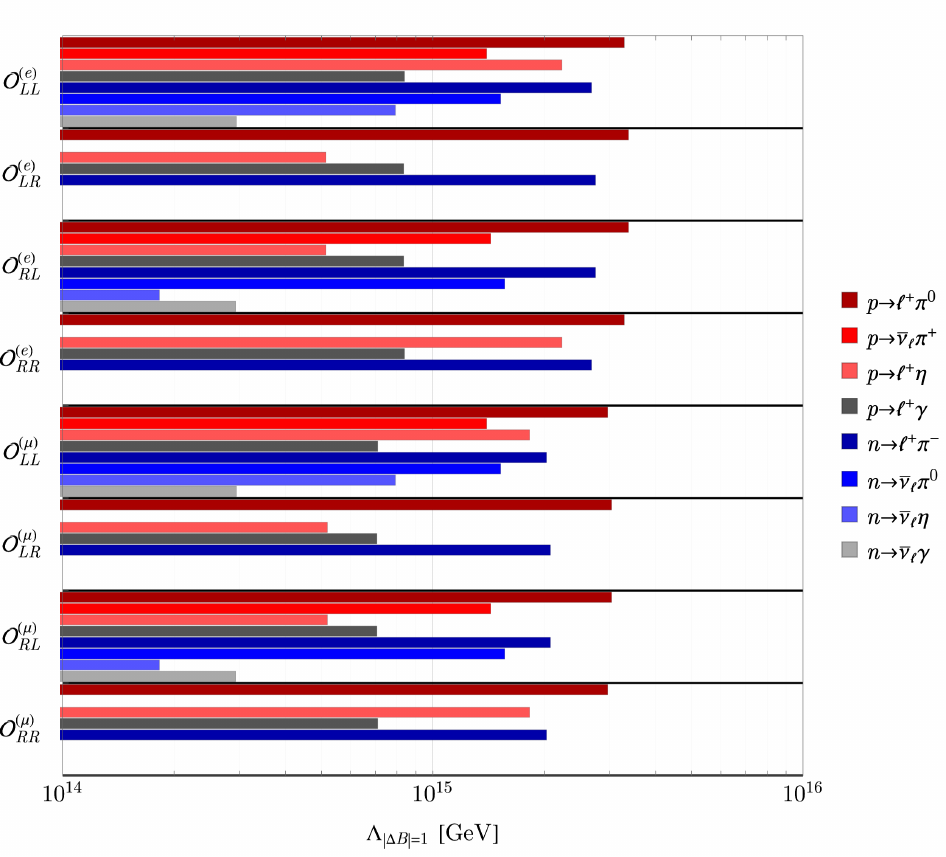}
    \caption{Representative constraints on operators with $|\Delta B| = 1$. We employ the lattice matrix elements, form factors, and low-energy constants of Refs.~\cite{Aoki:2017puj, Yoo:2021gql}. Note that the operators $\mathcal{O}^{(\ell)}_{\chi R}$ do not generate decays to antineutrinos; these are thus not constrained by these final states.}
    \label{fig:B1_constraints}
\end{figure}

The results of this analysis are shown in Fig.~\ref{fig:B1_constraints}. Allowing for one operator to be active at a time for simplicity and fixing the corresponding $c_{\chi\chi^\prime}^{(\ell)} = 1$, we obtain a constraint on $\Lambda$ at the level
\begin{equation}
    \Lambda \gtrsim \mathcal{O}(10^{15} - 10^{16})\text{ GeV.}
\end{equation}
Consequently, $|\Delta B| = 1$ processes are severely constrained; this is sufficient to rule out the simplest GUT models. We note, however, that there may be cancellations among the $c_{\chi\chi^\prime}^{(\ell)}$, suppressing rates of some subset of decay modes; this requires fine-tuning, but is logically possible. However, the nonobservation of \emph{at most one channel} can be explained by such a cancellation\footnote{The exception is the decays $p\to\overline{\ell}\gamma$ and $n\to\overline{\nu}_\ell \gamma$, which simultaneously vanish, at leading order in $\alpha_{\rm EM}$, if $\alpha c_{RL}^{(\ell)}=-\beta c_{LL}^{(\ell)}$ and $\alpha c_{LR}^{(\ell)} = - \beta c_{RR}^{(\ell)}$. However, one would generically expect that higher-order contributions would change the functional dependencies on these coefficients, so that this would no longer occur.} --- we are inexorably led to the conclusion that proton decay at mass-dimension six requires the scale of new physics to be not far below the Planck scale. That said, while the tower of operators starts at $d=6$, all higher mass dimensions also include $|\Delta B| = 1$ operators. It is logically possible that dimension-six contributions to proton decay could be suppressed in a given model and that, say, dimension-eight contributions could dominate. These latter contributions will depend on higher powers of $1/\Lambda_{|\Delta B| = 1}$, thereby allowing this scale to be significantly less than the $\sim10^{16}$ GeV figure we have found.

Nucleons may also decay into final states with kaons; there exist models \cite{Dimopoulos:1981dw, Ellis:1981tv, Ellis:1983qm} in which symmetries are invoked to suppress decays to first-generation fermions, thereby leading to decays involving the heavier mesons. These channels are only slightly less constrained than the strongest first-generation final states \cite{Zyla:2020zbs}. There are six operators that can engender these decays for each generation of lepton. The constraints on the associated scale turn out to be one or two orders of magnitude weaker than those shown above. We omit these due to space constraints, but, irrespective of the sensitivity of the experimental limits, we emphasize that the inferred scale of these operators need not be the same as the scale of the first-generation operators considered previously. In particular, if the coefficients of SMEFT, determined by the true high-scale theory of Nature, depend on quark flavor, then it might be reasonable to expect that processes involving the light quarks would be suppressed, and that new-physics effects are more likely to appear in heavy quark systems.

We note an important theorem regarding SMEFT \cite{Kobach:2016ami, Helset:2019eyc}: operators  with even (odd) mass dimension induce changes to $B-L$ in even (odd) multiples of two. The mass-dimension six operators in Eq.~\eqref{eq:d6lag} thus conserve $B-L$. However, it need not be the case that proton decay conserves $B-L$: there exist four BNV operators at mass-dimension seven that all violate $B-L$ by two units. The importance of the (non)conservation of $B-L$ in proton decay was recognized as early as the late 1970s~\cite{Wilczek:1979et}, where it was noted that $B-L$ breaking would imply the existence of a new scale between the weak and GUT scales. Here, we note that if proton decay were both $B$- and $B-L$-violating -- say, via $p\to e^-\pi^+\pi^+$ -- then the dependence on $1/\Lambda^3$ allows for the scale of proton decay to be much smaller than the values obtained in our analysis. We coarsely estimate this as
\begin{equation}
    \Lambda_{(d=7)} \sim \left( v \Lambda_{(d=6)}^2 \right)^{1/3} \sim \mathcal{O}(10^{10}) \text{ GeV.}
\end{equation}
This scale is far larger than directly accessible experimental energy scales, but is many orders of magnitude suppressed relative to typical GUT and Planck scales. Lastly, we note that one-unit violations of $B$ can manifest in processes besides nucleon decays: scattering processes such as $p e^- \to e^+ e^-$ or $pn \to e^+n$ may be generated at even mass dimension, whereas processes such as $nn\to p e^-$, which violate $B-L$, may occur at odd mass dimension.

We emphasize that there is no innate reason why baryon number should necessarily be violated by only one unit at a time. The study of $|\Delta B| = 1$ operators and the processes they engender is motivated by the observation that in the absence of non-SM states that there are no other means by which protons can decay. It would be a remarkable signature, to be sure, but it is not guaranteed that proton decay should occur if $B$ is explicitly violated --- the true theory of Nature may be such that it is dramatically suppressed relative to the intrinsic BNV scale or that it is for some reason completely forbidden. While the large scales discussed in this subsection may be discouraging from the point of view of phenomenological relevance, it is critical to keep in mind that proton decay is not the only way in which BNV can manifest at low energies. In the remainder of this section, we aim to demonstrate that $B$ violation may still be accessible in both the laboratory and in astrophysical settings and we turn to the consideration of processes with $|\Delta B|=2$.

\subsection{Processes with $|\Delta B|=2$}

We turn now to $|\Delta B| = 2$ operators. These first appear in SMEFT at mass-dimension nine and are of the form
\begin{equation}
    \label{eq:d9lag}
    \mathcal{L}^{(d=9)}_{|\Delta B|=2} \supset \sum_i \frac{c_i}{\Lambda_{|\Delta B| = 2}^5} (qqqqqq)_i \text{ + h.c.},
\end{equation}
The Lagrangian in Eq.~\eqref{eq:d9lag} differs from Eq.~\eqref{eq:d6lag} in that the former does not conserve $B-L$.  This is important because if $|\Delta B| = 2$ processes were subject to the same high scale as $|\Delta B| = 1$ processes, then the suppression by higher powers in $\Lambda$ would imply that the former would be impossible to observe, for all practical purposes. However, as we have noted, if $B-L$ were broken, then this would imply that new dynamics must be operative between the weak and unification scales \cite{Wilczek:1979et}. One may construct models \cite{Davidson:1978pm, Wilczek:1979et, Marshak:1979fm, Chang:1980ey, Mohapatra:1980qe, Babu:2001qr, Arnold:2012sd, Dev:2015uca, Berezhiani:2015afa, Fornal:2017xcj, Assad:2017iib, Helset:2021plg} that invoke a new scale at which $B-L$ is broken, such that $|\Delta B|=2$ processes are generated while generating small (or vanishing) contributions to proton decay. If Nature realized this scheme, then it would be quite a blow to the GUT paradigm --- the introduction of new states far below the GUT scale may spoil the promise of unification, or at least render such a framework hopelessly unpredictive. Still, the promise of lowering the scale of new physics to (potentially) accessible energies makes this a tantalizing possibility. This invites us to take seriously the possibility that $\Lambda_{|\Delta B|=2}$ might be much smaller than the assumed GUT scale deduced from experimental limits on $p$ decay. 

A crucial observation is that $|\Delta B| = 2$ operators do not contribute to free nucleon decay,\footnote{The exception is ``wrong-signed'' neutron decay, $n\to\overline{p}e^+\nu$. This process may occur, e.g., via $n \to \overline{n} \to \overline{p} e^+ \nu$ via $d=9$ operators or via a $d=13$ contact interaction, converting $n\to\overline{p}e^+\nu$ at a point. See, e.g., Ref.~\cite{He:2021sbl} for more discussion.} immediately eliminating a powerful set of potential constraints. The most well-studied effect of these operators at low energies is to mix neutrons and antineutrons, allowing for spontaneous \emph{oscillations} between the two. Ref.~\cite{Phillips:2014fgb} provides a comprehensive review of this subject (see also Refs.~\cite{Babu:2013yww, Addazi:2020nlz, Proceedings:2020nzz}); here, we only briefly touch upon the most phenomenologically relevant details. The operators of Eq.~\eqref{eq:d9lag} generate the following contribution to the nucleon-level Lagrangian:
\begin{equation}
    \label{eq:define_dm}
    \mathcal{L}_{|\Delta B|=2}^N \supset -\frac{\delta m}{2} \left( \overline{n}^c n + \overline{n} n^c\right),
\end{equation}
with $n$ the neutron field and $n^c = C \overline{n}^T$ its charge conjugate. This induces nondiagonal Hamiltonian matrix elements for the $n-\overline{n}$ system. Consequently, the probability $P_{n\overline{n}}(t)$ for a neutron to manifest as an antineutron after a propagation time $t$ is
\begin{equation}
    \label{eq:nn_prob}
    P_{n\overline{n}}(t) = \sin^2\left(t \times \delta m\right) \times e^{-t/\tau_n},
\end{equation}
where $\tau_n$ is the free neutron lifetime. We identify $(\delta m)^{-1}$ as the timescale of $n\overline{n}$ oscillations; we employ the standard notation $\tau_{n\overline{n}} = (\delta m)^{-1}$ \cite{Phillips:2014fgb, Babu:2013yww, Addazi:2020nlz, Proceedings:2020nzz}.

Searches for free $n\overline{n}$ oscillations have been performed at nuclear reactors dating back nearly forty years \cite{CERN-GRENOBLE-PADUA-RUTHERFORD-SUSSEX:1985fte, Bressi:1989zd, Bressi:1990zx}; the leading limit on the oscillation timescale using free neutrons comes from the ILL \cite{Baldo-Ceolin:1994hzw}:
\begin{equation}
    \tau_{n\overline{n}}^\text{(free)} > 8.6 \times 10^7 \text{ s \qquad (90\% C.L.)}
\end{equation}
To interpret this result in terms of the scale $\Lambda_{|\Delta B| = 2}$ in Eq.~\eqref{eq:d9lag}, we note that $\delta m$ can be parametrically written as
\begin{equation}
    \label{eq:n_mass_estimate}
    \delta m = \sum_i \frac{c_i}{\Lambda_{|\Delta B| = 2}^5} \langle \overline{n} | (qqqqqq)_i | n \rangle \sim \frac{C \Lambda_\text{QCD}^6}{\Lambda_{|\Delta B| = 2}^5};
\end{equation}
for $C \sim \mathcal{O}(1)$ and $\Lambda_\text{QCD} = 1$ GeV, this implies $\Lambda_{|\Delta B| = 2} \gtrsim \mathcal{O}(100)$ TeV. One can improve this estimate using more precise calculations of the $n-\overline{n}$ matrix elements, including those of lattice QCD \cite{Rinaldi:2018osy,Rinaldi:2019thf} or phenomenological models such as the MIT bag model \cite{Rao:1982gt}. These are useful for interpreting specific UV model predictions, but do not lead to qualitatively different conclusions, for our purposes. Therefore, we eschew an effective-operator analysis of the sort we have discussed for $|\Delta B|=1$ processes. It has been nearly 30 years since a new search for free $n\overline{n}$ oscillations has been performed. That said, the coming decade-plus promises an improvement of more than an order of magnitude in $\tau_{n\overline{n}}$ ($\sim3 \times 10^9$ s) from NNBAR at the European Spallation Source \cite{Addazi:2020nlz}; this implies a sensitivity to $\Lambda_{|\Delta B| = 2}$ at the PeV scale.

This scheme can be made arbitrarily more complicated:
\begin{itemize}
    \item Explicit and apparent BNV are not mutually exclusive. If mirror neutrons exist and $|\Delta B|=2$ dynamics are operative, then one can consider simultaneous oscillations among all four of $n, \, \overline{n}, \, n^\prime, \, \overline{n}^\prime$. It has been proposed that $n\overline{n}$ oscillations may operate through a ``shortcut'' \cite{Berezhiani:2020vbe} through the mirror neutron, via $n\to n^\prime/\overline{n}^\prime \to \overline{n}$: these transitions may occur even if the $n\overline{n}$ matrix element is small. This framework can be difficult to test due to the ill-constrained Hamiltonian involving the interactions of the mirror neutron. Proposals exist to probe this scenario at the European Spallation Source \cite{Addazi:2020nlz}, at Oak Ridge National Laboratory \cite{Broussard:2019tgw}, 
    and at the Paul Scherrer Institut~\cite{Ayres:2021zbh}.
    \item Eq.~\eqref{eq:nn_prob} assumes that the $n$ and $\overline{n}$ are exactly degenerate in the experimental environment. There are several reasons why this would not be the case; among these are the presence of external matter and magnetic fields \cite{Mohapatra:1980de, Cowsik:1980np}, which lift the degeneracy and suppress oscillations. While these suppress oscillations, they may also stimulate $|\Delta B|=2$ phenomena via other mechanisms, such as neutron-antineutron \emph{conversion} via scattering, such as $ne^-\to\overline{n}e^-$. This may be generated either by long-distance contributions (as in, e.g., Ref.~\cite{Gardner:2017szu}) or through short-distance contributions in SMEFT --- these processes can select different subsets of operators from those that generate $n\overline{n}$ oscillations directly. The latter come with higher inverse powers of the new scale; this may allow for $\Lambda_{|\Delta B|=2}$ to be lowered to the $\mathcal{O}(1-10)$ TeV scale without running aground of existing constraints.
\end{itemize}

While $|\Delta B|=2$ operators do not engender new nucleon decays, these can render some nuclei unstable through \emph{dinucleon decays}, $NN \to X$, where $X$ is some state comprised of leptons, mesons and photons \cite{Basecq:1983hi}. The \emph{nuclear} decay lifetimes can be related to the \emph{free} oscillation timescale via~\cite{Dover:1982wv}
\begin{equation}
    \label{eq:dinuc_decay}
    \tau_\text{dinuc.} = R_{\text{dinuc.},n\overline{n}} \times \tau_{n\overline{n}}^2,
\end{equation}
where $R_{\text{dinuc.},n\overline{n}}$ accounts for the effects of nuclear structure: ambient nuclear matter breaks the degeneracy between $n$ and $\overline{n}$, thereby suppressing the probability for a given neutron to convert. Searches for these decays have been performed using deuterium \cite{SNO:2017pha}, $^{16}$O \cite{Kamiokande:1986pyk,Super-Kamiokande:2011idx,Super-Kamiokande:2020bov} and $^{56}$Fe \cite{Frejus:1989csb,Chung:2002fx}; the resulting limits are typically of order $\sim\mathcal{O}(10^{32})$ years. Shell-model calculations of $R_{\text{dinuc.},n\overline{n}}$ in medium-$A$ nuclei \cite{Friedman:2008es} and separate effective field theory calculations of $R_{\text{dinuc.},n\overline{n}}$ for deuterium \cite{Oosterhof:2019dlo, Haidenbauer:2019fyd} all point to values of the order $\sim\mathcal{O}(10^{22}-10^{23})$ s$^{-1}$. Combining the limits on $\tau_{\rm dinuc.}$ and the calculations of $R_{\text{dinuc.},n\overline{n}}$, the strongest resulting limit on $\tau_{n\overline{n}}$ from searches for dinucleon decays comes from Super-Kamiokande \cite{Super-Kamiokande:2020bov},
\begin{equation}
    \tau_{n\overline{n}}^\text{(nucl.)} \gtrsim 4.7 \times 10^{8} \text{ s \qquad (90\% C.L.);}
\end{equation}
note that this is stronger than the exclusion from searches with free neutrons. DUNE proposes to measure the dinucleon decay lifetime of $^{40}$Ar at the level of $6.45 \times 10^{32}$ s$^{-1}$ for a 400 kt$\cdot$year exposure~\cite{DUNE:2020fgq}; using $R_{\text{dinuc.},n\overline{n}}^\text{Ar} = 0.56 \times 10^{23}$ s$^{-1}$~\cite{Barrow:2019viz}, this implies sensitivity to $\tau_{n\overline{n}} \lesssim 6.0\times10^{8}$ s.

\begin{figure}[!t]
    \centering
    \includegraphics[width=0.8\linewidth]{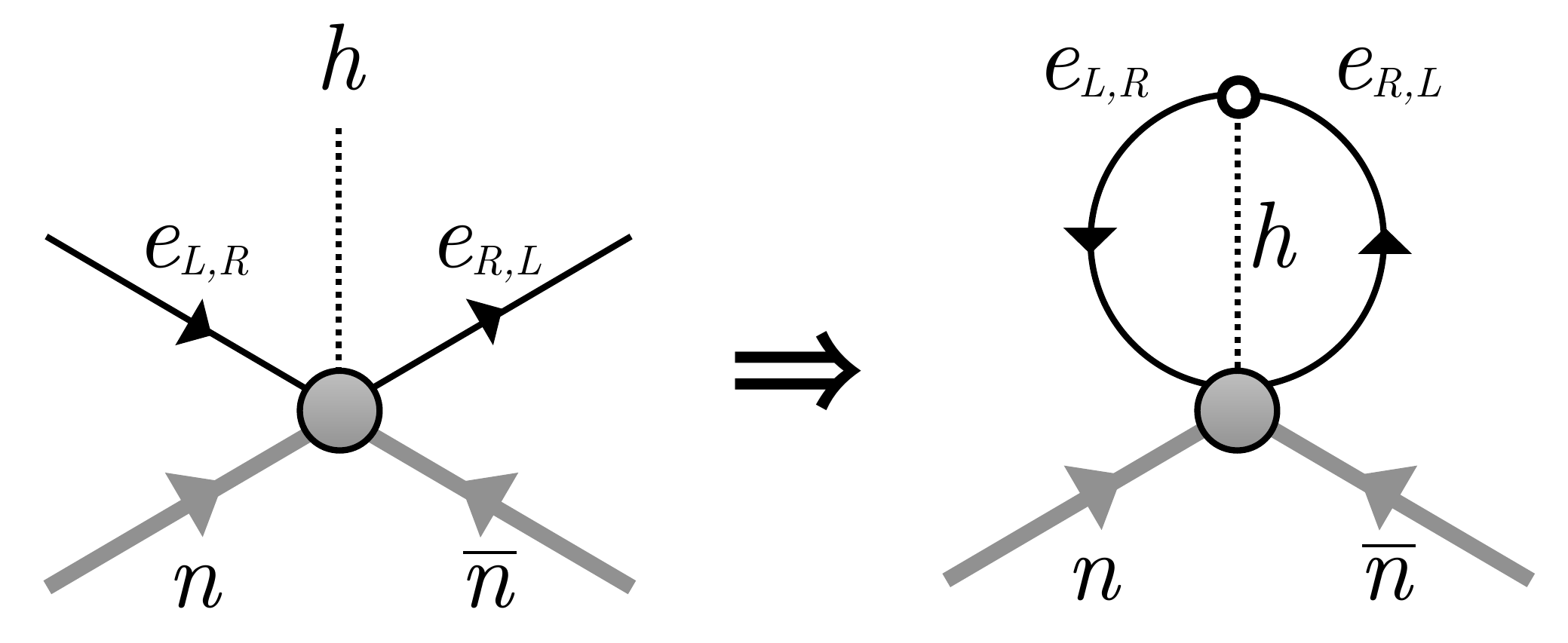}
    \caption{A sketch of a dimension-13 operator contribution to $n\overline{n}$ oscillations. On the left, we show a cartoon of an operator of the form $(udd)(udd) (h \overline{e}e)$. The Higgs field $h$ is required by gauge invariance --- this operator must flip the electron chirality in order to be Lorentz invariant, so that the Higgs is required to preserve all SM charges. On the right, we have closed the electron and Higgs legs to form a two-loop $n\overline{n}$ diagram. This diagram is suppressed by the electron Yukawa coupling, $y_e\approx 2\times10^{-6}$ (denoted by the white circle), as well as two loop factors, $(16\pi^2)^{-2}$. Therefore, for some scale of new physics $\Lambda_{|\Delta B|=2}$, $n\overline{n}$ oscillations would be suppressed relative to na{\"i}ve expectations.}
    \label{fig:dim13}
\end{figure}

An important caveat is that the dinucleon decay lifetime depends on how $|\Delta B|=2$ phenomena manifest at low energies. The values presented above assume that the dominant mechanism is $n\overline{n}$ oscillations, resulting in annihilation with a spectator nucleon. In truth, the connection between $\tau_\text{dinuc.}$ and $\tau_{n\overline{n}}$ depends on the (assumed) relative strengths of all operators that can generate $|\Delta B| = 2$ processes --- including those that do not generate $n\overline{n}$ oscillations at tree level. As an example of this, suppose the leading contribution to the SMEFT Lagrangian from high-scale BNV physics were of the form
\begin{equation}
    \mathcal{L} \sim \frac{1}{\Lambda_{(13)}^9} (udd)(udd) (h \overline{e}e);
\end{equation}
in other words, assume that some UV model gives rise to this dimension-13 contribution at tree level, but that the typical dimension-nine operators were, for some reason, absent. If so, then the process $nn \to e^+ e^-$ occurs at tree level, but $n\overline{n}$ oscillations only arise at two-loop level; this is demonstrated in Fig.~\ref{fig:dim13}. We estimate the contribution of this operator to $\delta m$ to be
\begin{equation}
    \delta m \sim \left(\frac{1}{16\pi^2}\right)^2 y_{e} \frac{\Lambda_\text{QCD}^6}{\Lambda_{(13)}^5},
\end{equation}
where $y_e \approx 2 \times 10^{-6}$ is the electron Yukawa coupling. The prefactor is significantly less than unity; if one tried to associate this with a dimension-nine operator as in  Eq.~\eqref{eq:n_mass_estimate}, then one would misinterpret the scale of the associated physics by about an order of magnitude --- to wit, current limits on $\tau_{n\overline{n}}$ imply $\Lambda_{(13)} \sim \mathcal{O}(10 \rm TeV)$ as opposed  $\mathcal{O}(100 \rm TeV$. Within the nuclear medium, $\tau_{\rm dinucleon}$ would receive contributions both from $n\overline{n}$ oscillations with subsequent annihilation, and from the direct reaction $n n \to e^+ e^-$. There is no reason to expect, a priori, that the latter should be suppressed relative to the former. We can generalize Eq.~\eqref{eq:dinuc_decay} to
\begin{equation}
    \label{eq:dinuc_decay2}
    \tau_\text{dinuc.} = \frac{\tau_{n\overline{n}}^2}{(R_{\text{dinuc.},n\overline{n}})^{-1} + (R_{\text{dinuc.},nn\to e^+ e^-})^{-1}};
\end{equation}
the precise relationship between $R_{\text{dinuc.},n\overline{n}}$ and $R_{\text{dinuc.},nn\to e^+ e^-}$ depends on the details of nuclear structure and on the precise connection between $nn\to e^+ e^-$ and $n\overline{n}$ oscillations.

Moreover, $|\Delta B|=2$ operators that conserve $B-L$ do appear starting at mass-dimension 12.\footnote{While it would be consistent with Kobach's theorem \cite{Kobach:2016ami}, there are no $|\Delta B|=2$ operators at mass-dimension ten. The reason for this is clear: to conserve $B-L$ with $|\Delta B|=2$, one must convert three quarks and an antilepton into three antiquarks and a lepton, requiring eight fermion operators, pushing the dimensionality to be no lower than 12.} This class of processes is interesting because they contain contributions that are second order in the $|\Delta B| = 1$ operators discussed in the previous subsection. If, for instance, $p\to e^+ \pi^0$ were present, then processes such as $pp\to e^+ e^+$ and $e^-p \to e^+ \overline{p}$ \cite{Bramante:2014uda}, as well as hydrogen-antihydrogen oscillations \cite{Feinberg:1978sd, Arnellos:1982nt, Grossman:2018rdg}, must be, too. However, the rates or cross sections for these processes scale as $\Lambda_{|\Delta B|=1}^{-8}$; given the strong constraints on the associated scale, these must be suppressed to a fantastic degree. However, while the existence of $p\to e^+ \pi^0$ is \emph{sufficient} to generate these processes, it is not \emph{necessary} --- the model building and phenomenology become much richer if one considers more general mechanisms of BNV. In particular, if these were connected to physics that innately provides for violations of baryon number by two units while conserving $B-L$, then these may be operative in the absence of proton decay. As a concrete realization of these ideas, we highlight the minimal scalar models studied in Ref.~\cite{Gardner:2018azu}. The catch is that because the associated operators are dimension-12, the rates must scale as $\Lambda_{|\Delta B|=2}^{-16}$. However, the upside is that nonobservation of any of these processes results in a constraint on $\Lambda$ only around the TeV scale \cite{Bramante:2014uda}; put more optimistically, there is still room for new, BNV physics to exist in a way that can be discovered at future collider experiments.

\subsection{Connections to Lepton Number Violation}

\begin{figure}[!t]
    \centering
    \includegraphics[width=0.9\linewidth,height=\textheight,keepaspectratio]{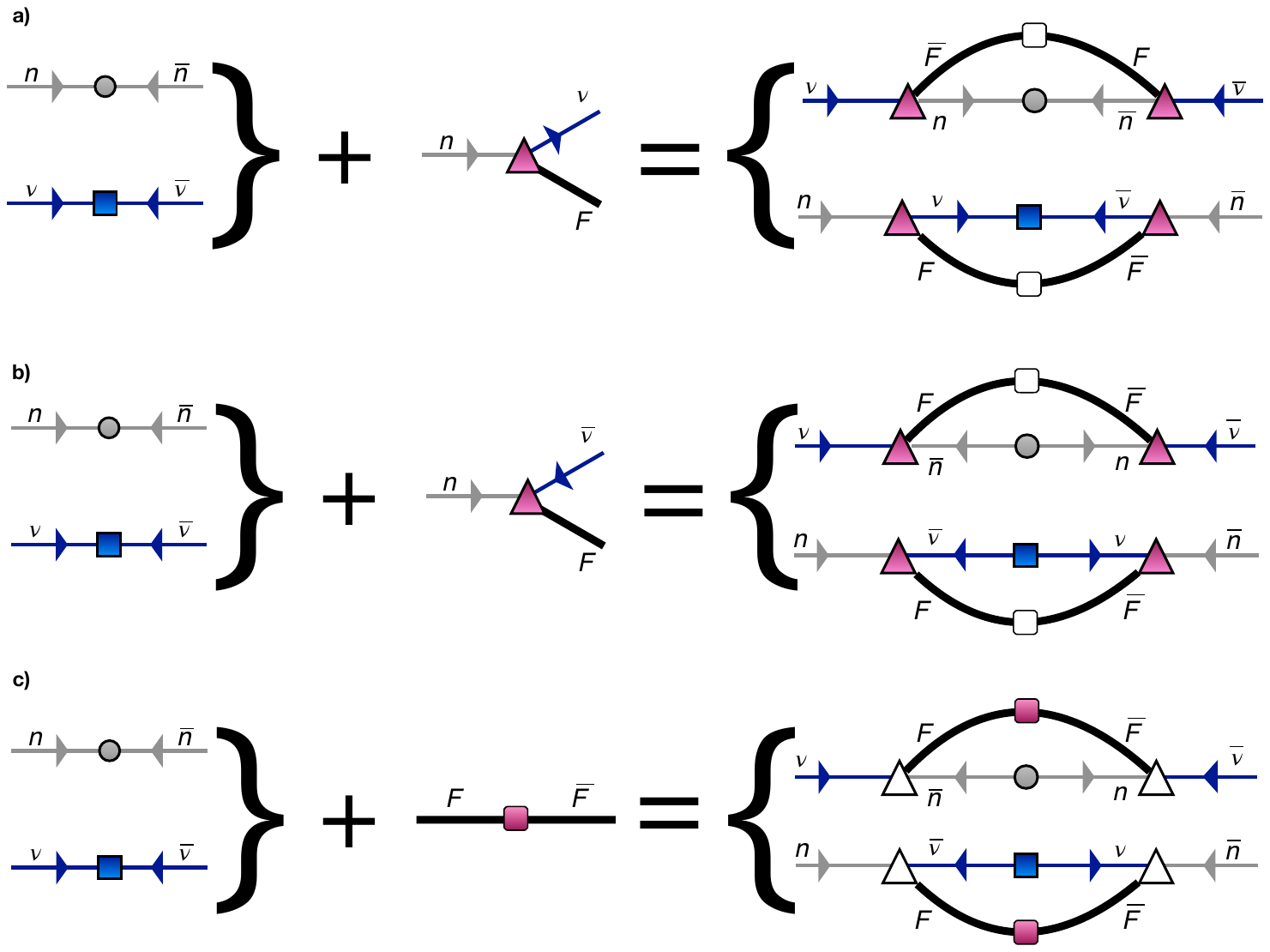}
    \caption{Sketches of the different ways in which possible neutron and neutrino Majorana masses could be connected, where the masses would be inferred from the detection of $n\overline{n}$ oscillations and $0\nu\beta\beta$ decay, respectively.  Connecting the two BSM observables would also require BSM physics, which is indicated through the appearance of shaded vertices, where $F$ is a final state comprised of SM particles, where in a) and b) $F$ has $B=L=0$ and in c) $F$ has $B-L=0$ only. a) With $n\to \nu F$, a $|\Delta(B-L)|=2$ process, with, e.g., $n\to e^- \pi^+ \to e^-e^+ \nu $ and $F=e^+e^-$~\cite{Babu:2014tra}; b) with $n\to \bar\nu F$, a $|\Delta(B-L)|=0$ process, with, e.g., $n\to e^+ \pi^- \to e^+e^- \bar\nu $ and $F=e^+e^-$ (Ref.~\cite{Babu:2014tra} considers the process $p\to e^+ \pi^0$, which is related by isospin.); c) with $F\bar F$ oscillations noting that $F$ is restricted to have $B-L=0$ only. If $F=e^- p$, then, e.g., $e^- p \to e^+ \bar p$ must occur for $n\bar n$ oscillations to imply the existence of $0\nu\beta\beta$~\cite{Gardner:2018azu}, or vice versa. In the context of a NS, these connections give rise to $nn \to \nu\nu\,(\bar\nu \bar\nu)$, or $nn \to FF$. See the text for further discussion.}
    \label{fig:majorana_sketch}
\end{figure}

We have seen that $|\Delta B| = 2$ operators may also violate $B-L$. This correspondence is not necessary, because the operator need not have an odd mass dimension~\cite{Kobach:2016ami}. Both $n\nu \to \overline{n} \overline{\nu}$ scattering and hydrogen-antihydrogen oscillations, e.g., arise from dimension-12 operators that violate both $B$ and $L$ by two units but preserve $B-L$; this suggests that there may be connections between BNV processes and particular lepton-number violating (LNV) processes. In particular, we are free to wonder if $n\overline{n}$ oscillations could possibly be connected to Majorana neutrino masses, similar to how the Schechter-Valle theorem connects neutrinoless double-$\beta$ decay ($0\nu\beta\beta$) to Majorana neutrino masses \cite{Schechter:1980gr}, because the Lagrangian in Eq.~\eqref{eq:define_dm} constitutes a Majorana mass term for the (anti)neutron. Although both processes violate $B-L$ by two units, no SM processes can connect them --- such a connection necessitates the existence of operators with both $B$ and $L$ violation. The effective operator analysis of Ref.~\cite{Babu:2014tra} determines that the observation of any two of $p\to e^+ \pi^0$, $n \to e^- \pi^+$, and $n \bar n$ oscillations would also show that $0\nu\beta\beta$ decay can occur. In contrast, in Ref.~\cite{Gardner:2018azu}, minimal scalar models from Ref.~\cite{Arnold:2012sd} are used to show that the observation of a $|\Delta B|= |\Delta L|=2$ scattering process, along with that of $n\bar n$ oscillations,  would also show that $0\nu\beta\beta$ can occur --- and if $0\nu\beta\beta$ decay can occur, then a Majorana neutrino mass also  exists~\cite{Schechter:1980gr}. These ideas can be generalized to realize three distinct mechanisms for connecting neutron and neutrino Majorana masses, illustrated in Fig.~\ref{fig:majorana_sketch}. That is, $n\bar{n}$ oscillations can combine with either $n\to \nu F$ decay, or $n\to \bar\nu F$ decay, where $F$ is a state of SM particles with total $B=L=0$ and zero electric charge in each case, or with $F-{\bar F}$ oscillations, where $F$ has $B-L=0$, to give rise to a Majorana neutrino mass. In the first instance,  we could have $n\to e^- \pi^+ \to e^- e^+ \nu $, so that $F=e^+ e^-$, but $F=\mu^+ \mu^-$ or $F=\pi, \eta$, or simply $\gamma \gamma$ are also possible, as is $F=K$. Analogously, in the second case, we could have $n\to e^+ \pi^- \to e^- e^+ \bar\nu $, to yield the same set of $F$ we have just enumerated. In the last case, with SM neutron $\beta$ decay we have $F = e^- p$, so that BSM physics is required to yield $e^- p \to e^+ \bar p$~\cite{Gardner:2018azu}. Ultimately, we see that the various connection processes would be signalled by the appearance of dineutron decay to $FF$ or $\bar\nu \bar \nu$ final states.

In conclusion, we emphasize that BNV processes can only be connected to LNV processes through operators that violate both. Although the connections we have noted could be discoverable, the resulting rates for LNV from BNV and vice versa are expected to be extremely small~\cite{Babu:2014tra}. 

\subsection{Effects in Neutron Stars}

For a neutron star containing $\sim \mathcal{O}(10^{57})$ neutrons with a BNV decay lifetime of $\sim \mathcal{O}(10^{30})$ years, corresponding to the most optimistic limit on the proton lifetime \cite{SNO:2018ydj}, one would expect $\sim \mathcal{O}(10^{27})$ such decays to occur in a year. This relatively small rate is unlikely to lead to observable changes in its macroscopic observables (mass, radius, moment of inertia, etc.), and any possible signature is expected to be too weak to observe. In addition to decays, though, there may be $|\Delta B|=1$ scattering processes operative within the star, e.g., $n e^-\to\overline{\nu} e^-$. While these processes would only expedite the conversion of neutrons into energy, it is unlikely that they would lead to observable signatures, either. Fundamentally, this is a consequence of the stringent constraints from proton decay experiments: the limits on $\Lambda_{|\Delta B|=1}$ are just too strong to allow for observable consequences at such low energies. Recall from our discussion in Sec.~\ref{sec:BV in NS}, however, that binary spin-down considerations tolerate much larger rates of BNV: taking $\alpha \approx 1$, $f \approx 1$ (for nucleons) in Eq.~\eqref{eq:BNV rate:binary limit} yields an upper limit of $\Gamma_\text{BNV} \lesssim 10^{-12}$ yr$^{-1}$. We therefore allow ourselves to contemplate rates this large by allowing for violations of $B$ by more than one unit, though apparent BNV could also act.

How $|\Delta B|=2$ physics manifests in neutron stars is expected to be qualitatively different from how it would manifest in the laboratory. Matter effects break the degeneracy between neutrons and antineutrons in neutron star matter, quenching free $n\overline{n}$ oscillations. Therefore, we do not expect to be able to associate any phenomena directly with the timescale $\tau_{n\overline{n}}$ for free neutrons. This is something of a pity --- the fact that pulsars with characteristic ages of $\mathcal{O}(10^8-10^{10})$ years have been observed~(see, e.g., the pulsar catalogs of Refs.~\cite{Perera:2019sca, NANOGrav:2020gpb, Reardon:2021gko}) would surely be able to probe phenomena that occur on the timescale $\tau_{n\overline{n}}$, which is only constrained to be $\gtrsim \mathcal{O}(3)$ years \cite{Baldo-Ceolin:1994hzw}. However, it seems this is just not the universe that we occupy. This is also the case in nuclei, as discussed above, but matter effects are stronger at the supranuclear densities present in the cores of neutron stars. As such, the rate of $n\overline{n}$ oscillations will be dramatically suppressed in neutron stars, even compared to nuclei.

These $|\Delta B|=2$ operators can manifest as two broad classes of processes:
\begin{itemize}
    \item \emph{Processes that destroy two nucleons.} These include processes such as $nn \to 2\gamma, \, 3\nu$, etc. This is completely analogous to dinucleon decay, discussed above.
    \item \emph{Processes that convert nucleons to antinucleons.} These include scattering processes such as $e^-n \to e^- \overline{n}$. 
\end{itemize}
These are not mutually exclusive --- the process $nn\to n\overline{n}$ belongs to both. In the remainder of this discussion, we focus on processes involving neutrons. Protons are present in the star at the $\sim\mathcal{O}(10\%)$ level; these may also participate in BNV interactions with the ambient matter, e.g., $e^-p \to e^+ \overline{p}$, but we expect neutrons to dominate the dynamics. On one hand, $|\Delta B| = 2$ processes should scale with $\delta m$ (see Eq.~\eqref{eq:define_dm}) at the amplitude level, assuming that they depend on the same underlying mechanism of $B$ violation. The partial lifetime for any such process thus generically scales as
\begin{equation}
    \tau_\text{NS} \sim R_\text{NS} \times \tau_{n\overline{n}}^2;
\end{equation}
$R_\text{NS}$ depends both on properties of the nuclear medium and, as in the case of dinucleon decays in nuclei, on which BNV processes are operative within the NS. On the other, $R_\text{NS}$ need not be of the same order of magnitude as its counterpart in nuclear matter, $\sim\mathcal{O}(10^{23})$ s$^{-1}$, given the larger densities, the lower proton fraction and the requirement of charge neutrality, allowing  charged species, such as electrons, muons, pions, and kaons to appear in appreciable numbers as well.

We can place weak limits on the size of $\tau_\text{NS}$, and thus on $R_\text{NS}$. Based on our upper limit from Eq.~\eqref{eq:BNV rate:binary limit}, $\Gamma_{\rm BNV} \lesssim \mathcal{O}(10^{-12})$ yr$^{-1}$, the associated timescale should be $\tau_{\rm NS} \sim \Gamma_{\rm BNV}^{-1} \gtrsim \mathcal{O}(10^{19})$ s --- isolated neutron stars should only be affected by BNV physics on, at least, trillion-year timescales. For $\tau_{n\overline{n}} \sim \mathcal{O}(10^8)$ s, this results in a loose constraint of $R_\text{NS}\gtrsim \mathcal{O}(10^3)$ s$^{-1}$. We also estimate a constraint on the rate of processes that heat the neutron star. The coldest known neutron star (PSR J2144–3933) has a temperature of no more than 42\,000 K \cite{Guillot:2019ugf}; this limits the rate of BNV-induced heating to be $\lesssim \mathcal{O}(10^{27})$ erg s$^{-1}$. This is equivalent to a rate of $\sim\mathcal{O}(10^{30})$ neutrons s$^{-1}$, implying a minimum timescale $\tau_\text{NS} \gtrsim \mathcal{O}(10^{27})$ s -- corresponding to $\Gamma_\text{BNV} \lesssim \mathcal{O}(10^{-20})$ yr$^{-1}$ -- and $R_\text{NS} \gtrsim \mathcal{O}(10^{11})$ s$^{-1}$. One would need a more complete analysis -- including the effects of the neutron star EoS and a concrete set of BNV interactions -- to derive more robust limits than this, but this estimate speaks to the power that the mere existence of cold, old neutron stars has on constraining this sort of new physics.

Let us emphasize a key finding: the present limit on the temperature of PSR J2144-3933 gives a stronger constraint than that derived from the nonobservation of anomalous binary spin-down. Indeed, if $B$ violation were operative and saturated the upper limit on the rate in Eq.~\eqref{eq:BNV rate:binary limit}, then this would imply an energy production rate of $\sim\mathcal{O}(10^{34}-10^{35})$ erg s$^{-1}$ from the destruction of baryons; this corresponds to $\sim\mathcal{O}(10-100)$ $L_{\astrosun}$. If most (or all) of this energy is trapped by the NS, then this would lead to significant heating: the corresponding asymptotic temperature would be $\gtrsim\mathcal{O}(10^6)$ K, in clear violation of the limit on J2144-3933. This constraint is only operative, however, if the products of the BNV reaction cannot escape the star. For species that interact electromagnetically, this is clearly the case. For neutrinos, this is less clear, a priori. However, recall that in the discussion surrounding Eq.~\eqref{eq:nu:mfp}, we concluded that a $\nu_e$ with energy $E_\nu \gtrsim 5$ MeV is likely to scatter on its way out of the core, where it is most likely to have been produced --- neutrinos produced in BNV processes have much higher energies than this, so that even these would need to deposit most of their energy before they can escape. As such, it would only be possible for BNV rates to be this large if the decay products were invisible to the rest of the star, as in our discussion of apparent $B$ violation in Sec.~\ref{sec:apparent BV}. Put more sharply, decay processes such as $n\to\chi\chi\chi$ \cite{Strumia:2021ybk} would evade constraints from neutron star heating.

Explicit BNV processes invariably lead to production of (anti)neutrinos from a combination of three sources:
\begin{enumerate}
    \item Those produced directly in the BNV reaction.
    \item Those produced by the weak reactions that restore chemical equilibrium after some BNV process has disrupted it (i.e., Urca reactions).
    \item Those emitted as a result of the heating of the star, via processes such as $NN\to NN\nu\overline{\nu}$.
\end{enumerate}
Those of the first category are directly sensitive to the relationship between $B$ and $L$ violation. To wit, if $L$ is conserved in $B$-violating processes, then these should produce neutrinos and antineutrinos in equal numbers. If, however, $B$ is violated in such a way that $B-L$ is conserved, then these reactions must produce antineutrinos in excess of neutrinos. The dependence of the second category on $L$ (non)conservation is more difficult to pin down --- it depends on the precise connection between the BNV processes and the weak disequilibrium they engender.\footnote{Recall that the neutron star possesses a total $L\gtrsim\mathcal{O}(10^{55})$; the $L$-(non)conserving nature of $B$-violating processes can interact nontrivially with this reservoir of lepton number.} The third type are, by assumption, SM processes, which conserve $L$ by default; neutrinos and antineutrinos must be emitted in equal numbers. We will discuss the observability of (anti)neutrino signals associated with $B$ violation in more detail in Sec.~\ref{sec:summary}.

Lastly, we note that BNV processes need not restrict themselves to protons and neutrons in such an environment. At high densities, hyperonic degrees of freedom may emerge within the cores of neutron stars; for recent reviews on the subject, see Refs.~\cite{Sedrakian:2021goo, Logoteta:2021iuy, Vidana:2021ppe}. These may also participate in BNV interactions, either among themselves or with the nucleons \cite{Basecq:1983hi}, that are, at best, poorly constrained. These cannot be directly probed in the laboratory, and only indirect comparisons with the operators controlling the nonobservation of $n\overline{n}$ oscillations or $NN \to$ kaons can be formed. Even less well understood are the contributions of the spin-3/2 $\Delta$ resonances, which may also appear in the cores of neutron stars \cite{Sedrakian:2021goo} (see also Ref.~\cite{Thapa:2021ifv}, which aggregates predictions of neutron star properties for several EoS involving $\Delta$s), and might decay to antinucleons (e.g., $\Delta \to \overline{N}\pi$). Appreciable amounts of non-nucleonic hadrons within neutron stars may yet allow for fantastic signatures of $B$ violation at the extremes of matter.

%%%%%%%%%%%%%%%%%%%%%%%%%%%%%%%%%%%%%%%%%%%%
\section{Implications of spontaneous BNV} 
\label{sec:spont bv}
%%%%%%%%%%%%%%%%%%%%%%%%%%%%%%%%%%%%%%%%%%%%

If baryon number were a \emph{gauge} symmetry of nature instead of an accidental one, then null results from fifth force searches (e.g., \cite{Adelberger:1990xq, Schlamminger:2007ht, Murata:2014nra}) imply that it is unlikely to be a symmetry of the vacuum that we occupy --- but it could be spontaneously broken. This notion has existed nearly as long as the concept of baryon number itself \cite{Lee:1955vk, Pais:1973mi, Rajpoot:1987yg, Foot1989PhRvD..40.2487F, Nelson:1989fx, He1990PhRvD..41.1636H, Bailey:1994qv, Carone1995PhRvL..74.3122C, Carone1995PhRvD..52..484C, Aranda1998PhLB..443..352A, FileviezPerez2010PhRvD..82a1901F, Dulaney:2010dj, Dong:2010fw, Ko:2010at, Buckley:2011vs, FileviezPerez:2011dg, FileviezPerez:2011pt, Graesser:2011vj, Duerr:2013dza, Dobrescu:2014fca, Tulin:2014tya, Fanelli:2016utb}. The mechanism of this breaking may leave an observable imprint on the low-energy spectrum of the theory. If an analogue of the Higgs mechanism within the SM or pion condensation in QCD were operative within the baryon-number sector, then there might be (fundamental or composite) scalars present. Even if the mechanism of breaking is not directly observable at low energies, the low-energy spectrum of the theory may contain more than simply the gauge boson. If one were to embed baryon number into a larger gauge group \cite{Fornal:2015boa, Fornal:2015one, FileviezPerez:2016laj}, or incorporate quark flavor into the gauge structure \cite{Berryman:2021wom}, then there may exist additional gauge states or fermions. In this article, we focus on the existence of a single, new gauge boson. New scalars and fermions are interesting in their own right, but these can exist irrespective of whether or not baryon number is gauged --- a massive new gauge state would be a smoking-gun signature of the gauging (and breaking) of $B$. We will call this state $X$: its gauge coupling is $g_X$, and its mass is $m_X$.

\subsection{Laboratory Constraints on a New Gauge Boson}

We are particularly interested in the effect of $X$ on two-nucleon interactions. The new vector state necessarily generates a repulsive contribution between nucleons; how this contribution compares to the strong nuclear force depends on $m_X$. At one extreme, the new force could be light enough that its range exceeds that of the nuclear force, the latter being set by $m_\pi^{-1} \sim \mathcal{O}$(1\, \rm fm).

\begin{figure}[!t]
    \centering
    \includegraphics[width=0.8\linewidth]{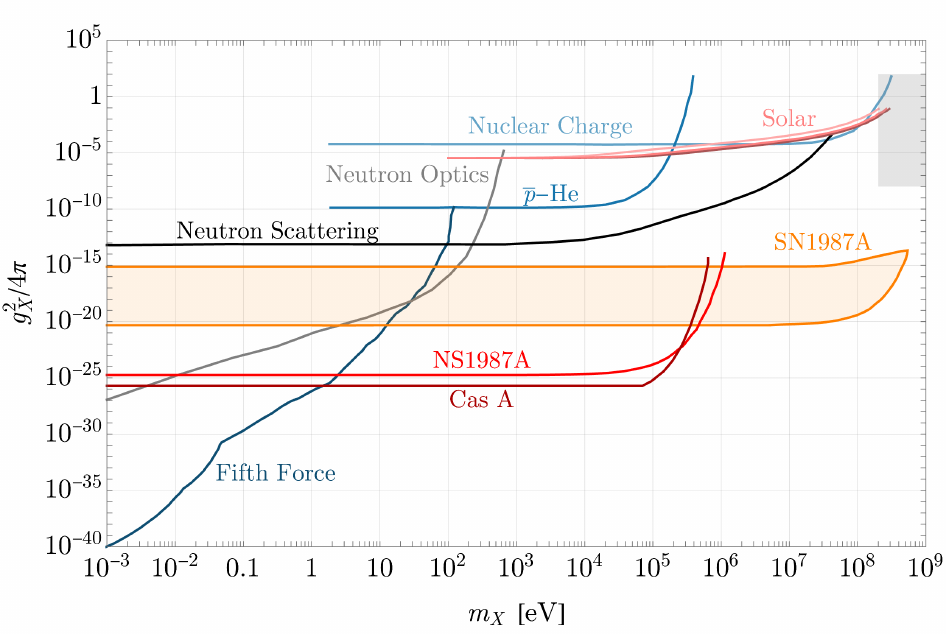}
    \caption{Constraints on the coupling and mass of a new boson associated with gauged baryon number, $U(1)_B$. Aggregate constraints from fifth-force searches (dark blue) are taken from the review of Ref.~\cite{Murata:2014nra}. Neutron scattering (black) and neutron optics (gray) constraints are from Ref.~\cite{Leeb:1992qf}. The $\overline{p}$-He constraint (blue) is from Ref.~\cite{Tanaka:2014pda}. The nuclear charge constraint (light blue) is from Ref.~\cite{Xu:2012wc}. Constraints from solar fusion \cite{Suliga:2020lir} are shown in various shades of pink; see the text for details. Constraints from the anomalous cooling and trapping from SN1987A are shown in orange \cite{Rrapaj:2015wgs}; we emphasize that only the shaded region is excluded. Anomalous cooling constraints from NS1987A \cite{Shin:2021bvz} and Cas A \cite{Hong:2020bxo} are given in red and dark red, respectively. The gray-shaded region corresponds to a new force whose range is comparable to the strong nuclear force; constraints in this region are presented in more detail in Fig.~\ref{fig:old_fig}.}
    \label{fig:constraints_u1B}
\end{figure}

Constraints on a new gauge boson in the range $m_X \in [10^{-3}, 10^9]$ eV from nucleon dynamics are shown in Fig.~\ref{fig:constraints_u1B}. In particular, we show the following:\footnote{We will discuss the bounds from SN1987A (orange), NS1987A (red), and Cas A (dark red) in Sec.~\ref{subsec:light_X}.}
\begin{itemize}
    \item A selection of fifth force searches have been aggregated in Ref.~\cite{Murata:2014nra}. We have converted Figs.~8 and 9 of this reference from the $\alpha-\lambda$ parameter space to $g_X^2/4\pi-m_X$ and plotted the result in dark blue.
    \item Neutron scattering and neutron optics constraints are taken from Fig.~2 of Ref.~\cite{Leeb:1992qf} (see also Ref.~\cite{Barbieri:1975xy}), and are shown in black and gray, respectively.
    \item This new interaction would also change energy levels, relative to QED predictions, of the antiproton-helium ($\overline{p}$-He) bound state. The constraint from Ref.~\cite{Tanaka:2014pda} is shown in blue.
    \item New forces also change the charge radii and binding energies of nuclei. Ref.~\cite{Xu:2012wc} studies the effects of new nuclear-range interactions on $^{48}$Ca, $^{120}$Sn, and $^{208}$Pb; their constraint is shown in light blue.
    \item This interaction would modify the long-distance potential between two protons in the Sun, thereby altering the rate of solar fusion. This could then (1) change the inferred age of the Sun to be inconsistent with the age of the solar system, and (2) modify solar neutrino production to be inconsistent with observations. Constraints have been derived in Ref.~\cite{Suliga:2020lir}. These are shown in shades of pink, corresponding to different proton energies: 10 (light), 50 (medium) and 100 (dark) keV.
\end{itemize}
At the other extreme, the new state might be so heavy -- that is, short ranged -- that it cannot contribute in any meaningful way to low-energy nucleon processes. If this is the case, then these interactions contribute to contact terms in the chiral Lagrangian, i.e., to the so-called low-energy constants (LECs) of chiral effective field theory ($\chi$EFT) \cite{Epelbaum:2008ga, Machleidt:2011zz, Hammer:2019poc, Drischler:2021kxf}. One would expect this if $m_X \gtrsim \Lambda_\text{QCD} \sim \mathcal{O}(1)$ GeV --- it may not make sense to talk about nucleons for momentum exchanges much larger than this scale, but these heavy states can be probed at colliders \cite{Buckley:2011vs, Graesser:2011vj, Dobrescu:2021vak}.

In between these domains -- that is, for $m_\pi \lesssim m_X \lesssim \Lambda_\text{QCD}$ -- is the regime in which the new interaction is not so short-range that it can be integrated out of the $NN$ force, but not so long-ranged that its contributions can be clearly distinguished from nuclear forces. This latter aspect is particularly confounding, because we lack a precise, first-principles description of the forces between nucleons, though lattice QCD may yet provide one~\cite{Drischler:2019xuo}.Historically, interactions among nucleons have been described using phenomenological models \cite{Wiringa:1984tg, Machleidt:1987hj, Wiringa:1994wb, Machleidt:2000ge}: one introduces a set of interactions with unknown coefficients that are fixed by low-energy nuclear data, including $NN$ phase shifts and deuterium data. Even with more modern approaches such as $\chi$EFT, the unknown coefficients of the theory must be fit to data in order to put the framework to use. If the new degrees of freedom are not explicitly included in the $NN$ potential, then their contributions could be inadvertently subsumed into some other part of the interaction --- these may contribute to the short-range pieces of, e.g., the Argonne $v_{18}$ potential \cite{Wiringa:1994wb}. In the context of $\chi$EFT, these would contribute to LECs; however, if these new states are not too much heavier than the pion, then one would expect these to become dynamical for not-too-large momentum exchanges, even if they do not appear for low-momentum exchanges. This would manifest as an apparent inconsistency in the effective theory.

One further complication is that if two-body nuclear forces are poorly understood from first principles, then three-body forces (and beyond) are even more so --- and these are not negligible. As with two-body forces, various phenomenological prescriptions exist for their inclusion \cite{Fujita:1957zz, McKellar:1968zza, Coon:1974vc, Coelho:1983woa, Carlson:1983kq, Robilotta:1986nv, Pieper:2001ap, Coon:2001pv} and they are naturally included within $\chi$EFT \cite{Friar:1998zt, Epelbaum:2005bjv, Epelbaum:2008ga, Machleidt:2011zz, Hammer:2012id, Drischler:2021kxf}. However, these contribute to the uncertainty in these potentials, further complicating extractions of new-physics contributions to nuclear processes. When discussing the impact of new states on neutron star structure, we assume that the new interaction is abelian, so that it does not innately generate new three-body interactions, making the SM contributions to $NNN$ forces overwhelmingly dominant.

\begin{figure}[!t]
    \centering
    \includegraphics[width=0.7\linewidth]{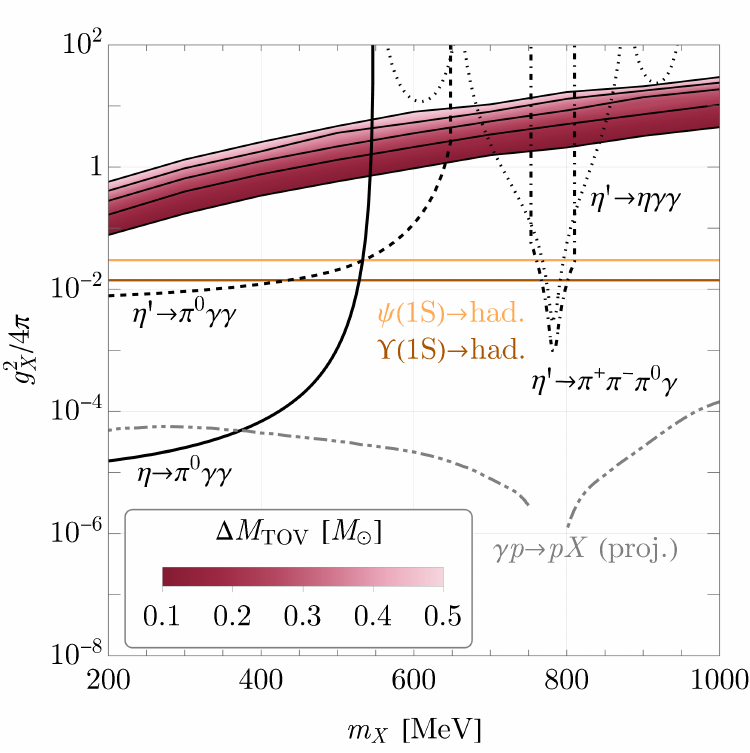}
    \caption{Constraints on a sub-GeV vector boson, adapted from Refs.~\cite{Tulin:2014tya, Fanelli:2016utb, Berryman:2021wom}. Black curves are from rare decays of pseudoscalar bosons: $\eta \to \pi^0 \gamma \gamma$ (solid) \cite{Zyla:2020zbs, Prakhov:2008zz}, $\eta^\prime \to \pi^0 \gamma \gamma$ (dashed) \cite{Ablikim:2016tuo}, $\eta^\prime \to \eta \gamma\gamma$ (dotted) \cite{Ablikim:2019wsb} and $\eta^\prime \to \pi^+ \pi^- \pi^0 \gamma$ (dot-dashed) \cite{Ablikim:2019wop}. The gray, double-dot-dashed curve represents the projected sensitivity to $X$ photoproduction ($\gamma p \to p X$) for the full design luminosity of GlueX Phase IV \cite{Adhikari:2020cvz}. The dark orange line is derived from the decay width for $\Upsilon(1S)\to$ hadrons \cite{Carone1995PhRvL..74.3122C, Carone1995PhRvD..52..484C, Aranda1998PhLB..443..352A}; the light orange line is from the hadronic decay width for $\psi(1S)$ \cite{Aranda1998PhLB..443..352A}. The red-pink-white band represents the loci of points for which the new interaction increases the maximum neutron star mass by $0.1-0.5$ $M_{\astrosun}$; see Sec.~\ref{subsec:spontBNV_NS} for details.
    }
    \label{fig:old_fig}
\end{figure}

Of course, the effects of such a new interaction would not be localized to nucleons. In particular, this new state vector contributes to radiative decays of light mesons \cite{Tulin:2014tya, Gan:2020aco, Berryman:2021wom}. These decays would essentially be two-step processes:
\begin{enumerate}
    \item The meson decays radiatively to $\gamma$ and $X$, e.g., $\eta \to \gamma X$. The computed rate of this process is scheme dependent: rates calculated at the quark level are different than rates calculated in, e.g., the vector meson dominance (VMD) scheme; see App.~A.1 of Ref.~\cite{Tulin:2014tya}.
    \item The $X$ then decays into some observable final state. At tree level, $X$ can decay to, e.g., $\pi^0\gamma$ or $3\pi$. If there exists some nontrivial kinetic mixing with $\gamma$, then $X$ may also decay into dilepton pairs, $e^+e^-$ or $\mu^+\mu^-$, even though these are uncharged under $B$. Additionally, because $X$ only couples to isoscalar currents, tree-level decays to $\pi^+\pi^-$ are absent, barring either (1) nonzero kinetic mixing, or (2) more complicated gauge structures.
\end{enumerate}
This width can then be compared against the width for the decay $\eta^{(\prime)} \to \gamma\gamma$. Constraints of this sort are shown in Fig.~\ref{fig:old_fig}. The black curves are constraints derived from $\eta \to \pi^0 \gamma \gamma$ (solid), $\eta^\prime \to \pi^0 \gamma \gamma$ (dashed), $\eta^\prime \to \eta \gamma\gamma$ (dotted) and $\eta^\prime \to \pi^+ \pi^- \pi^0 \gamma$ (dot-dashed); we collect the branching fractions in Table \ref{tab:decay_data}. Constraints may also be derived from decays of vector mesons \cite{Tulin:2014tya}; these turn out to be weaker than pseudoscalar meson constraints, so we will not consider these further.

\begin{table}[b!]
    \centering
    \begin{tabular}{|c|c|c|} \hline
        Decay Channel & Observed Branching Fraction & Limit \cite{Berryman:2021wom, Tulin:2014tya, Gan:2020aco} \\ \hline\hline
        $\eta \to \pi^0 \gamma \gamma$ & $\left(2.56\pm0.22\right) \times 10^{-4}$ \cite{Zyla:2020zbs, Prakhov:2008zz} & $<3\times10^{-4}$ \\ \hline
        $\eta^\prime \to \pi^0 \gamma \gamma$ & $\left(6.2\pm0.9\right) \times 10^{-4}$ \cite{Ablikim:2016tuo} & $<8 \times 10^{-4}$ \\ \hline 
        $\eta^\prime \to \eta \gamma\gamma$ & $<1.33\times10^{-4}$; 90\% C.L. \cite{Ablikim:2019wsb} & $<1.5\times10^{-4}$ \\ \hline 
        $\eta^\prime \to \pi^+ \pi^- \pi^0$ & $2.52\pm0.07$\% \cite{Ablikim:2019wop} & $<2.66$\% \\ \hline
    \end{tabular}
    \caption{The observed branching fractions for radiative $\eta^{(\prime)}$ decays and the assumed upper limit on contributions from a gauge boson of $U(1)_B$, given at the $2\sigma$ level (98\% C.L.). The limit for $\eta \to \pi^0 \gamma \gamma$ is taken from the PDG average $\eta$ decay measurements; we note that the measurement from CrystalBall@AGS \cite{Prakhov:2008zz} was the basis of the analyses in Refs.~\cite{Tulin:2014tya, Gan:2020aco}. The observed branching fraction in the $\eta^\prime \to \pi^+ \pi^- \pi^0 \gamma$ row corresponds to $\eta^\prime \to \omega \gamma$.}
    \label{tab:decay_data}
\end{table}

On one hand, these new contributions increase the decay rates relative to pure-SM production. Conservative constraints can be derived by insisting that the new contribution not exceed the total observed value at some significance; this is what has been presented in Refs.~\cite{Berryman:2021wom, Tulin:2014tya, Gan:2020aco}. One could try to strengthen these constraints by comparing with theoretical predictions of the rates \cite{Ametller:1991dp, Bijnens:1995vg, Oset:2002sh, Oset:2008hp, Escribano:2018cwg} --- in fact, there is ostensibly a discrepancy between measurement and the state-of-the-art prediction for $\eta \to \pi^0 \gamma \gamma$ in Ref.~\cite{Escribano:2018cwg}. However, one should be circumspect in so doing: Ref.~\cite{Escribano:2018cwg} employs a combination of the VMD and linear $\sigma$ models to calculate decay widths under the assumption of isospin conservation with empirically-derived meson coupling constants. It is difficult to assess the systematic uncertainty incurred by these model choices; a dead-reckoning between theory and measurement may not be entirely reliable.

On the other hand, one can sidestep the issue of theoretical predictions and use a more robust experimental observable to constrain the existence of such a state. If $X$ is a narrow state, then its decays induce narrow features in the invariant-mass distributions of the meson decays --- in other words, bumps. It has been previously proposed to use bump hunts to probe new gauge states in Ref.~\cite{Bjorken:2009mm} in the context of electron and proton beam-dump experiments; the same principles apply here, but with slightly different experimental configurations. One might be able to conduct such a search with the upcoming JLab Eta Factory experiment \cite{JEFproposal, JEFproposal2} or with the REDTOP proposal \cite{Gatto:2016rae, Gonzalez:2017fku, Gatto:2016rae}. Moreover, $X$ can also be produced via $\gamma p \to p X$ at GlueX \cite{Adhikari:2020cvz}, where such a search is also possible \cite{Berryman:2021wom, Fanelli:2016utb}.

\begin{figure}[!t]
    \centering
    \includegraphics[width=0.7\linewidth]{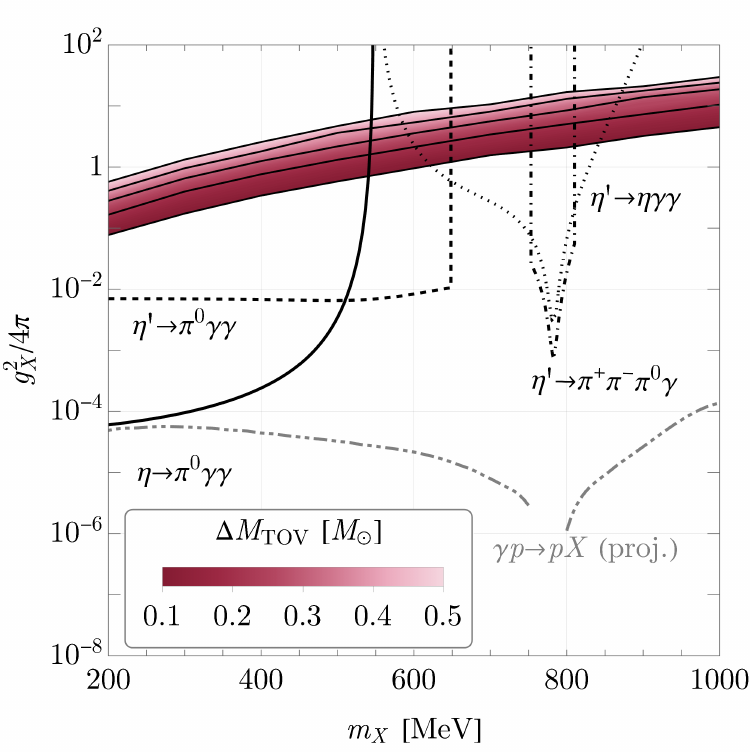}
    \caption{Similar to Fig.~\ref{fig:old_fig}, except that first-generation baryon number ($B_1$) has been gauged instead of total baryon number ($B$). The most important effect is to remove constraints from $\Upsilon(1S)$ and $\psi(1S)$ decays.}
    \label{fig:old_fig2}
\end{figure}

Also shown in Fig.~\ref{fig:old_fig} are constraints from hadronic decays of $\Upsilon(1S)$ \cite{Carone1995PhRvL..74.3122C, Carone1995PhRvD..52..484C, Aranda1998PhLB..443..352A} (dark orange) and of $\psi(1S)$ \cite{Aranda1998PhLB..443..352A} (light orange). These are sufficient to rule out new physics at the nuclear scale with strength comparable to electromagnetism ($g_X^2/4\pi \approx 1/137$). However, we note that one can circumvent these processes \cite{Berryman:2021wom} by insisting that the new physics only couples to first-generation baryon number, $B_1$. In so doing, one must rederive the radiative pseudoscalar meson constraints assuming no mixing to the strange quark; the results are shown in Fig.~\ref{fig:old_fig2}.

Apart from its tree-level couplings to quarks, the new vector state can kinetically mix with the SM photon. Limits on the kinetic mixing parameter $\varepsilon$ have been compiled in, e.g., Refs.~\cite{Ilten:2018crw, Caputo:2021eaa}. Usually, however, limits on $\varepsilon$ are derived from searches for minimal dark photons, in which the new vector only couples to the SM through this kinetic mixing. One must reinterpret these constraints with the tree-level couplings to quarks from the outset; Fig.~6 of Ref.~\cite{Ilten:2018crw} has recast these searches in terms of limits on $g_X$ and $m_X$. However, these limits assume that the kinetic mixing is given by $\varepsilon = e^2/(4\pi)^2$ --- otherwise, none of the constraints would be operative. We will not discuss these constraints in depth, but we note that, in the region 100 MeV $\lesssim m_X \lesssim$ 1 GeV, the kinetic mixing is most strongly probed by searches at LHCb for dimuon final states \cite{LHCb:2015nkv, LHCb:2017trq}.

We also note that baryon number is anomalous within the SM --- it is a symmetry of the \emph{Lagrangian}, but not of the corresponding \emph{action}. This is an acceptable state of affairs for global symmetries, but must be remedied for gauge symmetries by introducing additional fermions. From a model-building perspective, there is significant freedom in choosing how to resolve the anomalies, but in general, the existence of new fermions charged under baryon number (or some generalization thereof) can be probed at colliders \cite{Dobrescu:2014fca}. If these new fermions are heavier than the electroweak scale, then integrating them out of the theory at low energies leads to three-gauge-boson interaction terms, $X B B$, where $B$ is the gauge boson of hypercharge in this context \cite{Wess:1971yu, DHoker:1984izu, DHoker:1984mif, Anastasopoulos:2006cz, Dedes:2012me, Arcadi:2017jqd, Ismail:2017ulg}. These interactions enhance the emission of longitudinal $X$ in decays such as $Z \to X \gamma$ \cite{Dror:2017ehi, Dror:2017nsg, Michaels:2020fzj}. Aside from this, there are also terms involving the charged $W$s, $X W W$; these give rise to nonstandard flavor-changing neutral currents (FCNCs) such as $b\to s X$ at the quark level or $B \to K X$ at the hadron level \cite{Dror:2017ehi, Dror:2017nsg}. Limits from these anomalous decays, however, depend on $\varepsilon$, so we have not shown them in Figs.~\ref{fig:constraints_u1B}-\ref{fig:old_fig2}. We further note that FCNCs can also appear in models with generation-dependent couplings, such as the $U(1)_{B_1}$ model discussed above: if the three left-handed quark doublets are charged differently under the interaction, then $K-\overline{K}$ or $B-\overline{B}$ mixing contributions at odds with experimental constraints are induced. Ref.~\cite{Dobrescu:2014fca} estimates that in the absence of new fermions, the couplings should satisfy $g_X |z_{Q_3} - z_{Q_1}| \lesssim 10^{-5} (m_X/\text{GeV})$, where $z_{Q_1}$ ($z_{Q_3}$) is the charge of the first-generation (third-generation), left-handed quark doublet under the new interaction (in units of $g_X$). The only ways in which large couplings can exist without including additional fermions would be (1) to charge the third-generation quarks, thereby invoking the $\Upsilon(1S)$ constraint once again, or (2) to charge only the right-handed quarks, which are not constrained by FCNCs. However, in this latter scenario, the anomaly-cancellation conditions are sufficiently difficult to satisfy that it is all but required to invoke new fermions. The new states could, in principle, constitute some part of the dark matter; if so, then dark matter direct detection experiments predicated on \emph{nuclear} recoils might encounter an observable scattering rate.

We conclude this subsection by contrasting the scenario of gauged $U(1)_B$ symmetry with that of the related $U(1)_{B-L}$. Both symmetries are anomalous in the SM, but $U(1)_{B-L}$ can be easily rendered nonanomalous by introducing three right-handed neutrinos. This property has made $U(1)_{B-L}$ an attractive candidate for a gauge symmetry of nature, and it has been studied extensively as a result; see, e.g., Refs.~\cite{Ilten:2018crw, Harnik:2012ni, Bauer:2018onh} and references therein. Because $U(1)_{B-L}$ models couple to charged leptons at tree level, this incurs strong constraints from processes involving electrons and positrons, to which $U(1)_B$ models are not subject (in the absence of kinetic mixing). However, constraints derived for $U(1)_B$ from, e.g., fifth force searches will also apply to $U(1)_{B-L}$, albeit with some $\sim\mathcal{O}(1-10)$ differences; see, e.g., Fig.~5 of Ref.~\cite{Adelberger:2009zz}. To wit, the charge neutrality of matter implies that the numbers of protons and electrons must be equal; this implies that the $B$ charge of some laboratory probe cannot be less than its $B-L$ charge, though these are of the same order of magnitude. 

\subsection{Effects in Neutron Stars --- Heavy $X$}
\label{subsec:spontBNV_NS}

If new, repulsive contributions are incorporated into the $NN$ potential, then this \emph{stiffens} the nuclear EoS --- for a given (number) density of baryons, there is more energy density and pressure in a given fluid element than if these were not present. Conversely, attractive contributions \emph{soften} the EoS. Stiffening the EoS has two primary effects, for our purposes here:
\begin{enumerate}
    \item \emph{Neutron stars can be more massive.} The increase in the energy density of a given fluid element, relative to some nominal prediction, partially offsets some of the gravitational binding energy of the neutron star, resulting in a heavier star. The heaviest confirmed neutron star, PSR J0740+6620, was initially determined to have a mass $2.14^{+0.10}_{-0.09} \,M_{\astrosun}$ \cite{Cromartie:2019kug}, though this has since been refined to $2.08\pm0.07 \, M_{\astrosun}$ \cite{Fonseca:2021wxt}; any candidate EoS must be stiff enough to support neutron stars at least this heavy.
    \item \emph{Neutron stars have larger radii.} As the EoS is stiffened, the increase in pressure makes nuclear matter harder to compress. Therefore, a neutron star of a fixed gravitational mass will be physically larger with a stiffer EoS.
\end{enumerate}
The mass-radius relationship of neutron stars is therefore a powerful probe of the underlying EoS; see, e.g., Refs.~\cite{Landry:2020vaw, Legred:2021hdx, Landry:2018prl, Essick:2019ldf, Jiang:2019rcw, Raaijmakers:2019dks, Dietrich:2020efo, Raaijmakers:2021uju}. We note, in particular, the NICER mission, which can provide simultaneous estimates of a given neutron star's mass and radius \cite{Bogdanov:2019ixe, Bogdanov:2019qjb, Bogdanov:2021yip}. Measurements of this sort have been performed for PSR J0030+0451 \cite{Raaijmakers:2019qny, Miller:2019cac} and PSR J0740+6620 \cite{Miller:2021qha, Riley:2021pdl}; these indicate that the EoS is relatively stiff, supporting radii in the range 12-13 km over a wide range of possible masses (see, for instance, Fig.~11 of Ref.~\cite{Miller:2021qha}).

\begin{figure}[!t]
    \centering
    \includegraphics[width=0.7\linewidth]{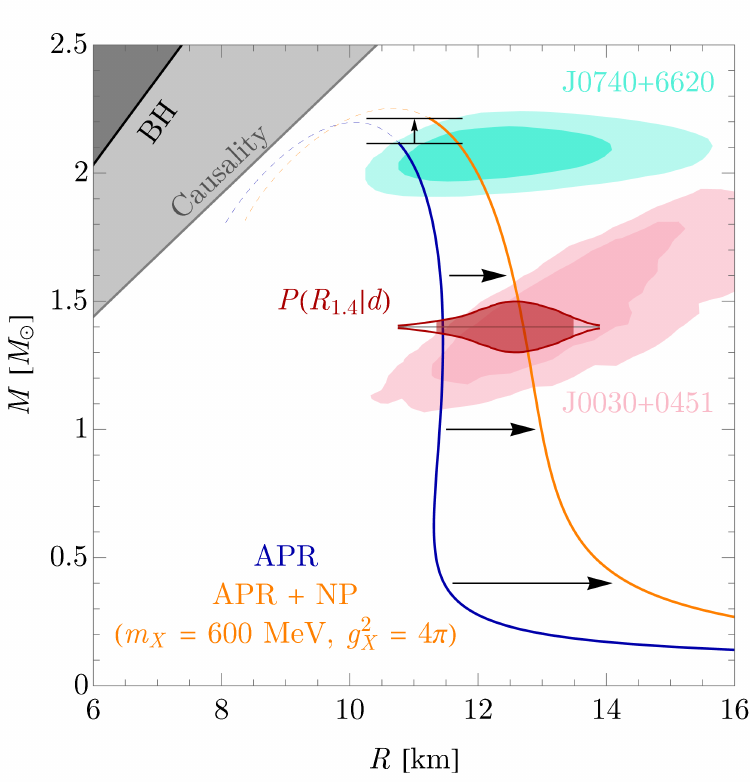}
	\caption{The neutron star mass-radius diagram. The blue line shows the predicted mass-radius relationship for the APR EoS \cite{Akmal:1998cf}; the orange line adds to this a new, repulsive interaction with $m_X = 600$ MeV and $g_X^2 = 4\pi$. The thin, dashed portion of either curve represents the points for which the sound speed in the core of the star exceeds $c$; these points are unphysical. The pink and cyan regions, respectively, represent inferences for J0030+0451 \cite{Miller:2019cac} and J0740+6620 \cite{Riley:2021pdl} from NICER and XMM-Newton; the dark (light) shading corresponds to 68\% (95\%) C.R. The red violin plot represents the posterior on the radius of a 1.4$M_{\astrosun}$ neutron star, calculated in Ref.~\cite{Legred:2021hdx}. The black curve represents the mass-radius relation for black holes (i.e., the Schwarzschild radius), while the gray curve represents a constraint from causality \cite{Koranda:1996jm}.}
    \label{fig:MR_plot}
\end{figure}

In Fig.~\ref{fig:MR_plot}, we show the neutron star mass-radius relationship for the  Akmal-Pandharipande-Ravenhall (APR) EoS \cite{Akmal:1998cf} -- a representative EoS with nucleonic degrees of freedom based on the Argonne $v_{18}$ two-body potential \cite{Wiringa:1994wb} and the Urbana IX three-body potential \cite{Carlson:1983kq}, including relativistic effects -- in blue. In orange, we have added a new vector interaction with coupling strength $g_X^2/4\pi = 1$ and mass $m_X = 600$ MeV. For each curve, the thin, dashed region denotes that the sound speed in the core of the star is greater than $c$; this is unphysical, and reflects that this EoS should not be applied for large central densities ($n \sim 5-10 \times n_\text{sat}$, where $n_\text{sat} \approx 0.16$ fm$^{-3}$ is the empirical nuclear saturation density). As described above, the mass-radius curve is shifted to larger radii, as reflected by the horizontal arrows, and the maximum neutron star mass is increased, as reflected by the vertical arrow. This figure also shows the 68\% (dark) and 95\% (light) credible regions (C.R.) for J0030+0451 \cite{Miller:2019cac} and J0470+6620 \cite{Riley:2021pdl} in cyan and pink shading, respectively. Moreover, the dark red violin plot depicts the posterior probability distribution \cite{Legred:2021hdx} for the radius of a $1.4M_{\astrosun}$ neutron star, $R_{1.4}$, conditioned on a combination of data (``$d$'') from (1) heavy pulsar masses, (2) the binary neutron star gravitational wave events GW170817 \cite{LIGOScientific:2017vwq} and GW190425 \cite{LIGOScientific:2020aai}, and (c) the same NICER observations of J0030+0451 and J0740+6620. The red shading represents the 90\% C.R.\footnote{The vertical extent of the violin plot does not reflect a mass constraint. The mass is \emph{fixed} to be $1.4M_{\astrosun}$, and the height of the curve is related to the (relative) likelihood of a given $R_{1.4}$. The reader can imagine that this curve extends into and out of the page or screen.} Taken together, these observations indicate that the neutron star EoS is required to be relatively stiff compared to the landscape of possible equations of state. The APR EoS that we have shown as an example is certainly compatible with existing data, but as more data become available in the coming decades from gravitational wave observatories, the nuclear EoS may end up being remarkably stiff. New repulsive interactions are one possible route to this outcome.

One effect not represented in Fig.~\ref{fig:MR_plot} is rotation: for a fixed baryonic mass, a rotating neutron star will have a larger gravitational mass than a nonrotating one \cite{Cook:1993qr, Lasota:1995eu, Breu:2016ufb}. For maximal uniform rotation, this effect may reach the level of $\sim20\%$ \cite{Breu:2016ufb}. Nonuniform rotation can lead to even larger enhancements \cite{Baumgarte:1999cq, Gondek-Rosinska:2016tzy, Weih:2017mcw}, but these configurations are not expected to be stable on long timescales. The fastest known NS spin is that of J1748-2446ad \cite{Hessels:2006ze, Grindlay:2006ky}, determined to be 716 Hz; the mass of this object is unknown, so it is unclear how close this object is to maximal rotation, and thus how large the contribution of its rotation to its total mass are. However, Ref.~\cite{Koliogiannis:2019rvh} has calculated how the mass-radius relationship for neutron stars with this rotational frequency differs from that of nonrotating stars; see Fig.~10 of that reference. The effect of rotation is to enhance the (equatorial) radius of the star, with the largest changes occurring for lower masses (around $\approx1M_{\astrosun}$). If the spin of the NS were unknown, then rotations can mimic the effect of an intrinsic stiffening of the EoS; this is a potentially important systematic in making extractions of the EoS.

It has long been suggested that new physics may be operative within neutron stars --- indeed, the appearance of a new, long-range force would be a realization of modified gravity \cite{Fujii:1971vv, Fujii:1989gg}. Some of the earliest work on explicitly determining the effects of a new interaction on neutron star structure comes from Ref.~\cite{Krivoruchenko2009PhRvD..79l5023K}. There, the authors determined that the new boson would modify the structure of neutron stars if the new coupling $g$ and boson mass $M$ satisfied
\begin{equation}
    \frac{g^2}{M^2} \gtrsim 25 \text{ GeV}^{-2}.
    \label{eq:howbig}
\end{equation}
The arguments of this paper are, however, approximate. In particular, this estimated limit arises from a comparison with the strength of the omega-exchange potential, $g_\omega^2/M_\omega^2\approx 400-500$ GeV$^{-2}$ \cite{Machleidt:1989tm}. Moreover, this treatment does not account for in-medium effects, which break the na\"ive dependence on $g^2/M^2$. Still, this work reflects the intuition that adding in a repulsive new interaction allows for larger, heavier neutron stars. Later work would show that, as expected, the precise modifications depend on the treatment of the baseline EoS considered; see, e.g., Refs.~\cite{Wen:2009av, Zhang:2011qj, Zheng:2011ya, Lin:2013rea, Zhang:2016tyb}.

Recently, Ref.~\cite{Berryman:2021wom} presented an analysis in which a new two-nucleon force was studied in pure neutron matter accounting for in-medium effects using Brueckner-Bethe-Goldstone theory (see, for instance, Refs.~\cite{HaftelThesis, Haftel:1970zz, HjorthJensen:1995ap} for details). In particular, this new interaction was studied in conjunction with the Argonne $v_{18}$ potential \cite{Wiringa:1994wb}; three-body and relativistic effects were studied through comparisons with the APR EoS \cite{Akmal:1998cf}. Within this framework, the shift in the maximum neutron star mass, $\Delta M_\text{TOV}$, was calculated as a function of the new boson mass $m_X$ and coupling $g_X$; the results are shown in Figs.~\ref{fig:old_fig} and \ref{fig:old_fig2} for the cases of gauged $B$ and $B_1$, respectively. Specifically, the red-pink-white band sweeps over contours of constant $\Delta M_\text{TOV}$, with the red side corresponding to $\Delta M_\text{TOV} = 0.1 M_{\astrosun}$ and the pink side to $\Delta M_\text{TOV} = 0.5 M_{\astrosun}$. As is evident from the figures, if the new boson has a mass in the range $\mathcal{O}(10^2-10^3)$ MeV, then the new coupling must be fairly strong, $g_X^2/4\pi \sim \mathcal{O}(0.1-10)$ to have a marked impact, though it can also be rather weaker than the strong force in this region. Nevertheless these findings are crudely compatible with the expectation shown in Eq.~\eqref{eq:howbig}. Most interestingly, if these effects were operative at this level, then they would contribute significantly to the rare meson decays we have discussed, allowing for a decisive test of this scenario.

We reiterate a point made previously. As of this writing, \emph{all usable nucleonic potentials are necessarily phenomenological} --- our knowledge of the nuclear force is fundamentally reliant on data. This state of affairs is necessary to make progress, but it is technically insufficient to add in a new interaction without considering how this changes the underlying parameters of the description. A useful analogy to this situation is muon decay. If one introduces a new interaction that contributes to $\mu^- \to e^- \nu_\mu \overline{\nu_e}$, then \emph{one cannot constrain this interaction through modifications to the rate.}\footnote{We emphasize that we are assuming this new interaction is subdominant to the SM.} This is because muon decays are used to \emph{define} the Fermi constant, $G_F$ --- this is one of a handful of parameters in the Standard Model that one must simply measure. Instead, one should probe consistency between a set of observables that all depend on $G_F$, even if these do not directly couple to the new interactions. Therefore, we should aspire to a similar scheme in the context of nucleon interactions: one must formally adjust the parameters of one's phenomenological prescription to accommodate the new interaction(s) and test for consistency among some subset of independent data. As such, it might be difficult for two- or many-nucleon interactions to definitively rule out the existence of new interactions --- but they can tell us where in parameter space to look.

Another issue on which we remark is the so-called masquerade problem \cite{Alford:2004pf}. In its original framing, this reflects the observation that hybrid stars, comprised of both nucleonic and quark degrees of freedom, may have a mass-radius relationship that can be emulated by a pure-nucleonic EoS. More broadly, a number of new phenomena may combine with the strong nuclear interactions in such a way that it is not possible to distinguish among them on the basis of the mass-radius relationship (though it may be possible to break this degeneracy using the $g$- modes of the NS \cite{Wei:2018tts, Jaikumar:2021jbw, Constantinou:2021hba}.) One of the conceptual advantages of introducing new interactions among nucleons, however, is that these can be directly probed in the laboratory. Indeed, we have seen that if new interactions are strong enough to meaningfully alter neutron star structure, then these can be emphatically probed with, e.g., rare meson decays. This is qualitatively different from, say, the presence of critical phenomena, such as a transition to hyperon matter or quark matter: these latter scenarios are extremely difficult to probe in the laboratory --- if not altogether impossible. Thus the \emph{falsifiability} of 
the former scheme is very much an asset, irrespective of whether the new interactions are connected to gauged $B$ or not.

We also remark that the presence of new interactions is not mutually exclusive with the appearance of other critical phenomena. Of particular interest is the so-called \emph{hyperon puzzle} (see, e.g., Refs.~\cite{Sedrakian:2021goo, Logoteta:2021iuy, Vidana:2021ppe}). In Sec.~\ref{sec:explicit BV}, we noted that for heavy neutron stars, the nucleon chemical potentials become large enough that it is energetically favorable to produce strange baryons ($\Lambda$, $\Sigma$, etc.). This significantly softens of the EoS to such an extent that, in the absence of new ingredients, EoSs with strange baryons can fail to support compact stars heavier than $\sim2 M_{\astrosun}$. It has been suggested that the appearance of hyperons can be pushed to higher densities and that their contribution to the EoS can be stiffened through (1) hyperon-hyperon repulsion, or (2) three-body forces involving hyperons (or both), among other possibilities; see Sec.~3.1 of Ref.~\cite{Vidana:2021ppe} for a review of the literature on these subjects.\footnote{We also note a recent study \cite{DelPopolo:2020pzh} in which the hyperon puzzle is addressed by the inclusion of dark matter within the neutron star core.} A new vector interaction between quarks may provide this additional stiffness. Additionally, the presence of the new interactions may modify the onset of certain critical phenomena by modifying, e.g., the relative energetics of the nucleonic and quark phases. This issue requires an in-depth treatment that is, to our knowledge, currently lacking in the literature.

\subsection{Effects in Neutron Stars --- (Ultra-)Light $X$}
\label{subsec:light_X}

If the new boson is light -- roughly below the MeV scale -- then it may be produced \emph{on-shell} in the neutron star. In this case, it may exist as a real state with finite energy within the neutron star, beyond simply contributing to the potential of nuclear matter. The effects of this new state on the long-term evolution of the neutron star depend on its coupling to the nucleons. If the couplings are too weak, then the new boson will be produced too infrequently to have a meaningful affect on the star, as, e.g., on its cooling. As the coupling strength is increased, the state may begin to overcool the star: enough bosons are produced to transport significant amounts of energy out of the core. This energy would then manifest as either
\begin{itemize}
    \item a flux of the new state, which one could hope to observe directly; or
    \item a flux of SM particles, produced via, e.g., decays of $X$ or conversion to photons in the star's magnetosphere \cite{Raffelt:1987im, Raffelt:1990yz, Fortin:2019npr}.
\end{itemize}
If, however, the new state is \emph{too} strongly coupled, then the new bosons are produced copiously but are trapped by the medium --- their mean-free path is too short to transport energy out of the star. In this case, there would be some accumulation of the new state within the nuclear medium. While these may modify the EoS, couplings in this regime do not lead to anomalous cooling. As an illustration of these ideas in a similar context, we show a constraint on the existence of a new gauge boson of $U(1)_B$ derived from observations of the neutrino signal of SN1987A from Ref.~\cite{Rrapaj:2015wgs} in orange in Fig.~\ref{fig:constraints_u1B}. Where the constraint is operative, the \emph{lower} bound represents the constraint from overcooling via $X$ emission: if too much energy had been radiated in the form of $X$, then the neutrino signal would have been substantially reduced. The \emph{upper} bound comes from trapping of $X$: if $X$ is sufficiently strongly coupled, then it is produced and reabsorbed in the explosion, so that it does not prevent neutrinos from transporting energy out of the supernova.

Constraints on anomalous neutron star cooling from new bosons have been derived for NS1987A (the remnant of SN1987A) \cite{Hong:2020bxo, Shin:2021bvz} and Cassiopeia A (Cas A) \cite{Hong:2020bxo}; Ref.~\cite{Lu:2021uec} presents constraints derived from SGR 0418+5729, Swift J1822.3-1606, and 1E 2259+586. The precise form of the constraints depends on the scenario considered; typically, these studies are framed in terms of either (1) a dark photon that mixes kinetically with the SM photon, or (2) a new gauge state for $U(1)_{B-L}$. The latter is closer to the scenario of gauged $B$ than the former, but we note that gauged $U(1)_{B-L}$ contains a tree-level coupling to the electron, which changes the constraint. Still, neutron stars are largely (though not entirely) insensitive to the differences between $B$ and $B-L$, so we expect constraints on the latter to be representative of constraints on the former. The constraints from NS1987A \cite{Shin:2021bvz} and Cas A \cite{Hong:2020bxo} are shown in red and dark red, respectively, in Fig.~\ref{fig:constraints_u1B}; we have omitted the constraints from Ref.~\cite{Lu:2021uec} because these depend on the kinetic mixing between the new gauge state and the SM photon, or are otherwise subdominant. Cas A is the stronger of the two, setting a limit of $g_X^2/4\pi \lesssim\mathcal{O}(10^{-26})$ for $m_X \lesssim 100$ keV.\footnote{Curiously, Ref.~\cite{Hong:2020bxo} reports a mild hint for anomalous cooling corresponding to a new gauge state with $m_X \approx 1$ eV with $g_X^2/4\pi \approx 3\times 10^{-27}$. This happens to coincide with the boundary of the constraint from fifth force searches \cite{Murata:2014nra}. We have shown the most conservative constraint from Ref.~\cite{Hong:2020bxo} here.} These energy-loss arguments are quite general and are not restricted to any particular source of BNV, nor to any particular astrophysical environment (though the form of the constraint is, of course, model dependent). To compare how these constraints fit in to the broader landscape of searches for new bosons, we refer the reader to, e.g., Refs.~\cite{Ilten:2018crw, Caputo:2021eaa, Harnik:2012ni, Knapen:2017xzo}.

\begin{figure}[!t]
    \centering
    \includegraphics[width=0.8\linewidth]{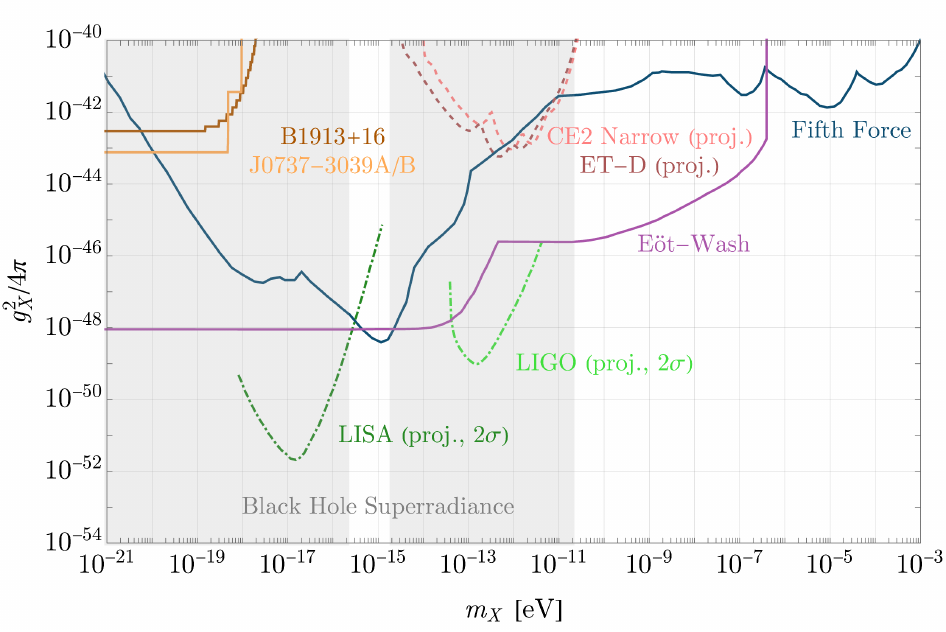}
    \caption{Constraints on an ultra-light new gauge boson. The dark blue line is the same set of fifth force constraints from Ref.~\cite{Murata:2014nra} as in Fig.~\ref{fig:constraints_u1B}, continued down to lower masses. The purple curve is the constraint from the E\"{o}t-Wash collaboration \cite{Schlamminger:2007ht}. The gray regions are excluded from nonobservations of black hole spin-down via superradiance \cite{Baryakhtar:2017ngi, Cardoso:2018tly, Stott:2020gjj, Ghosh:2021zuf} (see also Ref.~\cite{Caputo:2021eaa}). The dotted brown and pink curves represent the sensitivities of the third-generation gravitational wave observatories Einstein Telescope and Cosmic Explorer, respectively, to modifications to inspiral waveforms, adapted from Ref.~\cite{Alexander:2018qzg}. The dark and light orange contours represent constraints from anomalous energy loss via $X$ radiation from B1913+16 and J0737$-$3039A/B, respectively. The dot-dashed contours are projected sensitivities from LIGO (light green) and LISA (dark green) under the assumption that the gauge boson of $U(1)_B$ is the dark matter.}
    \label{fig:constraints2_u1B}
\end{figure}

If instead the new boson is ultra-light -- far below the eV scale -- then its interaction range may be of macroscopic size. In this mass regime, the constraints discussed above prevent the new interaction for making observable changes to the structure of the neutron star or to its cooling, but there is a mass regime in which the new state can mediate the interactions between astrophysical objects. Constraints on an ultra-light gauge boson are shown in Fig.~\ref{fig:constraints2_u1B}. The dark blue line represents the same set of constrains on a fifth force shown in Fig.~\ref{fig:constraints_u1B}, adapted from Ref.~\cite{Murata:2014nra}. To these, we have added the exclusion from the E\"ot-Wash torsion balance experiment from Ref.~\cite{Schlamminger:2007ht} in purple. The vertical gray bands represent constraints from \emph{black hole superradiance} \cite{Arvanitaki:2009fg, Brito:2015oca} --- ultra-light bosonic fields can condense in the region just outside of a black hole, thereby sapping it of its energy and angular momentum. Constraints from observations of astrophysical-scale black holes have been derived in Refs.~\cite{Baryakhtar:2017ngi, Cardoso:2018tly, Stott:2020gjj, Ghosh:2021zuf}; see also Ref.~\cite{Caputo:2021eaa}. We also note in passing that the conjecture that gravity should be the weakest force \cite{Arkani-Hamed:2006emk}, here loosely interpreted as $m_X \lesssim g_X M_\text{Planck}$, is satisfied throughout the parameter space that we have shown.

We have also included constraints from probes of nonstandard compact object interactions, on which we now elaborate. Firstly, if such a state exists, then it can induce modifications to the gravitational waveform of two merging compact objects. As we have discussed, a new vector state induces a repulsive interaction between nucleons. For ultra-light mediators, the contributions of the component baryons add coherently; neutron stars, having $B\sim\mathcal{O}(10^{57})$, will experience new interactions on macroscopic scales. These changes to the forces between neutron stars will manifest as apparent deviations from GR, inducing shifts to the gravitational waveform. To wit, a new vector interaction induces a long-range repulsion between neutron stars, thereby drawing out the merger over longer times. These shifts can be observable with gravitational wave interferometry \cite{Alexander:2018qzg, Sagunski:2017nzb, Croon:2017zcu, Choi:2018axi}. The dashed lines in Fig.~\ref{fig:constraints2_u1B} represent the projected sensitivities of third-generation gravitational wave observatories to non-Newtonian contributions to binary NS mergers, adapted from Ref.~\cite{Alexander:2018qzg}; brown is for the Einstein Telescope and pink is for the Cosmic Explorer. This reference explicitly concerns scenarios in which the non-Newtonian contribution arises from forces between dark matter cores, but we have reinterpreted their results in the context of gauged $U(1)_B$. While this mechanism would be an exquisite probe of dark forces, in the case of gauged baryon number, the parameter space of interest is already excluded at high significance, speaking to the exquisite sensitivity of the E{\"o}t-Wash measurements. 

Secondly, if the two compact objects have different ratios of $B/M$, then the orbiting stars constitute a time-varying dipole of baryon number. This results in radiation of $X$ as long as its mass is below the orbital frequency of the dipole \cite{Krause:1994ar}, and, as mentioned above, expediting the merger in addition to modifying the inspiral waveform. Interestingly, this effect is present even if the compact objects are not close to merging. In that case, this manifests as a nonstandard contribution to $\dot{P}_b/P_b$. We note Refs.~\cite{KumarPoddar:2019ceq, Dror:2019uea} as particularly interesting realizations of this idea: the authors have studied several such systems in order to constrain a boson associated with gauged $U(1)_{L_\mu-L_\tau}$, as opposed to $U(1)_B$. We have adapted the analyses performed in these works for the case of $U(1)_B$ for the Hulse-Taylor binary (B1913+16) and for J0737$-$3039A/B, previously discussed in Sec.~\ref{sec:BV in NS}; we briefly summarize our calculation below.

\begin{table}[b!]
    \centering
    \begin{tabular}{|c||c|c|} \hline
        Name & B1913+16 \cite{Weisberg:2016jye} & J0737$-$3039 \cite{PhysRevX.11.041050} \\ \hline\hline
        $M_1$ [$M_{\astrosun}$] & 1.438 & 1.3381 \\ \hline
        $M_2$ [$M_{\astrosun}$] & 1.392 & 1.2489 \\ \hline
        $B_1/M_1$ [$M_{\astrosun}^{-1}$] & $1.334\times 10^{57}$ & $1.322\times10^{57} $ \\ \hline
        $B_2/M_2$ [$M_{\astrosun}^{-1}$] & $1.328\times 10^{57} $ & $1.312\times10^{57} $ \\ \hline
        $\epsilon$ & 0.617 & 0.088 \\ \hline
    \end{tabular}
    \caption{The observables taken as input to calculate the contribution of ultra-light $X$ to binary spin-down. The values of $B/M$ have been calculated using the APR EoS \cite{Akmal:1998cf}. For reference, we note that $B/M \approx 1.19 \times 10^{57} M_{\astrosun}^{-1}$ for completely free nucleons; that the values in the table are larger than this is a reflection of the relatively large binding energy of matter in neutron stars.}
    \label{tab:more_binary_data}
\end{table}

The quantity of interest is the ratio of the time-averaged energy loss from $X$ emission, $\langle \dot E_X \rangle$, and that from GW emission, $\langle \dot E_\text{GW} \rangle$ \cite{Krause:1994ar, KumarPoddar:2019ceq, Dror:2019uea}:
\begin{eqnarray}
    & \dfrac{\langle \dot E_X \rangle}{\langle \dot E_\text{GW} \rangle} = \dfrac{5\pi}{12} \gamma \dfrac{g_X(m_X, \epsilon)}{g_\text{GR}(\epsilon)} \left( \dfrac{P_b}{2\pi G_N (M_1 + M_2)} \right)^{2/3}, & \\
    & \gamma \equiv \dfrac{g_X^2}{4\pi G_N} \left( \dfrac{B_1}{M_1} - \dfrac{B_2}{M_2} \right)^2, & \\
    \label{eq:gXsum}
    & g_X(m_X, \epsilon) \equiv \sum_{n>n_0} \left\{\left( \left[J^\prime_n(n\epsilon)\right]^2 + \dfrac{1-\epsilon^2}{\epsilon^2} \left[J_n(n\epsilon)\right]^2 \right) \sqrt{1-\dfrac{n_0^2}{n^2}} \left( 2n^2 + n_0^2 \right) \right\}, & \\
    & n_0 \equiv \dfrac{m_X P_b}{2\pi}, & \\
    &g_\text{GR}(\epsilon) = \dfrac{1 + (73/24)\epsilon^2 + (37/96)\epsilon^4}{(1-\epsilon^2)^{7/2}}, &
\end{eqnarray}
where $M_{1,2}$ are the masses of the compact objects, $B_{1,2}$ are their respective total baryon numbers, $\epsilon$ is the eccentricity of the binary, and $J_n(x)$ is the $n^\text{th}$-order Bessel function. These observables have been summarized in Table~\ref{tab:more_binary_data}; the $B/M$ ratios have been calculated using the APR EoS \cite{Akmal:1998cf}. We then compare against the ratio of the intrinsic and GR-predicted spin-down rates of the binaries,
\begin{equation}
    \frac{\dot P^\text{GR}_b}{\dot P^\text{int}_b} = 1 - \frac{\langle \dot E_X \rangle}{\langle \dot E_\text{GW} \rangle},
\end{equation}
using the values presented in Table~\ref{tab:psrbinary}. Results for B1913+16 and J0737$-$3039A/B are shown in dark orange and light orange, respectively, in Fig.~\ref{fig:constraints2_u1B};\footnote{The step-like features of these constraints are physical in origin. Because the binary orbits are eccentric, multipole radiation beyond dipole order would occur, with more eccentric orbits having proportionally more energy radiated at higher multipolarity; this is the meaning of the sum over $n$ in Eq.~\eqref{eq:gXsum}. The step features arise because lower-$n$ modes will be cut off as $m_X$, and thus $n_0$, increases, for a fixed $P_b$. If we included the uncertainties on the parameters of the binary systems, then these features would have been smeared out as in, e.g., Fig.~5 of Ref.~\cite{Dror:2019uea}. However, it is sufficient to demonstrate that binary spin-down constraints are not competitive with fifth force searches without including such refinements.} we have not included J1713+0747 as a part of this study, since its poorly constrained $\dot P^\text{GR}_b/\dot P^\text{int}_b$ ratio does not lend itself to a competitive constraint. Similar to the projected sensitivities to modifications to the inspiral waveform, these regions are entirely excluded by fifth force searches --- precision studies of binary spin-down are well suited for some classes of new mediators, but are seemingly overwhelmed by terrestrial experiments whenever tree-level couplings to $npe$ matter are present.

We note an interesting, complementary capability of the global GW observation program. If the new vector is ultra-light, then it may constitute some or all of the dark matter \cite{Hu:2000ke, Nelson:2011sf, Graham:2015rva, Agrawal:2018vin}, though its mass is required to be no smaller than $\mathcal{O}(10^{-22}-10^{-21})$ eV \cite{Hui:2016ltb, Schutz:2020jox, Benito:2020avv, Gardner:2021ntg} (see also Ref.~\cite{Hui:2021tkt} and references therein). Such a light dark matter candidate would coherently oscillate over long length scales, producing a nearly monochromatic, stochastic gravitational wave signature. Ref.~\cite{Pierce:2018xmy} has studied the sensitivity of LIGO and LISA to such a signature; Fig.~\ref{fig:constraints2_u1B} reproduces their findings for two years of observation, specific to the case of gauged $U(1)_B$, in light green and dark green, respectively. These curves are presented in dot-dashing to remind the reader that these sensitivities assume that the new state constitutes the entirety of the dark matter. We note analyses of LIGO O1 data performed in Ref.~\cite{Guo:2019ker,Miller:2020vsl}; the resulting exclusions are between one and two orders of magnitude weaker than the projection shown here. Similar projections have been made for other future GW observatories \cite{Michimura:2021hwr, Chen:2021apc} as well as pulsar timing arrays \cite{Nomura:2019cvc, Xue:2021xts}; see also Sec.~V of Ref.~\cite{Bertone:2019irm}.

\begin{figure}[t!]
    \centering
    \includegraphics[width=0.85\linewidth]{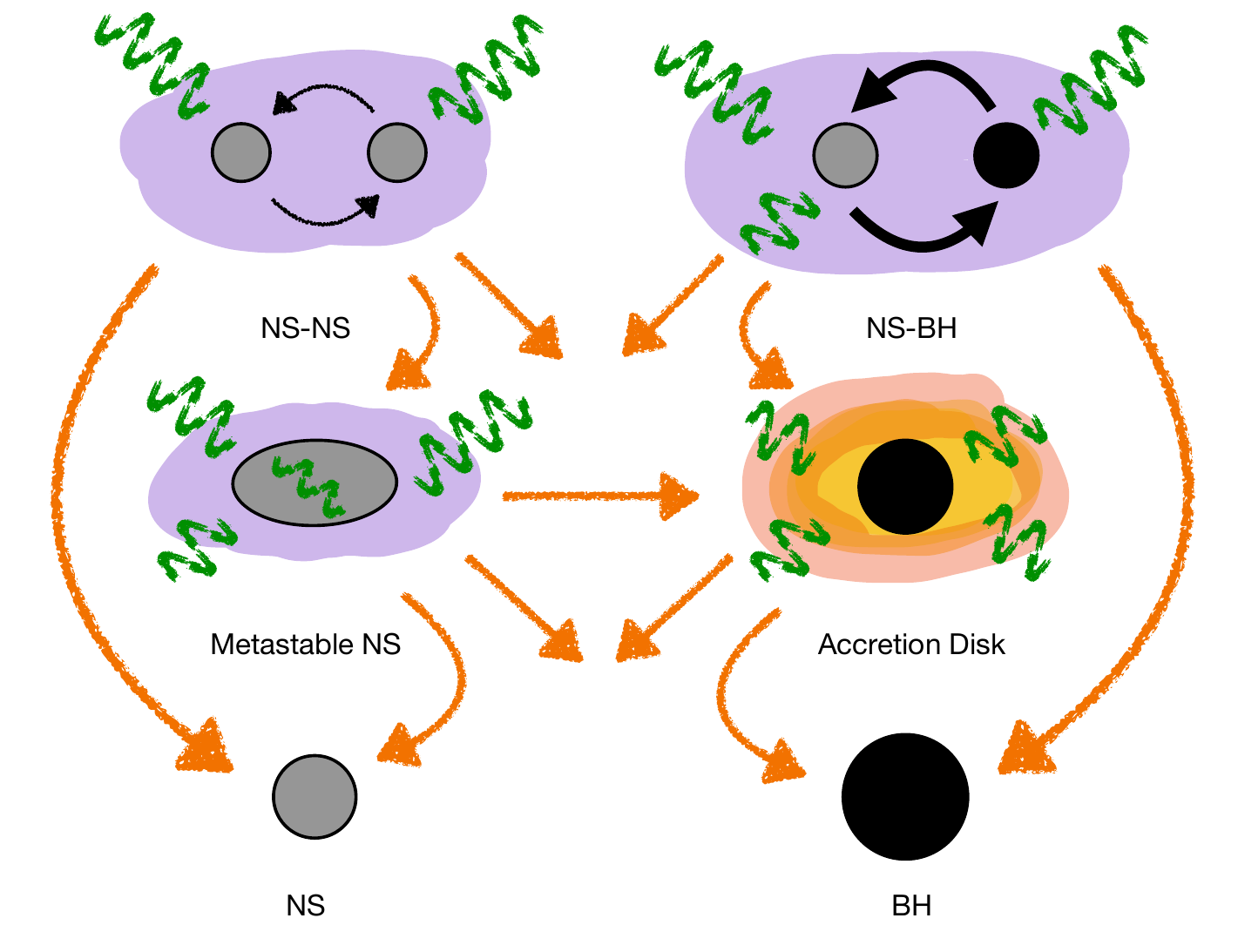}
    \caption{Evolutionary steps in a NS-NS or a NS-BH merger, after Ref.~\cite{Ascenzi:2020xqi}, with the possibility of $X$ emission, as indicated by the green squiggles. Ultra-light $X$ emission can appear throughout the merger event and can modify the inspiral wave form. In some portions of parameter space, $X$ can be trapped in the interior, leading to anomalous cooling or transport of SM particles. If $X$ stiffens the EoS of nuclear matter, then intermediate, metastable super-massive or hyper-massive neutron stars could potentially be heavier and longer-lived. These various modifications can modify the likelihood of the merger event following a particular evolutionary path and can also modify multi-messenger signals of the merger event at later times.}
    \label{fig:NS_inspiral}
\end{figure}

We conclude this section by synthesizing some of the possibilities we have discussed in Fig.~\ref{fig:NS_inspiral}, where we show a generic NS-NS and NS-BH merger. We have not yet commented on how $X$ might participate in such a merger; we briefly sketch this here. The green sinusoids represent the emission of $X$, which may occur (1) as a result of the time-varying $B$ dipole, either before or during the collision, or (2) through thermal processes operative in the hot remnant. If $X$ is not too strongly coupled, then either of these effects might lead to increased cooling rate of the remnant, or may lead to novel electromagnetic signatures. For instance, Ref.~\cite{Diamond:2021ekg} has recently presented a calculation of the rate of dark photon production and decay after merger; this produces a flash of gamma radiation potentially as energetic as $\sim\mathcal{O}(10^{46})$ ergs within the first second. More broadly, the new state may alter the cooling of the remnant in an observable way, or modify the kilonova signature in the hours, days and weeks after the event \cite{Metzger:2019zeh}. Moreover, if $X$ is in the regime in which it can modify the $NN$ potential, then the additional stiffness can modify the dynamics of the merger itself, or modify the intermediate state(s) that can appear in the merger. If the remnant is a super-massive or hyper-massive neutron star, then additional stiffness in the EoS may allow for heavier remnants to persist for longer times before collapsing to either a neutron star or black hole. There remains a significant amount of uncovered territory in understanding how these new states can manifest in extreme astrophysical environments; the age of multi-messenger astronomy promises a new avenue by which to study light, weakly coupled physics.

%%%%%%%%%%%%%%%%%%%%%%%%%%%%%%%%%%%%%%%%%%%%
\section{Other imprints of DM physics on NSs and their mergers}
\label{sec:DM}
%%%%%%%%%%%%%%%%%%%%%%%%%%%%%%%%%%%%%%%%%%%%

In this article we have delved into the ways different manifestations of BNV can impact the structure and evolution of a NS. Yet there are still broader ways in which dark sector and dense matter physics can intersect, and we offer a brief survey of the sweep of the possibilities here. These should largely be available to dark sector particles that carry baryon number, but could be possible even if they do not. Here we distinguish between DM and hidden-sector force mediators, where we assume for simplicity that such mediators are not sufficiently long-lived to be a component of DM. 

Generally, DM can be produced within the NS, or it can accumulate within the NS through capture onto the star~\cite{Goldman:1989nd}. We consider each scenario in turn. The DM production rate can be extremely slow, in which case chemical equilibrium is never attained, as studied in Ref.~\cite{Ellis:2018bkr}, or it can be fast enough so that DM reaches chemical equilibrium with baryonic matter. The models proposed for the neutron lifetime anomaly are of this latter class, and they are severely constrained by the existence of NSs. 

In the case that DM is captured on the NS, the exothermic nature of the reaction leads to DM thermalization and with the possibility of DM annihilation as well, can lead to significant heating of the star~\cite{Kouvaris:2007ay}. In such stars, if DM annihilation does not occur, or if DM carries an internal quantum number such as baryon number, then a DM core can form, and the possibility that this can induce collapse to a black hole serves as a constraint on the DM model, as studied in the case of bosonic Antisymmetric Dark Matter (ADM) models~\cite{Kouvaris:2011fi, McDermott:2011jp, Bell:2013xk}. If the DM is also self-interacting, then the collapse into a black hole can be avoided, and asymmetric dark stars can form~\cite{Kouvaris:2015rea}. In this later case, the capture and accretion of ordinary baryonic matter on these dark objects could presumably occur, forming an outer skin of baryonic matter.

If significant amounts of DM can be either produced or accumulated in the NS, then the structure of the NS can be modified. In the presence of a NS with a dark core, the star can be become more compact with a smaller $M_{\rm max}$, modifying the $M-R$ relationship~\cite{Ciarcelluti:2010ji,Ellis:2018bkr}. 
This scenario has also been discussed in the context of
the possibility of mirror neutron dark cores~\cite{Sandin:2008db,Berezhiani:2020zck,Yang:2021bpe}.
%Also, 
Moreover, new observables can appear in NS mergers, such as additional peaks in the postmerger frequency spectrum~\cite{Ellis:2017jgp}, and the tidal deformability ($\Lambda$) can decrease (increase) in the case of dark cores~\cite{Ellis:2018bkr} (halos~\cite{Nelson:2018xtr}); see also Ref.~\cite{Karkevandi:2021ygv}. Moreover, the DM content of a NS can impact its EoS, so that, e.g., an EoS that was ruled out from the GW observation of the upper bound on $\Lambda$ could be revived if a small admixture, say 5\%,  of DM were present. On the other hand, an assessment of the minimal value of $\Lambda$, as in $\Lambda_{\rm min}\approx 400$ from  Ref.~\cite{Radice:2017lry}, can also act as a constraint on the EoS with a dark core admixture. 

The existence of dark, or hidden, sector mediators can also impact the structure and evolution of a NS. The mediator can couple to quarks and modify the EoS, making it either stiffer, as we have described in Sec~\ref{subsec:spontBNV_NS}, or softer, with concomitant implications for the tidal deformability. The existence of the mediator can also impact the nature of the critical phenomena that may occur with increasing density. In addition to this, Ref.~\cite{Reddy:2021rln} has considered the possibility of a dark lepton condensate at the core of a NS, noting that this can modify the neutrino transport properties in and evolution of the star. 

%%%%%%%%%%%%%%%%%%%%%%%%%%%%%%%%%%%%%%%%%%%%
\section{Summary}
\label{sec:summary}
%%%%%%%%%%%%%%%%%%%%%%%%%%%%%%%%%%%%%%%%%%%%

The advent of the gravitational era has opened the nearby cosmos to us in new ways. We have considered the mechanisms by which BNV can exist and how such effects can combine with the physics of hidden sectors, whose inner workings are presumably key to the resolution of the dark-matter problem, or {\it not} to realize new ways of probing this physics through the study of neutron stars. Motivated by the neutron lifetime anomaly, we have observed that it is possible for apparent BNV to appear at rates not very much slower than $\sim 1$\% of the neutron lifetime and that these possibilities can be constrained through astrometric measurements. We emphasize that there is vast difference between the ``global'' BNV limits we have set in Eq.~(\ref{eq:BNV rate:binary limit}) from measurements of the decaying orbital period of binaries with at least one neutron star, presuming our assumptions of Sec.~\ref{sec:BV in NS} hold, and the far more stringent constraints that appear on single-nucleon or dinucleon decays through either direct or indirect searches. These disparate limits are compatible in that only a fraction of the ``star'' may be active in regards to BNV processes, as particular local densities may be required for them to occur. Nevertheless, we have concluded that BNV, both real and apparent, is somewhat slower than the effective weak scale within the environment of a neutron star, thus making our studies of the thermodynamics of neutron stars with BNV both viable and concrete. Consequently the observables whence we can realize new insights into BNV and dark sectors, which emerge from our study, are: 
\begin{itemize}
    \item In the presence of BNV, the distribution of neutron star masses to be found through gravitational wave studies can be expected to change with lookback time. We note that over the local volume available to us with present and next-generation gravitational wave detectors, the population of stars available to form neutron stars should differ little, making shifts in the mass distribution of the ensemble sensitive to the possibility of BNV effects, albeit likely apparent ones. 

    \item It is possible that BNV can produce unbearably light neutron stars, leading to explosions with detectable signatures in X-rays or soft gamma rays~\cite{1990SvA....34..595B}. This may be difficult to realize, however, as common mechanisms of neutron star formation favor roughly ${\cal O}(1\, M_{\astrosun})$ stars and as BNV may become inefficient, due to its possible density dependence, in the lightest mass neutron stars.

    \item We can hope to detect neutron stars of sub-solar mass. This may speak to new mechanisms for neutron star formation and possibly, too, to BNV.

    \item The possibility of compact objects that are bright in X-ray or neutrino emission may allow the detection of these objects individually, or more probably, through their additional contributions to the diffuse supernova neutrino background, which may soon be detected at Super-K~\cite{Beacom:2003nk}. This effect has also been suggested from considerations of binary-star interactions~\cite{Horiuchi:2020jnc}. 

    \item BNV processes with final states involving photons, mesons, and charged leptons (e.g., $n\to e^\mp\pi^\pm$) can be expected to dump all of their energy back into the NS, raising the temperature of the NS to a potentially detectable level, given upcoming observational possibilities, both in X-ray and the optical~\cite{national2022pathways}. Ground-based follow-up optical studies of targets of opportunity from gravitational wave observations, given their expected sensitivity~\cite{Andreoni:2021epw}, may also yield new surprises. Put more pithily, old neutron stars should be cold; if they are not, then this would really be quite a coup.
\end{itemize}

We look forward to future discoveries!

\bigskip

{\it Acknowledgements.} 
We thank our colleagues in the Network for Neutrinos, Nuclear Astrophysics, and Symmetries (\href{https://n3as.berkeley.edu}{N3AS}) for an inspiring environment. J.M.B. further thanks the Institute for Nuclear Theory at the University of Washington for its kind hospitality and stimulating research environment. J.M.B. acknowledges support from the National Science Foundation, Grant PHY-1630782, and the Heising-Simons Foundation, Grant 2017-228, and S.G. and M.Z. acknowledge partial support from the U.S. Department of Energy under contract DE-FG02-96ER40989. 

{%\footnotesize
\bibliography{Nstars_BNV_DM}}

\end{document}